\documentclass[a4paper,11pt]{article}
\usepackage{jheppub}

\usepackage{silence}
\usepackage[compatibility=false]{caption}

\usepackage{lineno}
\usepackage{indentfirst}
 \usepackage{lmodern}
 \usepackage{graphics}
 \usepackage{float}
 \usepackage{tikz-cd} 
\usepackage{tikz}
\tikzset{transition/.style={->,line width=1pt}}
\usepackage{url}
\usepackage{float}
\usepackage{standalone}
\usepackage{bm}
\usepackage{amssymb}
\newcommand{\id}{\mathbb{1}}
\usepackage{cleveref}
\usepackage{subcaption}\usepackage{longtable}
\usepackage{booktabs}
\usepackage{graphicx} 
\usepackage{array}
\usepackage{paralist}
\usepackage{amsmath}
\usepackage{physics}
\usepackage{float}
\usepackage{xcolor}
\usepackage{amsfonts}
\usepackage{amssymb}
\usepackage{verbatim}
 \usepackage{indentfirst}
\usepackage{tikz}
\usepackage{soul}
\usepackage{pifont}
\usepackage{forest}
\usepackage{tikz-cd} 
\tikzcdset{
	arrow style=tikz,
	diagrams={>={Straight Barb[scale=0.6]}},
    every arrow/.append style={line width=0.4pt}
}
\usetikzlibrary{calc}
\tikzset{curve/.style={settings={#1},to path={(\tikztostart)
    .. controls ($(\tikztostart)!\pv{pos}!(\tikztotarget)!\pv{height}!270:(\tikztotarget)$)
    and ($(\tikztostart)!1-\pv{pos}!(\tikztotarget)!\pv{height}!270:(\tikztotarget)$)
    .. (\tikztotarget)\tikztonodes}},
    settings/.code={\tikzset{quiver/.cd,#1}
        \def\pv##1{\pgfkeysvalueof{/tikz/quiver/##1}}},
    quiver/.cd,pos/.initial=0.25,height/.initial=0}
\tikzset{
	solid node/.style={circle,draw,inner sep=1.2,fill=black},
	hollow node/.style={circle,draw,inner sep=1.2},
	left label/.style={above left,midway},
	right label/.style={above right,midway}
}
\usetikzlibrary{decorations.markings}
\usetikzlibrary{arrows.meta}
\tikzset{->-/.style={decoration={
  markings,
  mark=at position #1 with {\arrow{>}}},postaction={decorate}}}

 \usepackage{tikz}
\usepackage{multirow} 
\usepackage{makecell}
\usepackage{longtable}

\def\id{\mathbf{1}}

\let\a=\alpha   \let\d=\delta 
  \let\th=\theta  \let\k=\kappa
\let\l=\lambda     \let\r=\rho
\let\s=\sigma \let\t=\tau    
    \let\D=\Delta  
  \let\S=\Sigma

\def\CA{{\cal A}}
\def\CB{{\cal B}}
\def\CC{{\cal C}}

\def\CN{{\cal N}}

\def\CZ{{\cal Z}}

\def\Wb{\widetilde{b}}

\def\Wq{\widetilde{q}}

\def\Wu{\widetilde{u}}
\def\Wv{\widetilde{v}}

\def\Wx{\widetilde{x}}
\def\Wy{\widetilde{y}}

\def\WB{\widetilde{B}}

\def\WN{\widetilde{N}}

\def\WP{\widetilde{P}}
\def\WQ{\widetilde{Q}}

\def\WX{\widetilde{X}}
\def\WY{\widetilde{Y}}

\title{Two-node Dualities from Elementary Dualities in 3d and 4d }

\author[a]{Yuanyuan Fang,}
\author[b]{Sungjoon Kim}
\affiliation[a]{Shing-Tung Yau Center and School of Physics, Southeast University, Nanjing 210096, China}
\affiliation[b]{Department of Physics, Korea Advanced Institute of Science and Technology, Daejeon 34141, Korea}
\emailAdd{yuanyuan\_fang@seu.edu.cn}
\emailAdd{sungjoon@kaist.ac.kr}

\abstract{
A new 3d $\mathcal{N}=2$ two-node Seiberg-like duality, analogous to the 4d $\mathcal{N}=1$ Intriligator--Leigh--Strassler~(ILS) duality, is proposed. As strong evidence, we check the agreement between the 3d $\mathcal{N}=2$ ellipsoid partition functions for numerous cases using s-confinement method. Surprisingly, our derivation assumes only two elementary 3d $\mathcal{N}=2$ dualities: Aharony and Benini--Benvenuti--Pasquetti, which reinforces the idea dualities from dualities. We also revisit the original ILS duality and provide a derivation based on nodewise dualities. In particular, we apply a sequence of Seiberg and Kutasov--Schwimmer dualities to individual gauge nodes, with Higgsing at an intermediate step, and subsequently integrate out the massive fields.}

\begin{document} 
\maketitle
\flushbottom
\section{Introduction}
\noindent
Dualities provide different perspectives on the same underlying physics, offering complementary ways to understand quantum field theories (QFTs). In particular, numerous examples have been discovered in supersymmetric theories~\cite{Seiberg:1994pq,Kutasov:1995ss,Kutasov:2014yqa,Bajc:2019vbp,Fang:2024nqy,Fang:2026ncq,Brodie:1996vx,Giveon:2008zn,Benini:2011mf,Kapustin:2011gh,Hwang:2011qt,Hwang:2011ht,Hanany:2015via,Benvenuti:2016wet,Amariti:2018gdc,Amariti:2018wht,Hwang:2018uyj,Hwang:2020ddr,Benvenuti:2024mpn,Benvenuti:2024seb,Benvenuti:2025huk,Benvenuti:2025qnq,Hayashi:2025guk,Okazaki:2021pnc,Okazaki:2021gkk,Cho:2026tqf,Benvenuti:2026usm,Amariti:2019rhc,Hwang:2022jjs,Jia:2025koz}. Seiberg duality is a pioneering example in four-dimensional $\CN=1$ SQCD, providing a non-Abelian generalization of the electric-magnetic duality~\cite{Seiberg:1994pq}.

In this paper, we revisit the 4d $\CN=1$ Intriligator--Leigh--Strassler (ILS) duality~\cite{Intriligator:1995ax}, a two-node extension of the Seiberg duality, and propose its 3d $\CN=2$ counterpart. We provide strong evidence for the proposed duality by matching the ellipsoid partition functions of the dual pair, extending the recently discussed special confining case~\cite{Amariti:2026pqh}. Our derivation uses the s-confinement of the 3d $\CN=2$ $\CC_p[U(N)]$ theory~\cite{Benvenuti:2024glr,Hwang:2024hhy} and successive applications of Aharony duality~\cite{Aharony:1997gp}, the 3d counterpart of 4d Seiberg duality. Notably, the s-confinement itself follows solely from Aharony and Benini--Benvenuti--Pasquetti (BBP) dualities, where the latter is a monopole superpotential deformation version of the former~\cite{Benini:2017dud}. Therefore, these two dualities serve as elementary building blocks, and our result reinforces the spirit of {\it dualities from elementary dualities}~\cite{Pasquetti:2019uop,Pasquetti:2019tix,Benvenuti:2020gvy,Benvenuti:2020wpc,Bottini:2022vpy,Amariti:2022iaz,Bottini:2021vms,Hwang:2021ulb,Comi:2022aqo,Comi:2026lcj}.

A common approach to constructing quiver gauge theory dualities in both 3d and 4d is to apply successive elementary dualities to individual gauge nodes. We refer to these as \textit{nodewise dualities}. In contrast, one can also transform multiple gauge groups at the same time \cite{Intriligator:1995ax,Brodie:1997sz,Brodie:1996vx,Fang:2026ncq}. Those dualities are referred to as {\it simultaneous dualities}. It is then natural to ask whether simultaneous dualities can also be derived from a sequence of elementary nodewise dualities.


In general, a key obstacle to successive nodewise dualizations is the generation of adjoint chirals at neighboring nodes through meson contributions, which obstruct further dualizations. 
This is where three dimensions become particularly useful: these adjoints can be replaced by linear quiver gauge theories $\CC_p[U(N)]$~\cite{Benvenuti:2024glr,Hwang:2024hhy}. In contrast, the 4d counterpart $D_p[SU(N)]$ is non-Lagrangian, though the same s-confinement has been observed~\cite{Cecotti:2013lda,Maruyoshi:2023mnv}.

Nevertheless, as a second part of the paper, we show that the original ILS duality can still be derived from nodewise Seiberg~\cite{Seiberg:1994pq} and Kutasov-Schwimmer (KS) dualities~\cite{Kutasov:1995ss} together with the Higgs mechanism induced by the F-term equations. The resulting Higgsing removes the mesonic adjoint contributions, allowing successive nodewise dualizations to reach the expected magnetic theory.

\medskip
\noindent{\bf Intriligator-Leigh-Strassler duality}.
We begin by reviewing the ILS duality. The gauge groups and superpotential of the electric side theory are 
\begin{equation}
 G_E=SU(N_1)\times SU(N_2),\qquad
 W_E=\Tr[(X\widetilde X)^{k+1}],
 \qquad k\geq1
 \,,
\end{equation}
while those of the magnetic theory are
\begin{align}
 G_M\!=SU(\widetilde N_1)\times SU(\widetilde N_2),\;\;\widetilde N_1=(k\!+\!1)F_1\!+\!kF_2\!-\!N_1,\widetilde N_2&=kF_1\!+\!(k\!+\!1)F_2\!-\!N_2,
\end{align}
\begin{align}
 W_M={}&\Tr[(x\widetilde x)^{k+1}]
 +\sum_{j=0}^{k}\left[
 M_{1,j}\widetilde q_1(x\widetilde x)^{k-j}q_1
 +M_{2,j}\widetilde q_2(\widetilde x x)^{k-j}q_2\right]
 \nonumber\\
 &+\sum_{j=0}^{k-1}\left[
 P_{12,j}\widetilde q_2(\widetilde x x)^{k-1-j}\widetilde xq_1
 +P_{21,j}\widetilde q_1(x\widetilde x)^{k-1-j}xq_2
 \right].
\label{eq: ILS WM}
\end{align}
For $i=1,2$, the electric fields $Q_i$ and $\widetilde Q_i$ transform as fundamentals of $SU(F_i)_L$ and $SU(F_i)_R$, respectively, while their magnetic counterparts $\widetilde q_i$ and $q_i$ transform as anti-fundamentals. Meson singlets $M_{1,j},M_{2,j},P_{12,j},P_{21,j}$ inherit the flavor representations of their electric composites, with flavor indices implicitly contracted in the superpotentials.  The field representations and charges are summarized in Table~\ref{tab:4d-ils-charge}.

The chiral ring elements of the duality are mapped as
\begin{equation}
\begin{alignedat}{2}
    \hphantom{\text{composite}}\Tr\big( (X\WX)^i \big)
    \;&\longleftrightarrow
    \Tr\big( (x\Wx)^i \big)&
    \\
    \text{composite}\;\left\{
    \begin{aligned}
        \WQ_1 (X\WX)^i Q_1 \\
        \WQ_2 (\WX X)^i Q_2 \\
        \WQ_1 X(\WX X)^j Q_2 \\
        \WQ_2 \WX (X\WX)^j Q_1
    \end{aligned}
    \right.
    \;&\longleftrightarrow\;
    \left.
    \begin{aligned}
        M_{1,i} \\
        M_{2,i} \\
        P_{12,j} \\
        P_{21,j}
    \end{aligned}
    \right\}\;\text{Singlet,}&
\end{alignedat}
\end{equation}
with
$i=0,\ldots,k$ and $j=0,\ldots,k-1$.
\begin{table}[tbp]
\centering
\begin{minipage}[t]{\linewidth}
\centering\vspace{0pt}
\scriptsize
\setlength{\tabcolsep}{2pt}
\renewcommand{\arraystretch}{1}
\begin{adjustbox}{max width=\linewidth}
\begin{tabular}{|c|c|c|c|c|}
\toprule
Field
& \tiny $SU(N_1)\!\times\!SU(N_2)$
& \tiny $SU(F_1)_L\!\times\!SU(F_1)_R\!\times\!SU(F_2)_L\!\times\!SU(F_2)_R$
& \tiny $U(1)_{B_1}\!\times\!U(1)_{B_2}\!\times\!U(1)_X$
& \tiny $U(1)_R$ \\
\hline
$X$ & $(\square,\overline{\square})$
& $(\mathbf1,\mathbf1,\mathbf1,\mathbf1)$
& $(\frac1{N_1},-\frac1{N_2},1)$ & $\Delta$ \\
$\widetilde X$ & $(\overline{\square},\square)$
& $(\mathbf1,\mathbf1,\mathbf1,\mathbf1)$
& $(-\frac1{N_1},\frac1{N_2},-1)$ & $\Delta$ \\
$Q_1$ & $(\square,\mathbf1)$
& $(\square,\mathbf1,\mathbf1,\mathbf1)$
& $(\frac1{N_1},0,0)$ & $r_1$ \\
$\widetilde Q_1$ & $(\overline{\square},\mathbf1)$
& $(\mathbf1,\square,\mathbf1,\mathbf1)$
& $(-\frac1{N_1},0,0)$ & $r_1$ \\
$Q_2$ & $(\mathbf1,\square)$
& $(\mathbf1,\mathbf1,\square,\mathbf1)$
& $(0,\frac1{N_2},0)$ & $r_2$ \\
$\widetilde Q_2$ & $(\mathbf1,\overline{\square})$
& $(\mathbf1,\mathbf1,\mathbf1,\square)$
& $(0,-\frac1{N_2},0)$ & $r_2$ \\
\hline\hline
Field
& \tiny $SU(\widetilde N_1)\!\times\!SU(\widetilde N_2)$
& \tiny $SU(F_1)_L\!\times\!SU(F_1)_R\!\times\!SU(F_2)_L\!\times\!SU(F_2)_R$
& \tiny $U(1)_{B_1}\!\times\!U(1)_{B_2}\!\times\!U(1)_X$
& \tiny $U(1)_R$ \\
\hline
$x$ & $(\square,\overline{\square})$
& $(\mathbf1,\mathbf1,\mathbf1,\mathbf1)$
& $(-\frac1{\widetilde N_1},\frac1{\widetilde N_2},1-\frac{kF_2}{\widetilde N_1}-\frac{kF_1}{\widetilde N_2})$ & $\Delta$ \\
$\widetilde x$ & $(\overline{\square},\square)$
& $(\mathbf1,\mathbf1,\mathbf1,\mathbf1)$
& $(\frac1{\widetilde N_1},-\frac1{\widetilde N_2},-(1-\frac{kF_2}{\widetilde N_1}-\frac{kF_1}{\widetilde N_2}))$ & $\Delta$ \\
$q_1$ & $(\square,\mathbf1)$
& $(\mathbf1,\overline{\square},\mathbf1,\mathbf1)$
& $(-\frac1{\widetilde N_1},0,-\frac{kF_2}{\widetilde N_1})$ & $\Delta-r_1$ \\
$\widetilde q_1$ & $(\overline{\square},\mathbf1)$
& $(\overline{\square},\mathbf1,\mathbf1,\mathbf1)$
& $(\frac1{\widetilde N_1},0,\frac{kF_2}{\widetilde N_1})$ & $\Delta-r_1$ \\
$q_2$ & $(\mathbf1,\square)$
& $(\mathbf1,\mathbf1,\mathbf1,\overline{\square})$
& $(0,-\frac1{\widetilde N_2},\frac{kF_1}{\widetilde N_2})$ & $\Delta-r_2$ \\
$\widetilde q_2$ & $(\mathbf1,\overline{\square})$
& $(\mathbf1,\mathbf1,\overline{\square},\mathbf1)$
& $(0,\frac1{\widetilde N_2},-\frac{kF_1}{\widetilde N_2})$ & $\Delta-r_2$ \\
\hline
$M_{1,i}$ & $(\mathbf1,\mathbf1)$
& $(\square,\square,\mathbf1,\mathbf1)$
& $(0,0,0)$ & $2r_1+2i\Delta$ \\
$M_{2,i}$ & $(\mathbf1,\mathbf1)$
& $(\mathbf1,\mathbf1,\square, \square)$
& $(0,0,0)$ & $2r_2+2i\Delta$ \\
$P_{12,j}$ & $(\mathbf1,\mathbf1)$
& $(\mathbf1,\square,\square,\mathbf1)$
& $(0,0,1)$ & $r_1+r_2+(2j+1)\Delta$ \\
$ P_{21,j}$ & $(\mathbf1,\mathbf1)$
& $(\square,\mathbf1,\mathbf1,\square)$
& $(0,0,-1)$ & $r_1+r_2+(2j+1)\Delta$ \\
\bottomrule
\end{tabular}
\end{adjustbox}
\end{minipage}
\caption{Representations and charges in the 4d $\mathcal N=1$ ILS duality.
The magnetic singlets correspond to electric mesons, with
$i=0,\ldots,k$ and $j=0,\ldots,k-1$.
The $R$-charges are the anomaly-free assignments compatible with the
superpotential. The variables are \(\Delta=\frac{1}{k+1},
r_1=1-\frac{N_1-k\Delta N_2}{F_1},
r_2=1-\frac{N_2-k\Delta N_1}{F_2}\).}
\label{tab:4d-ils-charge}
\vspace{-12pt}
\end{table}

\section{New 3d \texorpdfstring{$\CN=2$}{N=2} two-node duality}
\vspace{-5pt}
\noindent
In this section, we propose a new 3d $\CN=2$ Seiberg-like duality. We also provide a partition function derivation using identities from Aharony duality and s-confinement of $\CC_p[U(N)]$.

\vspace{-5pt}
\subsection{Duality proposal}
\vspace{-4pt}
\noindent
We propose a new 3d $\CN=2$ Seiberg-like duality, analogous to the 4d $\CN=1$ ILS duality~\cite{Intriligator:1995ax}, which can be presented as a pair of quivers:
\begin{equation}
\begin{tikzpicture}[thick,scale=0.8, every node/.style={scale=0.8}]
  \tikzset{vertex/.style={circle,fill=white!25,minimum size=12pt,inner sep=2pt}}
  \tikzset{every loop/.style={}}
    \node at (-2.2,2.2) { (e)};
    \node[vertex] (eN1) at (0,0) [shape=circle,draw=black,minimum size=2em] {$N_1$};
    \node[vertex] (eN2) at (2,0) [shape=circle,draw=black,minimum size=2em] {$N_2$};
    \node[vertex] (eNf1t) at (-1.5,1.3) [shape=rectangle,draw=black,minimum height=2em, minimum width=2em] {$F_1$};
    \node[vertex] (eNf1b) at (-1.5,-1.3) [shape=rectangle,draw=black,minimum height=2em, minimum width=2em] {$F_1$};
    \node[vertex] (eNf2t) at (3.5,1.3) [shape=rectangle,draw=black,minimum height=2em, minimum width=2em] {$F_2$};
    \node[vertex] (eNf2b) at (3.5,-1.3) [shape=rectangle,draw=black,minimum height=2em, minimum width=2em] {$F_2$};

    \draw[->-=.5] ([yshift= 2pt] eN1.east) to ([yshift= 2pt] eN2.west);
    \draw[->-=.5] ([yshift= -2pt] eN2.west) to ([yshift= -2pt] eN1.east);
    \draw[->-=.5] ([yshift= 2pt] eNf1b.east) to ([xshift= 5pt,yshift= -10pt] eN1.west);
    \draw[->-=.5] ([xshift= 5pt,yshift= 10pt] eN1.west) to ([yshift= -2pt] eNf1t.east);
    \draw[->-=.5] ([xshift= -5pt,yshift= -10pt] eN2.east) to ([yshift= 2pt] eNf2b.west);
    \draw[->-=.5] ([yshift= -2pt] eNf2t.west) to ([xshift= -5pt,yshift= 10pt] eN2.east);

    \node at (1,0.5) {\scriptsize $X$};
    \node at (1,-0.5) {\scriptsize $\WX$};
    \node at (-0.4,1) {\scriptsize $Q_1$};
    \node at (-0.4,-1) {\scriptsize $\WQ_1$};
    \node at (2.4,-1) {\scriptsize $Q_2$};
    \node at (2.4,1) {\scriptsize $\WQ_2$};
    \node at (1,-2.5) {$W_{\rm e}$};
    
    \node at (5.5,0.2) {$\overset{\text{Dual}}{\Longleftrightarrow}$};

    \node at (9-2.2,2.2) { (m)};
    \node[vertex] (mN1) at (9,0) [shape=circle,draw=black,minimum size=2em] {\!$\WN_1$\!};
    \node[vertex] (mN2) at (11,0) [shape=circle,draw=black,minimum size=2em] {\!$\WN_2$\!};
    \node[vertex] (mNf1t) at (7.5,1.3) [shape=rectangle,draw=black,minimum height=2em, minimum width=2em] {$F_1$};
    \node[vertex] (mNf1b) at (7.5,-1.3) [shape=rectangle,draw=black,minimum height=2em, minimum width=2em] {$F_1$};
    \node[vertex] (mNf2t) at (12.5,1.3) [shape=rectangle,draw=black,minimum height=2em, minimum width=2em] {$F_2$};
    \node[vertex] (mNf2b) at (12.5,-1.3) [shape=rectangle,draw=black,minimum height=2em, minimum width=2em] {$F_2$};

    \draw[->-=.5] ([yshift= 2pt] mN1.east) to ([yshift= 2pt] mN2.west);
    \draw[->-=.5] ([yshift= -2pt] mN2.west) to ([yshift= -2pt] mN1.east);
    \draw[->-=.5] ([xshift= 5pt,yshift= -10pt] mN1.west) to ([yshift= 2pt] mNf1b.east);
    \draw[->-=.5] ([yshift= -2pt] mNf1t.east) to ([xshift= 5pt,yshift= 10pt] mN1.west);
    \draw[->-=.5] ([yshift= 2pt] mNf2b.west) to ([xshift= -5pt,yshift= -10pt] mN2.east);
    \draw[->-=.5] ([xshift= -5pt,yshift= 10pt] mN2.east) to ([yshift= -2pt] mNf2t.west);
    \draw[dashed,->-=.5] (mNf1b.north) to (mNf1t.south);
    \draw[dashed,->-=.5] (mNf2t.south) to (mNf2b.north);
    \draw[dashed,->-=.5] ([yshift= 4pt] mNf2t.west) to ([yshift= 4pt] mNf1t.east);
    \draw[dashed,->-=.5] ([yshift= -4pt] mNf1b.east) to ([yshift= -4pt] mNf2b.west);

    \node at (10,0.5) {\scriptsize $x$};
    \node at (10,-0.5) {\scriptsize $\Wx$};
    \node at (9-0.4,1) {\scriptsize $\Wq_1$};
    \node at (9-0.4,-1) {\scriptsize $q_1$};
    \node at (11.4,-1) {\scriptsize $\Wq_2$};
    \node at (11.4,1) {\scriptsize $q_2$};
    \node at (10,1.8) {\scriptsize $P_{21,j}$};
    \node at (10,-1.8) {\scriptsize $P_{12,j}$};
    \node at (7,0) {\scriptsize $M_{1,j}$};
    \node at (13,0) {\scriptsize $M_{2,j}$};

    \node at (10,-2.5) {$W_{\rm m}$};

\end{tikzpicture}
    \label{eq: new dual}
\end{equation}
with dual gauge ranks
\begin{align}
    \WN_1 = (k+1)F_1 + k F_2 - N_1
    \;,\;\;\;
    \WN_2 = k F_1 + (k+1)F_2 - N_2
    \,,
\end{align}
and superpotentials
\begin{align}
    & W_{\rm e} = 
    \text{Tr} (\WX X)^{k+1}
    \,,
    \nonumber\\
    & W_{\rm m} =
    \text{Tr} (\Wx x)^{k+1}
    +
    \sum_{j=0}^{k-1}
    \Big(
    P_{12,j} \Wq_2 \Wx (x \Wx)^{k\texttt{-}j\texttt{-}1} q_1
    +
    P_{21,j} \Wq_1 x (\Wx x)^{k\texttt{-}j\texttt{-}1} q_2
    \Big)
    \nonumber\\
    &\quad
    +
    \sum_{j=0}^k\!
    \Big(\!
    M_{1,j} \Wq_1 (x \Wx)^{k\texttt{-}j} \!q_1
    +
    M_{2,j} \Wq_2 (\Wx x)^{k\texttt{-}j} q_2
    \!\Big)
    \!+\!
    \sum_{s=\pm}\!
    \Big(\!
    V_{s0} \hat v_{s0}
    +
    V_{0s} \hat v_{0s}
    +
    \sum_{j=0}^{k-1}
    V_{ss}^{(j)} \hat v_{ss}^{(k\texttt{-}j\texttt{-}1)}
    \Big)
    \,,
\end{align}
where $(\WQ_1,Q_1)$ and $(\WQ_2,Q_2)$ denote $F_1$ and $F_2$ pairs of flavor chirals, respectively, and $(X,\WX)$ are bi-fundamentals. The magnetic theory similarly contains flavor chirals $(\Wq_1,q_1)$ and $(\Wq_2,q_2)$, bi-fundamentals $(x,\Wx)$, and singlets $(P_{12,j},P_{21,j},M_{1,j},M_{2,j})$ charged under the flavor symmetries as collectively denoted by dashed arrows.

In the electric theory, we denote monopole operators with magnetic charges $(s,0)$, $(0,s)$, and $(s,s)$ under $U(1)_{T_1}\times U(1)_{T_2}$ by $\hat V_{s0}$, $\hat V_{0s}$, and $\hat V_{ss}^{(j)}$, respectively, where the last one is dressed by $(X\WX)^j$:
\begin{align}
    \hat V_{ss}^{(j)}
    :=
    \text{tr}
    \big(
    (X\WX)^j \hat V_{ss}^{(0)}
    \big)
    \,.
\end{align}
The trace $\text{tr}(\cdots)$ is over the unbroken gauge groups in the monopole background. The operators $\hat V_{s0}$ and $\hat V_{0s}$ admit no independent bi-fundamental dressings, as the flux-carrying components of $X$ and $\widetilde X$ become massive. The dual magnetic monopoles are denoted by $\hat v_{s0}$, $\hat v_{0s}$, and $\hat v_{ss}^{(j)}$, coupled to singlets $V_{s0}$, $V_{0s}$, and $V_{ss}^{(j)}$ mapped from their electric counterparts. We adopt opposite sign convention for the electric and dual magnetic $U(1)_{T_1}\times U(1)_{T_2}$ charges to ensure neutral singlet-monopole couplings.

\noindent{\bf Chiral ring map}.
The chiral ring elements of the duality are mapped as
\begin{align}
    \text{Tr}\big( (X \WX)^i \big)
    \quad &\longleftrightarrow \quad
    \text{Tr} \big( (x\Wx)^i \big)
    \nonumber\\
    \WQ_1 (X\WX)^i Q_1
    \quad &\longleftrightarrow \quad
    M_{1,i}
    \nonumber\\
    \WQ_2 (\WX X)^i Q_2
    \quad &\longleftrightarrow \quad
    M_{2,i}
    \nonumber\\
    \WQ_1 X(\WX X)^j Q_2
    \quad &\longleftrightarrow \quad
    P_{12,j}
    \nonumber\\
    \WQ_2 \WX (X\WX)^j Q_1
    \quad &\longleftrightarrow \quad
    P_{21,j}
    \nonumber\\
    \hat V_{s0}
    \quad &\longleftrightarrow \quad
    V_{s0}
    \nonumber\\
    \hat V_{0s}
    \quad &\longleftrightarrow \quad
    V_{0s}
    \nonumber\\
    \hat V_{ss}^{(j)}
    \quad &\longleftrightarrow \quad
    V_{ss}^{(j)}
    \,,
\end{align}
where $i=0,\cdots,k$, and $j=0,\cdots,k-1$, while $s = \pm$. We conjecture the $k$-th dressed monopole is a composite by quantum relation
\begin{align}
    \hat V_{ss}^{(k)} \sim \hat V_{s0}\hat V_{0s}
    \,.
\end{align}
The global charges of both sides agree. This relation implements our proposed tower of dressed monopole operators and we will illustrate it shortly via the singlet spectrum in an example. Further verification, for instance through superconformal index computations, remains to be done.

\noindent{\bf Charges}.
Global symmetry charges of the elementary fields are summarized in Table \ref{tab: 3d charge}.  $R$-charges of the monopole operators are computed by counting fermionic zero modes:
\begin{align}
    &R[\hat V_{s0}] = R[V_{s0}] = F_1(1-r_1) + k \D N_2 - N_1 + 1
    \,,
    \nonumber\\
    &R[\hat V_{0s}] = R[V_{0s}] = F_2(1-r_2) + k \D N_1 - N_2 + 1
    \,,
    \nonumber\\
    &R[\hat V_{ss}^{(j)}] = R[V_{ss}^{(j)}] = F_1(1-r_1) + F_2(1-r_2) -\D(N_1+N_2-2) + 2\D j
    \,,
    \label{eq: R[V]}
\end{align}
\begin{align}
    &R[\hat v_{s0}] = F_1(1-\D+r_1) + k \D \WN_2 - \WN_1 + 1
    \,,
    \nonumber\\
    &R[ \hat v_{0s}] = F_2(1-\D+r_2) + k \D \WN_1 - \WN_2 + 1
    \,,
    \nonumber\\
    &R[ \hat v_{ss}^{(j)}] = F_1(1-\D+r_1) + F_2(1-\D+r_2) -\D(\WN_1+\WN_2-2) + 2\D j
    \,.
    \label{eq: R[v]}
\end{align}
\begin{table}[tbp]
\centering
\begin{minipage}[t]{1\linewidth}
\centering\vspace{0pt}
\scriptsize\setlength{\tabcolsep}{1.5pt}\renewcommand{\arraystretch}{1.1}
\begin{adjustbox}{max width=\linewidth}
\begin{tabular}{|c|c|c|c|c|}\toprule
Field 
& \tiny $U(N_1)\!\!\times\!\! U(N_2)$ 
& \tiny
$SU(F_1)_{\!L} \!\!\times\!\! SU(F_1)_{\!R} \!\!\times\!\! SU(F_2)_{\!L} \!\!\times\!\! SU(F_2)_{\!R} $
& \tiny
$U(1)_{\!A_1} \!\!\times\!\! U(1)_{\!A_2} \!\!\times\!\! U(1)_{\!X} \!\!\times\!\! U(1)_{\!T_1} \!\!\times\!\! U(1)_{\!T_2}$ 
&\tiny
$U(1)_R$
\\
\hline
$X$ & $(\square,\overline{\square})$ 
& $(\mathbf1,\mathbf1,\mathbf1,\mathbf1)$ 
& $(0,0,1,0,0)$ & $\D$ 
\\
$\WX$ & $(\overline{\square},\square)$ 
& $(\mathbf1,\mathbf1,\mathbf1,\mathbf1)$ 
& $(0,0,-1,0,0)$ & $\D$ 
\\
$Q_1$ & $(\square,\mathbf1)$ 
& $( \square,\mathbf1,\mathbf1,\mathbf1)$ 
& $(1,0,0,0,0)$ & $r_1$ 
\\
$\WQ_1$ & $(\overline{\square},\mathbf1)$ 
& $(\mathbf1,\square,\mathbf1,\mathbf1)$ 
& $(1,0,0,0,0)$ & $r_1$ 
\\
$Q_2$ & $(\mathbf1,\square)$ 
& $(\mathbf1,\mathbf1,\square,\mathbf1)$ 
& $(0,1,0,0,0)$ & $r_2$ 
\\
$\WQ_2$ & $(\mathbf1,\overline{\square})$ 
& $(\mathbf1,\mathbf1,\mathbf1,\square)$ 
& $(0,1,0,0,0)$ & $r_2$ 
\\
\hline
$\hat V_{s0}$ & $(\mathbf1,\mathbf1)$ 
& $(\mathbf1,\mathbf1,\mathbf1,\mathbf1)$ 
& $(-F_1,0,0,s,0)$ &  
\\
$\hat V_{0s}$ & $''$ 
& $''$ 
& $(0,-F_2,0,0,s)$ &  \eqref{eq: R[V]}
\\
$\hat V_{ss}^{(j)}$ & $''$
& $''$
& $(-F_1,-F_2,0,s,s)$ &  
\\
\hline\hline
Field 
& \tiny $U(\WN_1)\!\!\times\!\! U(\WN_2)$ 
& \tiny
$SU(F_1)_{\!L} \!\!\times\!\! SU(F_1)_{\!R} \!\!\times\!\! SU(F_2)_{\!L} \!\!\times\!\! SU(F_2)_{\!R} $
& \tiny
$U(1)_{\!A_1} \!\!\times\!\! U(1)_{\!A_2} \!\!\times\!\! U(1)_{\!X} \!\!\times\!\! U(1)_{\!T_1} \!\!\times\!\! U(1)_{\!T_2}$ 
&\tiny
$U(1)_R$
\\
\hline
$x$ & $(\square,\overline{\square})$ 
& $(\mathbf1,\mathbf1,\mathbf1,\mathbf1)$ 
& $(0,0,1,0,0)$ & $\D$ 
\\
$\Wx$ & $(\overline{\square},\square)$ 
& $(\mathbf1,\mathbf1,\mathbf1,\mathbf1)$ 
& $(0,0,-1,0,0)$ & $\D$ 
\\
$q_1$ & $(\square,\mathbf1)$ 
& $(\mathbf1,\overline{\square},\mathbf1,\mathbf1)$ 
& $(-1,0,0,0,0)$ & $\D-r_1$ 
\\
$\Wq_1$ & $(\overline{\square},\mathbf1)$ 
& $(\overline{\square},\mathbf1,\mathbf1,\mathbf1)$ 
& $(-1,0,0,0,0)$ & $\D-r_1$ 
\\
$q_2$ & $(\mathbf1,\square)$ 
& $(\mathbf1,\mathbf1,\mathbf1,\overline{\square})$ 
& $(0,-1,0,0,0)$ & $\D-r_2$ 
\\
$\Wq_2$ & $(\mathbf1,\overline{\square})$ 
& $(\mathbf1,\mathbf1,\overline{\square},\mathbf1)$ 
& $(0,-1,0,0,0)$ & $\D-r_2$ 
\\
\hline
$\hat v_{s0}$ & $(\mathbf1,\mathbf1)$ 
& $(\mathbf1,\mathbf1,\mathbf1,\mathbf1)$ 
& $(F_1,0,0,-s,0)$ &  
\\
$\hat v_{0s}$ & $''$ 
& $''$ 
& $(0,F_2,0,0,-s)$ &  \eqref{eq: R[v]}
\\
$\hat v_{ss}^{(j)}$ & $''$
& $''$
& $(F_1,F_2,0,-s,-s)$ &  
\\
\bottomrule\end{tabular}
\end{adjustbox}
\end{minipage}
\caption{Charge table of 3d $\CN=2$ two-node duality.  $R$-charges of the monopole operators are given in \eqref{eq: R[V]} and \eqref{eq: R[v]}. }
\label{tab: 3d charge}
\end{table}

\subsection{Low \texorpdfstring{$k$}{k} cases}
\noindent
As a first duality check, let us consider the cases with small $k$. For $k=0$, all the bi-fundamental fields $X$, $\WX$, $x$, and $\Wx$ become massive and can be integrated out. The resulting dual gauge ranks and superpotentials are then given by
\begin{align}
    &\qquad\qquad\qquad \WN_1 = F_1 - N_1
    \;,\;\;\;
    \WN_2 = F_2 - N_2
    \,,
    \nonumber\\
    &W_{\rm e} = 0
    \;,\;\;
    W_{\rm m} = M_{1,0} \Wq_1 q_1 + M_{2,0} \Wq_2 q_2
    +
    \sum_{s=\pm} \Big( V_{s0} \hat v_{s0} + V_{0s} \hat v_{0s} \Big)
    \,,
\end{align}
which reproduce a pair of Aharony dualities.

\begin{figure}[tbp]
\centering
\begin{tikzpicture}[thick,scale=0.7, every node/.style={scale=0.7}]
  \tikzset{vertex/.style={circle,fill=white!25,minimum size=12pt,inner sep=2pt}}
  \tikzset{every loop/.style={}}
    \node at (-2.2,2.2) { (e)};
    \node[vertex] (eN1) at (0,0) [shape=circle,draw=green,minimum size=2em] {$N_1$};
    \node[vertex] (eN2) at (2,0) [shape=circle,draw=black,minimum size=2em] {$N_2$};
    \node[vertex] (eNf1t) at (-1.5,1.3) [shape=rectangle,draw=black,minimum height=2em, minimum width=2em] {$F_1$};
    \node[vertex] (eNf1b) at (-1.5,-1.3) [shape=rectangle,draw=black,minimum height=2em, minimum width=2em] {$F_1$};
    \node[vertex] (eNf2t) at (3.5,1.3) [shape=rectangle,draw=black,minimum height=2em, minimum width=2em] {$F_2$};
    \node[vertex] (eNf2b) at (3.5,-1.3) [shape=rectangle,draw=black,minimum height=2em, minimum width=2em] {$F_2$};

    \draw[->-=.5] ([yshift= 2pt] eN1.east) to ([yshift= 2pt] eN2.west);
    \draw[->-=.5] ([yshift= -2pt] eN2.west) to ([yshift= -2pt] eN1.east);
    \draw[->-=.5] ([yshift= 2pt] eNf1b.east) to ([xshift= 5pt,yshift= -10pt] eN1.west);
    \draw[->-=.5] ([xshift= 5pt,yshift= 10pt] eN1.west) to ([yshift= -2pt] eNf1t.east);
    \draw[->-=.5] ([xshift= -5pt,yshift= -10pt] eN2.east) to ([yshift= 2pt] eNf2b.west);
    \draw[->-=.5] ([yshift= -2pt] eNf2t.west) to ([xshift= -5pt,yshift= 10pt] eN2.east);

    \node at (1,0.5) {\scriptsize $X$};
    \node at (1,-0.5) {\scriptsize $\WX$};
    \node at (-0.4,1) {\scriptsize $Q_1$};
    \node at (-0.4,-1) {\scriptsize $\WQ_1$};
    \node at (2.4,-1) {\scriptsize $Q_2$};
    \node at (2.4,1) {\scriptsize $\WQ_2$};
    
    \node at (5.3,0) {$\Longleftrightarrow$};
    \node at (2,-2.5) {\textcolor{green}{$\downarrow$}};
    \node at (1,-2.5) {\textcolor{green}{Aharony}};
    \node at (9,-2.5) {\textcolor{red}{$\uparrow$}};
    \node at (10,-2.5) {\textcolor{red}{Aharony}};
    \node at (5.3,-6) {\textcolor{blue}{$\rightarrow$}};
    \node at (5.3,-5.6) {\textcolor{blue}{Aharony}};

    \node at (9-2.2,2.2) {(m)};
    \node[vertex] (mN2) at (9,0) [shape=circle,draw=black,minimum size=2em] {\!$\WN_2$\!};
    \node[vertex] (mN1) at (11,0) [shape=circle,draw=black,minimum size=2em] {\!$\WN_1$\!};
    \node[vertex] (mNf1t) at (7.5,1.3) [shape=rectangle,draw=black,minimum height=2em, minimum width=2em] {$F_1$};
    \node[vertex] (mNf1b) at (7.5,-1.3) [shape=rectangle,draw=black,minimum height=2em, minimum width=2em] {$F_1$};
    \node[vertex] (mNf2t) at (12.5,1.3) [shape=rectangle,draw=black,minimum height=2em, minimum width=2em] {$F_2$};
    \node[vertex] (mNf2b) at (12.5,-1.3) [shape=rectangle,draw=black,minimum height=2em, minimum width=2em] {$F_2$};

    \draw[-to, min distance=0.8cm,orange]  (mN1) edge [out=120, in=60] node {} (mN1);
    \draw[->-=.5,orange] ([yshift= -6pt] mNf2t.west) to ([xshift= -2pt,yshift= 8pt] mN1.east);
    \draw[->-=.5,orange] ([xshift= -2pt,yshift= 8pt] mN1.east) to ([yshift= -6pt] mNf2t.west);
    \draw[->-=.5,orange] ([xshift= -2pt,yshift= -8pt] mN1.east) to ([yshift= 6pt] mNf2b.west);
    \draw[->-=.5,orange] ([yshift= 6pt] mNf2b.west) to ([xshift= -2pt,yshift= -8pt] mN1.east);
   
    \draw[->-=.5] ([yshift= 2pt] mN1.west) to ([yshift= 2pt] mN2.east);
    \draw[->-=.5] ([yshift= -2pt] mN2.east) to ([yshift= -2pt] mN1.west);
    \draw[->-=.5] ([xshift= 5pt,yshift= -10pt] mN1.west) to [bend left=20] ([yshift= 2pt] mNf1b.east);
    \draw[->-=.5] ([yshift= -2pt] mNf1t.east) to [bend left=20]([xshift= 5pt,yshift= 10pt] mN1.west);
    \draw[->-=.5] ([yshift= 2pt] mNf2b.west) to [bend left=20]([xshift= -5pt,yshift= -10pt] mN2.east);
    \draw[->-=.5] ([xshift= -5pt,yshift= 10pt] mN2.east) to [bend left=20]([yshift= -2pt] mNf2t.west);
    \draw[dashed,->-=.5] (mNf1b.north) to (mNf1t.south);
    \draw[dashed,->-=.5] (mNf2t.south) to (mNf2b.north);
    \draw[dashed,->-=.5] ([yshift= 4pt] mNf2t.west) to ([yshift= 4pt] mNf1t.east);
    \draw[dashed,->-=.5] ([yshift= -4pt] mNf1b.east) to ([yshift= -4pt] mNf2b.west);

    \node at (10,0.3) {\scriptsize $x$};
    \node at (10,-0.3) {\scriptsize $\Wx$};
    \node at (9-0.4,1) {\scriptsize $\Wq_1$};
    \node at (9-0.4,-1) {\scriptsize $q_1$};
    \node at (11.4,-1) {\scriptsize $\Wq_2$};
    \node at (11.4,1) {\scriptsize $q_2$};
    \node at (10,1.8) {\scriptsize $P_{21,0}$};
    \node at (10,-1.8) {\scriptsize $P_{12,0}$};
    \node at (7,0.2) {\scriptsize $M_{1,0}$};
    \node at (7,-0.2) {\scriptsize $M_{1,1}$};
    \node at (13,0.2) {\scriptsize $M_{2,0}$};
    \node at (13,-0.2) {\scriptsize $M_{2,1}$};

    \begin{scope}[shift={(0,-5.5)}]
    \node at (-2.2,2.2) { (e$'$)};
    \node[vertex] (eN1) at (0,0) [shape=circle,draw=black,minimum size=2em] {\!$N_1'$\!};
    \node[vertex] (eN2) at (2,0) [shape=circle,draw=blue,minimum size=2em] {$N_2$};
    \node[vertex] (eNf1t) at (-1.5,1.3) [shape=rectangle,draw=black,minimum height=2em, minimum width=2em] {$F_1$};
    \node[vertex] (eNf1b) at (-1.5,-1.3) [shape=rectangle,draw=black,minimum height=2em, minimum width=2em] {$F_1$};
    \node[vertex] (eNf2t) at (3.5,1.3) [shape=rectangle,draw=black,minimum height=2em, minimum width=2em] {$F_2$};
    \node[vertex] (eNf2b) at (3.5,-1.3) [shape=rectangle,draw=black,minimum height=2em, minimum width=2em] {$F_2$};

    \draw[-to, min distance=0.8cm,orange]  (eN2) edge [out=120, in=60] node {} (eN2);

    \draw[->-=.5] ([yshift= 2pt] eN1.east) to ([yshift= 2pt] eN2.west);
    \draw[->-=.5] ([yshift= -2pt] eN2.west) to ([yshift= -2pt] eN1.east);
    \draw[->-=.5] ([xshift= 5pt,yshift= -10pt] eN1.west) to ([yshift= 2pt] eNf1b.east);
    \draw[->-=.5] ([yshift= -2pt] eNf1t.east) to ([xshift= 5pt,yshift= 10pt] eN1.west);
    \draw[->-=.5] ([xshift= -5pt,yshift= -10pt] eN2.east) to ([yshift= 2pt] eNf2b.west);
    \draw[->-=.5] ([yshift= -2pt] eNf2t.west) to ([xshift= -5pt,yshift= 10pt] eN2.east);
    \draw[->-=.5,dashed] (eNf1b.north) to (eNf1t.south);
    \draw[->-=.5] ([yshift= -4pt] eNf1b.east) to [bend right=20] ([xshift= 5pt,yshift= -10pt] eN2.west);
    \draw[->-=.5] ([xshift= 5pt,yshift= 10pt] eN2.west) to [bend right=20]([yshift= 4pt] eNf1t.east);

    \node at (1,0.5) {\scriptsize $Y$};
    \node at (1,-0.5) {\scriptsize $\WY$};
    \node at (-0.4,1) {\scriptsize $\Wq'_1$};
    \node at (-0.4,-1) {\scriptsize $q'_1$};
    \node at (2.4,-1) {\scriptsize $Q_2$};
    \node at (2.4,1) {\scriptsize $\WQ_2$};
    \node at (-2,0) {\scriptsize $M_{1,0}$};
    \node at (0.7,1.5) {\scriptsize $B$};
    \node at (0.7,-1.5) {\scriptsize $\WB$};
    \end{scope}

    \begin{scope}[shift={(0,-5.5)}]
    \node at (9-2.2,2.2) {(m$'$)};
    \node[vertex] (mN2) at (9,0) [shape=circle,draw=red,minimum size=2em] {\!$N_1'$\!};
    \node[vertex] (mN1) at (11,0) [shape=circle,draw=black,minimum size=2em] {\!$\WN_1$\!};
    \node[vertex] (mNf1t) at (7.5,1.3) [shape=rectangle,draw=black,minimum height=2em, minimum width=2em] {$F_1$};
    \node[vertex] (mNf1b) at (7.5,-1.3) [shape=rectangle,draw=black,minimum height=2em, minimum width=2em] {$F_1$};
    \node[vertex] (mNf2t) at (12.5,1.3) [shape=rectangle,draw=black,minimum height=2em, minimum width=2em] {$F_2$};
    \node[vertex] (mNf2b) at (12.5,-1.3) [shape=rectangle,draw=black,minimum height=2em, minimum width=2em] {$F_2$};

    \draw[-to, min distance=0.8cm,orange]  (mN2) edge [out=120, in=60] node {} (mN2);
    
    \draw[->-=.5] ([xshift= -2pt,yshift= 8pt] mN1.east) to ([yshift= -6pt] mNf2t.west);
    
    \draw[->-=.5] ([yshift= 6pt] mNf2b.west) to ([xshift= -2pt,yshift= -8pt] mN1.east);
   
    \draw[->-=.5] ([yshift= 2pt] mN1.west) to ([yshift= 2pt] mN2.east);
    \draw[->-=.5] ([yshift= -2pt] mN2.east) to ([yshift= -2pt] mN1.west);
    \draw[->-=.5] ([xshift= 5pt,yshift= -10pt] mN1.west) to [bend left=20] ([yshift= 2pt] mNf1b.east);
    \draw[->-=.5] ([yshift= -2pt] mNf1t.east) to [bend left=20]([xshift= 5pt,yshift= 10pt] mN1.west);
    \draw[->-=.5] ([xshift= -5pt,yshift= -10pt] mN2.east) to [bend right=20]([yshift= 2pt] mNf2b.west);
    \draw[->-=.5] ([yshift= -2pt] mNf2t.west) to [bend right=20]([xshift= -5pt,yshift= 10pt] mN2.east);
    \draw[dashed,->-=.5] (mNf1b.north) to (mNf1t.south);
    \draw[dashed,->-=.5] (mNf2t.south) to (mNf2b.north);
    \draw[dashed,->-=.5] ([yshift= 4pt] mNf2t.west) to ([yshift= 4pt] mNf1t.east);
    \draw[dashed,->-=.5] ([yshift= -4pt] mNf1b.east) to ([yshift= -4pt] mNf2b.west);

    \draw[->-=.5,orange] ([yshift= -6pt] mNf1t.east) to ([xshift= 2pt,yshift= 8pt] mN2.west);
    \draw[->-=.5,orange] ([xshift= 2pt,yshift= 8pt] mN2.west) to ([yshift= -6pt] mNf1t.east);
    \draw[->-=.5,orange] ([xshift= 2pt,yshift= -8pt] mN2.west) to ([yshift= 6pt] mNf1b.east);
    \draw[->-=.5,orange] ([yshift= 6pt] mNf1b.east) to ([xshift= 2pt,yshift= -8pt] mN2.west);

    \node at (10,0.3) {\scriptsize $y$};
    \node at (10,-0.3) {\scriptsize $\Wy$};
    \node at (9-0.4,1) {\scriptsize $\Wq_1$};
    \node at (9-0.4,-1) {\scriptsize $q_1$};
    \node at (11.4,-1) {\scriptsize $b$};
    \node at (10.3,1.2) {\scriptsize $\Wb$};
    \node at (11.4,0.8) {\scriptsize $q'_2$};
    \node at (11.8,-0.4) {\scriptsize $\Wq'_2$};
    \node at (10,1.8) {\scriptsize $P_{21,0}$};
    \node at (10,-1.8) {\scriptsize $P_{12,0}$};
    \node at (7,0.2) {\scriptsize $M_{1,0}$};
    \node at (7,-0.2) {\scriptsize $M_{1,1}$};
    \node at (13,0) {\scriptsize $M_{2,0}$};
    \end{scope}

\end{tikzpicture}
\caption{\label{fig: 3d k=1} $k=1$ two-node duality from three successive Aharony dualities.}
\end{figure}
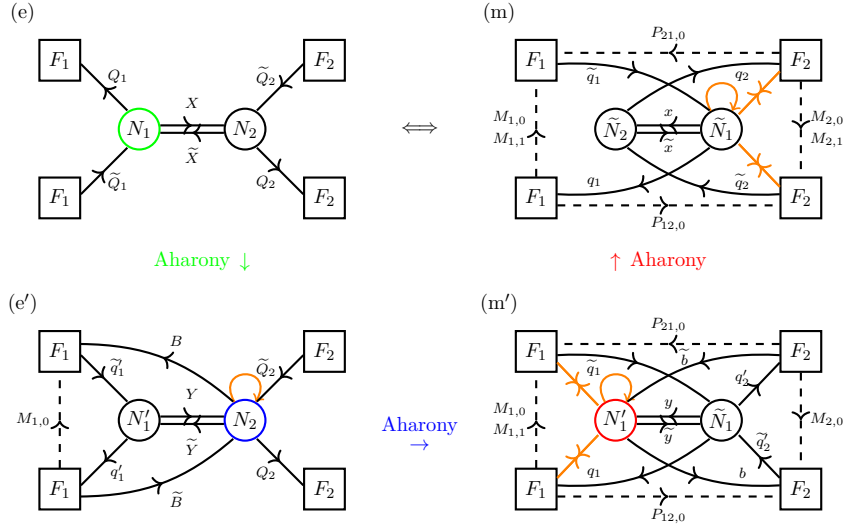
For $k=1$, the duality can be derived through three successive applications of Aharony duality, as illustrated in Figure~\ref{fig: 3d k=1}. Starting from quiver (e) with superpotential $W_{\rm e}=\text{Tr}(X\WX)^2$, we obtain (e$'$) by applying Aharony duality to the $U(N_1)$ node, highlighted in green, with dual gauge rank $N_1'=F_1+N_2-N_1$ and superpotential
\begin{align}
    W_{\rm e '} 
    =
   \text{Tr}  (Y\WY)^2
    +
    B \Wq_1' Y + \WB \WY q_1'
    +
    M_{1,0} \Wq_1' q_1'
    +
    \sum_{s=\pm} V_{s0} \hat{V}'_{s0}
    \,,
\end{align}
where all gauge and flavor indices are implicitly contracted. The monopole operators in this frame are denoted by $\hat V_{ss'}'$, while the singlets $V_{s0}$ are mapped from the monopole operators $\hat V_{s0}$ in the previous frame (e). The massive adjoint at the $U(N_2)$ node, highlighted in orange, is then integrated out. Applying another Aharony duality to the $U(N_2)$ node, highlighted in blue, we obtain (m$'$) with one of the expected dual gauge ranks, $\WN_1=2F_1+F_2-N_1$, and superpotential
\begin{align}
    W_{\rm m '} 
    &=
    \text{Tr} (\Wy y)^2 
    +
    \Wb y q_2' + b \Wq_2' \Wy + P_{21,0} \Wq_1 q_2' + P_{12,0} \Wq_2' q_1
    \nonumber\\
    &
    + M_{1,0} \Wq_1 \Wy y q_1 + M_{1,1} \Wq_1 q_1 + M_{2,0} \Wq_2' q_2'
    +
    \sum_{s=\pm} \Big( V_{s0} \hat v_{ss}' + V_{0s}' \hat v_{0s}' \Big)
    \,.
\end{align}
Here, we have integrated out the massive adjoint and bi-fundamental fields, indicated by the orange lines. We denote the monopole operators in this frame by $\hat v_{ss'}'$, while the coupled singlets $V_{0s}'$ are mapped from the monopole operators in frame (e$'$). Finally, applying Aharony duality to the $U(N_1')$ node, highlighted in red, we obtain the expected magnetic theory (m) with superpotential
\begin{align}
    W_{\rm m}
    =
    & \text{Tr}(x\Wx)^2 
    + \sum_{j=0}^1 \Big( 
    M_{1,1-j} \Wq_1 (x \Wx)^j q_1 + M_{2,1-j} \Wq_2 (\Wx x)^j q_2
    \Big)
    \nonumber\\
    &
    +
    P_{21,0} \Wq_1 x q_2 + P_{12,0} \Wq_2 \Wx q_1
    + \sum_{s=\pm}
    \Big(
    V_{s0} \hat v_{0s}
    +
    v'_{s0} \hat v_{s0}
    +
    V_{0s}' \hat v_{ss}
    \Big)
    \,,
\end{align}
where the massive adjoint at the $U(\WN_1)$ node and the bi-fundamental fields are integrated out. The monopole operators in this frame are denoted by $\hat v_{ss'}$. We summarize the monopole maps along the sequence of dualities
\begin{equation}
\begin{array}{c@{\quad}c@{\quad}c@{\quad}c@{\quad}c@{\quad}c@{\quad}c}
(\mathrm e)&&(\mathrm e')&&(\mathrm m')&&(\mathrm m)\\[7pt]
\hat V_{s0}&\longleftrightarrow&V_{s0}
 &\longleftrightarrow&V_{s0}&\longleftrightarrow&V_{s0}\\
\hat V_{0s}&\longleftrightarrow&\hat V'_{ss}
 &\longleftrightarrow&\hat v'_{s0}&\longleftrightarrow&v'_{s0}\\
\hat V_{ss}&\longleftrightarrow&\hat V'_{0s}
 &\longleftrightarrow&V'_{0s}&\longleftrightarrow&V'_{0s}
\end{array}
\end{equation}
Thus, we would rewrite the last two singlets in frame (m) as $v'_{s0}\equiv V_{0s}$ and $V_{0s}'\equiv V_{ss}$. By exchanging the positions of the $U(\WN_1)$ and $U(\WN_2)$ nodes in (m), which implements $\hat v_{ss'}\to\hat v_{s's}$, we finally get the expected magnetic superpotential $W_{\rm m}$
\begin{align}
    W_{\rm m}
    =
    & \text{Tr}(x\Wx)^2 
    + \sum_{j=0}^1 \Big( 
    M_{1,1-j} \Wq_1 (x \Wx)^j q_1 + M_{2,1-j} \Wq_2 (\Wx x)^j q_2
    \Big)
    \nonumber\\
    &
    +
    P_{21,0} \Wq_1 x q_2 + P_{12,0} \Wq_2 \Wx q_1
    + \sum_{s=\pm}
    \Big(
    V_{s0} \hat v_{s0}
    +
    V_{0s} \hat v_{0s}
    +
    V_{ss} \hat v_{ss}
    \Big)
    \,,
\end{align}
which proves the proposed duality at $k=1$.

In what follows, we verify the duality for general $k$ by matching the 3d $\CN=2$ ellipsoid partition functions of the dual pair.

\begin{figure}[tbp]
\centering
\begin{tikzpicture}[thick,scale=0.3,every node/.style={scale=0.3},
  >=Stealth]
\tikzset{vertex/.style={circle,fill=white!25,minimum size=5pt,inner sep=1pt}}
\tikzset{every loop/.style={}}
\tikzset{gn/.style={vertex,shape=circle,draw=black,minimum size=2em},
         fn/.style={vertex,shape=rectangle,draw=black,minimum height=2em,minimum width=2em},
         note/.style={font=\Large},
         gr/.style={draw=green!70!black},
         re/.style={draw=red},
         bl/.style={draw=blue}}

\newcommand{\flavors}{%
  \node[fn] (fl) at (-2.5,3) {};
  \node[fn] (fr) at ( 2.5,3) {};
  \node[fn] (bl) at (-2.5,-3) {};
  \node[fn] (br) at ( 2.5,-3) {};
}
\newcommand{\framefour}{%
  \draw[dashed] (fl)--(fr);
  \draw[dashed] (fl)--(bl);
  \draw[dashed] (fr)--(br);
  \draw[dashed] (bl)--(br);
}
\newcommand{\topdashed}{\draw[dashed] (fl)--(fr);}
\newcommand{\adjN}[1]{%
  \draw[black] ($(#1)+(155:0.35)$)
    .. controls ($(#1)+(120:1.1)$) and ($(#1)+(60:1.1)$)
    .. ($(#1)+(25:0.35)$);
}
\newcommand{\adjNE}[1]{%
  \draw[black] ($(#1)+(145:0.35)$)
    .. controls ($(#1)+(105:1.1)$) and ($(#1)+(45:1.1)$)
    .. ($(#1)+(15:0.35)$);
}
\newcommand{\adjNW}[1]{%
  \draw[black] ($(#1)+(165:0.35)$)
    .. controls ($(#1)+(135:1.1)$) and ($(#1)+(75:1.1)$)
    .. ($(#1)+(35:0.35)$);
}
\newcommand{\adjredN}[1]{%
  \draw[red] ($(#1)+(155:0.35)$)
    .. controls ($(#1)+(120:1.1)$) and ($(#1)+(60:1.1)$)
    .. ($(#1)+(25:0.35)$);
}
\newcommand{\adjredE}[1]{%
  \draw[red] ($(#1)+(55:0.35)$)
    .. controls ($(#1)+(20:1.1)$) and ($(#1)+(-20:1.1)$)
    .. ($(#1)+(-55:0.35)$);
}
\newcommand{\adjredS}[1]{%
  \draw[red] ($(#1)+(-155:0.35)$)
    .. controls ($(#1)+(-120:1.1)$) and ($(#1)+(-60:1.1)$)
    .. ($(#1)+(-25:0.35)$);
}
\newcommand{\adjE}[1]{%
  \draw[black] ($(#1)+(55:0.35)$)
    .. controls ($(#1)+(20:1.1)$) and ($(#1)+(-20:1.1)$)
    .. ($(#1)+(-55:0.35)$);
}
\newcommand{\adjW}[1]{%
  \draw[black] ($(#1)+(145:0.35)$)
    .. controls ($(#1)+(160:1.1)$) and ($(#1)+(200:1.1)$)
    .. ($(#1)+(215:0.35)$);
}
\newcommand{\gtickarrow}[4]{%
  \draw[green!65!black] ($(#1,#2)!4.2pt!90:(#3,#4)$)--($(#1,#2)!4.2pt!-90:(#3,#4)$);
  \draw[green!65!black,->] (#1,#2)--(#3,#4);
}
\path[use as bounding box] (-0.5,-0.2) rectangle (47.5,32.3);

\draw (10.8,21.0) rectangle (34.1,31.9);
\begin{scope}[shift={(15.05,26.0)}]
  \node at (-3.4,4.3) {\Large (e)};
  \flavors
  \node[gn] (up) at (0,1.5) {};
  \node[gn] (dn) at (0,-1.5) {};
  \draw (up)--(dn);
  \draw ([xshift=-10pt,yshift=-4pt]up.north)--([yshift=-5pt]fl.east);
  \draw ([xshift=10pt,yshift=-4pt]up.north)--([yshift=-5pt]fr.west);
  \draw ([xshift=-10pt,yshift=4pt]dn.south)--([yshift=5pt]bl.east);
  \draw ([xshift=10pt,yshift=4pt]dn.south)--([yshift=5pt]br.west);
\end{scope}
\node at (21.6,26.2) {\huge $\Longleftrightarrow$};
\begin{scope}[shift={(29.0,26.0)}]
  \node at (-3.4,4.3) {\Large (m)};
  \flavors
  \node[gn] (up) at (0,1.5) {};
  \node[gn] (dn) at (0,-1.5) {};
  \draw (up)--(dn);
  \draw ([xshift=-10pt,yshift=-4pt]up.north)--([yshift=-5pt]fl.east);
  \draw ([xshift=10pt,yshift=-4pt]up.north)--([yshift=-5pt]fr.west);
  \draw ([xshift=-10pt,yshift=4pt]dn.south)--([yshift=5pt]bl.east);
  \draw ([xshift=10pt,yshift=4pt]dn.south)--([yshift=5pt]br.west);
  \framefour
\end{scope}

\begin{scope}[shift={(4.0,22.2)}]
  \node at (-3.2,4.4) {\Large (a.1)};
  \flavors \topdashed
  \node[gn] (up) at (0,1.25) {};
  \node[gn] (dn) at (0,-1.5) {};
  \draw (up)--(dn);
  \draw (fl)--(up)--(fr);
  \draw (fl)--(dn)--(fr);
  \draw (dn)--(bl) (dn)--(br);
  \adjredE{dn}
\end{scope}
\node[green!65!black] at (9.0,26.2) {\Large $A$};
\node[green!65!black,scale=0.85,inner sep=0pt] at (9.0,25.08) {\Huge $\swarrow$};
\node[red,scale=0.85,inner sep=0pt] at (4.0,18.0) {\Huge $\downarrow$};
\node[red,anchor=west] at (4.4,18.0) {\large deconf};

\begin{scope}[shift={(4.0,13.35)}]
  \node at (-3.2,4.3) {\Large (a.2)};
  \flavors \topdashed
  \node[gn] (up) at (0,1.25) {};
  \node[gn] (dn) at (0,-1.5) {};
  \draw (up)--(dn) (fl)--(up)--(fr);
  \draw (fl)--(dn)--(fr);
  \draw (dn)--(bl) (dn)--(br);
  \node[gn] (q1) at (1.4,-1.5) {};
  \node[gn] (q2) at (4.0,-1.5) {};
  \node[gn] (q3) at (5.1,-1.5) {};
  \draw (dn)--(q1) (q2)--(q3);
  \draw[densely dotted] (q1)--(q2);
  \adjN{q1}\adjN{q2}\adjN{q3}
\end{scope}
\draw[green!65!black] (4.1,8.42)--(4.1,8.87);
\draw[green!65!black,->] (4.1,8.65)--(9.25,8.65);
\draw[green!65!black] (8.30,7.09)--(8.30,7.55);
\draw[green!65!black,->] (8.30,7.32)--(3.35,7.32)--(3.35,8.53);
\gtickarrow{8.10}{8.22}{8.10}{7.48}
\node[green!65!black,anchor=west] at (8.30,8.2) {\large $A$-deconf};

\begin{scope}[shift={(4.0,3.4)}]
  \node at (-3.2,4.25) {\Large (a.3)};
  \flavors\framefour
  \node[gn] (up) at (0,0.95) {};
  \node[gn] (dn) at (0,-1.50) {};
  \node[gn] (external) at (5.10,-1.50) {};
  \draw (fl) .. controls +(2.5,-0.08) and +(-1.0,1.08) .. (external);
  \draw (fr) .. controls +(1.15,-0.40) and +(0.28,1.1) .. (external);
  \draw (up)--(external);
  \draw (up)--(bl) (up)--(dn) (up)--(br);
  \draw (dn)--(bl) (dn)--(br);
  \node[gn] (q1) at (1.45,-1.50) {};
  \node[gn] (q2) at (3.75,-1.50) {};
  \node[gn] (q3) at (3.75,-2.70) {};
  \draw (dn)--(q1);
  \draw[densely dotted] (q1)--(q2);
  \draw (q2)--(q3);
  \adjNE{dn}\adjN{q1}\adjN{q2}\adjE{q3}

  \draw[blue,semithick] (-0.70,-2.20)--(-0.70,-0.25)--(4.55,-0.25)
      --(4.55,-3.76)--(2.90,-3.76)--(2.90,-2.20)--cycle;
  \node[blue] at (5.5,1.56) {\large conf};
  \node[green!65!black] at (-1.95,-0.75) {\Large $A$};
  \draw[green!65!black,->] (-1.7,-0.90)--(-0.18,-1.46);
\end{scope}

\node[blue,scale=0.85,inner sep=0pt] at (9.65,4.2) {\Huge $\longrightarrow$};

\begin{scope}[shift={(12.9,3.4)}]
  \node at (-3.1,4.25) {\Large (b.1)};
  \flavors\framefour
  \node[gn] (up) at (0,1.15) {};
  \node[gn] (dn) at (0,-1.5) {};
  \node[gn] (external) at (5.1,-1.45) {};
  \draw (fl) .. controls +(1.30,-0.10) and +(-1.10,0.72) .. (external);
  \draw (fr) .. controls +(0.72,-0.52) and +(-0.68,1.20) .. (external);
  \draw (up)--(external) (up)--(dn);
  \draw (dn)--(bl) (dn)--(br);
  \adjredN{up}
\end{scope}
\node[red] at (17.5,6.2) {\large deconf};
\node[red,scale=0.85,inner sep=0pt] at (17.7,5.4) {\Huge $\longrightarrow$};

\begin{scope}[shift={(22.2,3.4)}]
  \node at (-3.1,4.25) {\Large (b.2)};
  \flavors\framefour
  \node[gn] (up) at (0,1.15) {};
  \node[gn] (dn) at (0,-1.5) {};
  \node[gn] (external) at (5.1,-1.45) {};
  \draw (fl) .. controls +(1.30,-0.10) and +(-1.10,0.72) .. (external);
  \draw (fr) .. controls +(0.72,-0.52) and +(-0.68,1.20) .. (external);
  \draw (up)--(external) (up)--(dn);
  \draw (dn)--(bl) (dn)--(br);
  \node[gn] (q1) at (1.5,1.15) {};
  \node[gn] (q2) at (4.0,1.15) {};
  \node[gn] (q3) at (5.1,1.15) {};
  \draw (up)--(q1) (q2)--(q3);
  \draw[densely dotted] (q1)--(q2);
  \adjN{q1}\adjN{q2}\adjN{q3}
\end{scope}
\draw[green!65!black] (27,10.00)--(27,10.45); 
\draw[green!65!black,->] (27,10.22)--(22.25,10.22)--(22.25,8.98);
\draw[green!65!black] (22.35,7.98)--(22.35,8.43); 
\draw[green!65!black,->] (22.35,8.20)--(27.72,8.20);
\gtickarrow{26.95}{8.82}{26.95}{9.62}
\node[green!65!black,anchor=west] at (27.24,9.35) {\large $A$-deconf};
\node[green!65!black,scale=0.85,inner sep=0pt] at (28,2.6) {\Huge $\longrightarrow$};

\begin{scope}[shift={(31.3,3.4)}]
  \node at (-3.1,4.25) {\Large (b.3)};
  \flavors\framefour 

  \coordinate (b3n1) at (3.92,2.48);
  \coordinate (b3n2) at (0,1.10);
  \coordinate (b3n3) at (1.63,1.10);
  \coordinate (b3n4) at (3.92,1.10);
  \coordinate (b3n5) at (5.08,1.00);
  \coordinate (b3n6) at (0,-1.50);
  \coordinate (b3n7) at (5.08,-1.50);

  \draw (fl.east) .. controls +(2.30,-0.05) and +(-1.25,0.90) .. (b3n7);
  \draw (fr.east) .. controls +(1.20,-0.45) and +(0.42,1.10) .. (b3n7);

  \draw (bl.north east) .. controls +(.95,1.05) and +(-1.35,-0.82) .. (b3n5);
  \draw (br.north east) .. controls +(.80,0.90) and +(-0.62,-1.20) .. (b3n5);

  \draw (b3n5)--(b3n7);   
  \draw (b3n1)--(b3n4);   
  \draw (b3n2)--(b3n3);   
  \draw (b3n2)--(b3n6);   
  \draw (b3n2)--(bl) (b3n2)--(br); 
  \draw (b3n6)--(bl) (b3n6)--(br); 
  \draw (b3n6) .. controls +(2.15,0.16) and +(-1.15,-0.85) .. (b3n5); 

  \draw[densely dotted] (b3n3)--(b3n4);

  \draw[blue,semithick] (-0.77,0.43)--(-0.77,2.29)--(3.09,2.29)
     --(3.09,3.67)--(4.60,3.67)--(4.60,0.43)--cycle;
  \node[blue] at (2.35,3.90) {\large conf};

  \foreach \k in {1,2,3,4,5,6,7} {
    \node[gn,fill=white] at (b3n\k) {};
  }
  \adjN{b3n2}\adjN{b3n3}\adjN{b3n1}

  \node[green!65!black] at (-2,0.17) {\Large $A$};
  \draw[green!65!black,->] (-1.8,0.19)--(-0.29,1.05);
\end{scope}

\node[blue,scale=0.85,inner sep=0pt] at (36.55,5.5) {\Huge $\longrightarrow$};

\begin{scope}[shift={(40.3,3.4)}]
  \node at (-3.2,4.25) {\Large (c.1)};
  \flavors\framefour
  \node[gn] (up) at (0,1.1) {};
  \node[gn] (dn) at (0,-1.5) {};
  \node[gn] (r1) at (3.6,1.1) {};
  \node[gn] (r2) at (3.6,-1.5) {};
  \draw (fl)--(r1) (fr)--(r1) (r1)--(r2);
  \draw (r2)--(br) (r2)--(bl);
  \draw (up)--(dn) (up)--(bl) (up)--(br);
  \adjredS{dn}
\end{scope}
\node[red,anchor=west] at (39.15,8.6) {\large deconf $\uparrow$};

\begin{scope}[shift={(40.3,13.3)}]
  \node at (-3.2,4.25) {\Large (c.2)};
  \flavors\framefour
  \node[gn] (up) at (0,1.05) {};
  \node[gn] (r1) at (3.6,1.1) {};
  \node[gn] (r2) at (3.6,-1.5) {};
  \draw (fl)--(r1) (fr)--(r1) (r1)--(r2);
  \draw (r2)--(bl) (r2)--(br);
  \draw (up)--(bl) (up)--(br);
  \node[gn] (q1) at (-4.55,-1.5) {};
  \node[gn] (q2) at (-3.12,-1.5) {};
  \node[gn] (q3) at (-1.10,-1.5) {};
  \node[gn] (q4) at (0,-1.5) {};
  \draw (q1)--(q2) (q3)--(q4) (up)--(q4);
  \draw[densely dotted] (q2)--(q3);
  \adjN{q1}\adjN{q2}\adjN{q3}
  \draw[green!65!black] (-4.55,-1.1) ellipse (0.63 and 1.02);
  \node[green!65!black] at (-5.1,0.65) {\large cancel};
  \gtickarrow{0}{-3.65}{-4.95}{-3.65}
  \node[green!65!black] at (-2.9,-4.2) {\Large $A$};
\end{scope}
\node[green!65!black,scale=0.85,inner sep=0pt] at (40.5,18.3) {\Huge $\uparrow$};

\begin{scope}[shift={(40.9,22.65)}]
  \node at (-3.2,4.25) {\Large (c.3)};
  \flavors\framefour
  \node[gn] (up) at (0,1.08) {};
  \node[gn] (r1) at (3.6,1.1) {};
  \node[gn] (r2) at (3.6,-1.5) {};
  \draw (fl)--(r1) (fr)--(r1) (r1)--(r2);
  \draw (r2)--(bl) (r2)--(br);
  \draw (up)--(bl) (up)--(br);
  \node[gn] (q1) at (-3.25,-1.50) {};
  \node[gn] (q2) at (-1.48,-1.50) {};
  \node[gn] (q3) at (0,-1.50) {};
  \draw[densely dotted] (q1)--(q2);
  \draw (q2)--(q3) (up)--(q3);
  \adjN{q1}\adjNE{q2}\adjN{q3}\adjW{up}

  \draw[blue,semithick] (-4.05,-2.28)--(-4.05,-0.20)
      --(-2.10,-0.20)--(-2.10,2.38)--(1.00,2.38)
      --(1.00,-2.28)--cycle;
  \node[green!65!black] at (2.10,1.03) {\Large $A$};
  \draw[green!65!black,->] (1.63,1.02)--(0.38,1.08);
\end{scope}

\node[blue] at (35.20,24.15) {\large conf};
\node[blue,scale=0.85,inner sep=0pt] at (35.02,25.08) {\Huge $\nwarrow$};

\end{tikzpicture}
\caption{\label{fig: 3d derivation} Quiver diagrams forming a loop of dualities, providing a derivation of the two-node duality between (e) and (m). For simplicity, gauge ranks, arrow directions, and singlet contributions are omitted. Mesons are represented by dashed lines. Deconfinement of an adjoint is highlighted in red, while confinement of a quiver tail into an adjoint is highlighted in blue. Green arrows indicate successive applications of Aharony duality. Note that adjoint deconfinement via Aharony duality is denoted by `A-deconf'.}
\end{figure}
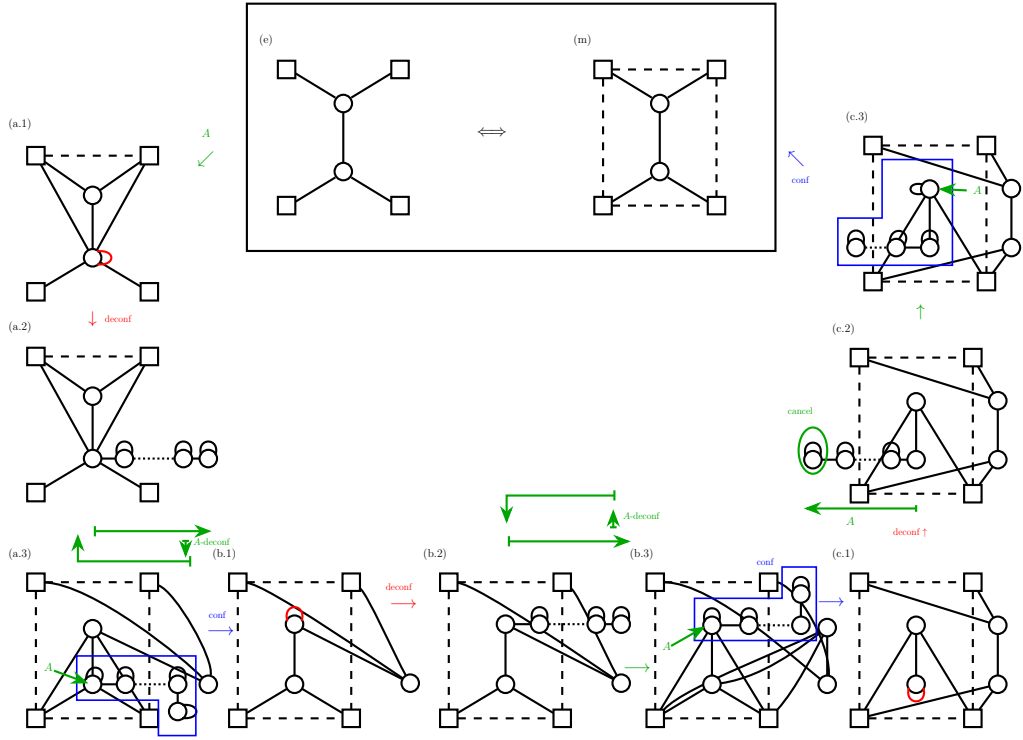
\subsection{Generic \texorpdfstring{$k$}{k} : the partition function derivation}
\noindent
An overview of the 3d $\CN=2$ ellipsoid partition function derivation is illustrated in Figure~\ref{fig: 3d derivation}. The derivation forms a loop of dualities involving only Aharony duality and the confinement of $\CC_p[U(N)]$ into an adjoint, where the latter itself follows from Aharony and BBP dualities. Thus, the entire derivation relies solely on these two elementary dualities: Aharony and BBP.\footnote{These two dualities are not independent, as Aharony duality can be obtained from BBP duality through a mass deformation.} We adopt the quiver representation of the partition function developed in~\cite{Hwang:2024hhy,Benvenuti:2026usm} to streamline the discussion; see Appendix~\ref{app: 3d ptf-quiver} for a brief review.

The derivation consists of {\it three iterations of a common procedure}, labeled by (a.i), (b.i), and (c.i), each involving deconfinement (i=1), successive Aharony dualities (i=2), and confinement (i=3), as illustrated in Figure~\ref{fig: 3d derivation}. Let us briefly outline the procedure.

Starting from (e), we apply Aharony duality to obtain (a.1), where an adjoint chiral arises from the meson contribution, highlighted in red. We deconfine this adjoint into a linear quiver tail as in (a.2) and successively apply Aharony dualities along the tail toward its end node. We then deconfine the adjoint at the second-to-last node via Aharony duality --- whose inverse reproduces the adjoint --- and apply another sequence of Aharony dualities in the reverse direction from the second-to-last node, back toward the original node, as highlighted in green.

This produces (a.3), containing a confining tail enclosed in the blue box. Its confinement generates an adjoint that cancels the existing one, allowing another Aharony duality to reach (b.1).

The same procedure applies to the second iteration. Deconfining the adjoint highlighted in red produces the linear quiver (b.2), and successive Aharony dualities lead to (b.3). The tail enclosed in the blue box then confines into an adjoint, which again cancels the existing one. We can thus apply Aharony duality to obtain (c.1) and deconfine the newly generated adjoint into a linear quiver tail as in (c.2).

In the final iteration, successive Aharony dualities along the tail eliminate its end node, leaving the confining tail enclosed in the blue box in (c.3). Its confinement cancels the existing adjoint. The remaining gauge node has the same rank as the flavor node and therefore confines upon applying Aharony duality, giving the expected magnetic quiver (m).

\begin{figure}[tbp]
\centering
\includegraphics[width=0.95 \textwidth]{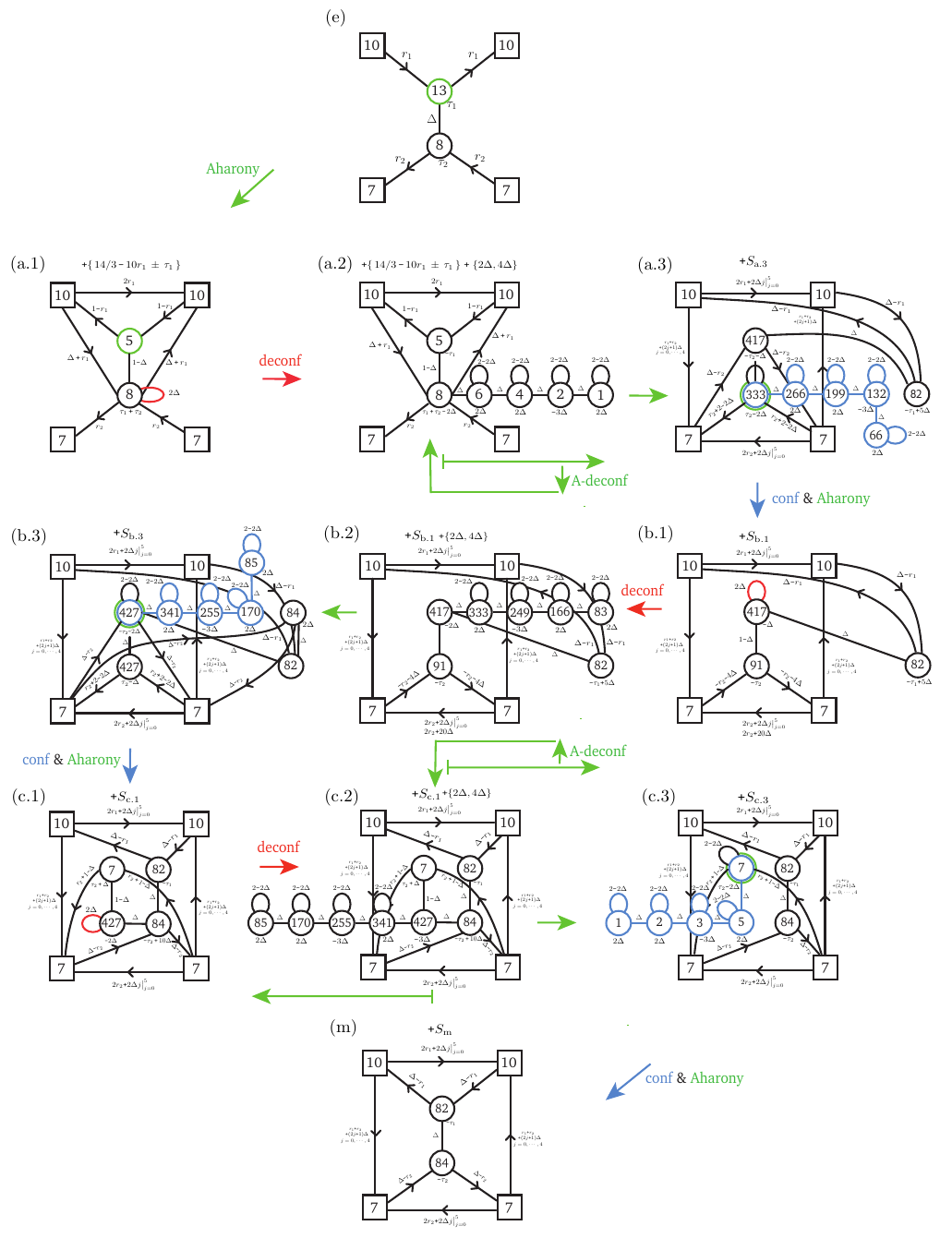}
\caption{\label{fig: 3d ex} Ellipsoid partition function derivation for $(k,F_1,F_2,N_1,N_2) = (5,10,7,13,8)$. }
\end{figure}
\medskip\noindent{\bf Example}. Let us explicitly illustrate the procedure for $(k,F_1,F_2,N_1,N_2)=(5,10,7,13,8)$, as shown in Figure~\ref{fig: 3d ex}. We denote Aharony duality by $\CA$ for brevity and set $\D=\frac{1}{k+1}=\frac{1}{6}$. Starting from quiver (e), applying $\CA$ to the $U(13)$ node produces (a.1), where an adjoint of  $R$-charge $2\D$ arises from the meson contribution, highlighted in red. Deconfining this adjoint gives a length-5 tail as in (a.2). We then successively apply $\CA$ from the $U(8)$ node to the $U(2)$ node, deconfine the adjoint at the second-to-last node via $\CA$, and apply another sequence of $\CA$ in the reverse direction up to the $U(5)$ node, obtaining (a.3). 

The tail highlighted in blue confines into an adjoint of  $R$-charge $2\D$ at the $U(333)$ node, which cancels the existing adjoint of  $R$-charge $2-2\D$, allowing us to apply $\CA$ at $U(333)$ again. This produces (b.1), where another adjoint appears, highlighted in red. Its deconfinement gives a length-5 tail as in (b.2). We successively apply $\CA$ from the $U(417)$ node to the $U(83)$ node, deconfine the adjoint at the second-to-last node via $\CA$, and then apply $\CA$ in the reverse direction up to the $U(91)$ node. The resulting quiver (b.3) contains a confining tail highlighted in blue, whose confinement produces an adjoint that cancels the existing one of  $R$-charge $2-2\D$ at the $U(427)$ node. 

We can thus apply $\CA$ to the upper $U(427)$ node to obtain (c.1), where the newly generated adjoint at $U(427)$ node, highlighted in red, is deconfined into a linear quiver tail as in (c.2). Successive applications of $\CA$ from the $U(427)$ node to the $U(85)$ node eliminate the last node. The remaining tail, together with the $U(7)$ node, forms the confining tail highlighted in blue in (c.3). Its confinement produces an adjoint that cancels the existing one of  $R$-charge $2-2\D$. Finally, applying $\CA$ to the $U(7)$ node gives the expected magnetic quiver (m). The singlet contributions in each frame are organized as follows:
\begin{align}
    &S_{\rm a.3}
    =
    S_{V_{s0}} + S_{V_{ss}}
    +
    \Big\{
    \frac{199}{3} - 7 r_2 \pm (\t_2 + \frac{1}{6}),
    \frac{1}{3},\frac{2}{3}
    \Big\}
    +
    \Big\{
    \frac{167}{3} + \frac{j}{3}
    \Big\}_{j=0}^5
    +
    \Big\{
    \frac{168}{3} + \frac{j}{3}
    \Big\}_{j=0}^3
    \,,
    \nonumber\\
    &S_{\rm b.1}
    =
    S_{\rm a.3} + S_{V_{0s}} + \Big\{ \frac{5}{3},\frac{4}{3} \Big\}
    ,
    \;\;\;\;
    S_{\rm c.1}
    =
    S_{\rm b.3}
    +\Big\{
    \frac{5}{3},\frac{4}{3}
    \Big\}
    +\Big\{
    7 r_2 - \frac{193}{3}
    \pm
    (\t_2 + \frac{1}{6})
    \Big\}
    \,,
    \nonumber\\
    &S_{\rm b.3}
    =
    S_{\rm m} 
    + \! \Big\{
    \frac{199}{3} \!-\! 7 r_2 \!\pm\! (\t_2 \!+\! \frac{1}{6}),
    \frac{41}{6} \!+\! 7 r_2 \!\pm\! (\t_2 \!-\! \frac{1}{6}),
    \frac{1}{3},\frac{2}{3}
    \Big\}
    \!+\!
    \Big\{ \frac{427}{6} \!+\! \frac{j}{3} \Big\}_{j=0}^6
    \!\!\!+\!
    \Big\{ \frac{431}{6} \!+\! \frac{j}{3} \Big\}_{j=0}^2
    \,,
    \nonumber\\
    &S_{\rm c.3}
    =
    S_{\rm c.1}
    + 
    \Big\{
    \frac{1}{3},\frac{2}{3}
    \Big\}
    +
    \Big\{
    -\frac{415}{6}-\frac{j}{3}
    \Big\}_{j=0}^{6}
    +
    \Big\{
    -\frac{419}{6}-\frac{j}{3}
    \Big\}_{j=0}^{2}
    \,,
\end{align}
and the resulting singlet contribution $S_{\rm m}=S_{V_{s0}} + S_{V_{0s}} + S_{V_{ss}}$ in the magnetic theory (m) with
\begin{align}
    &S_{V_{s0}} \!=\!
    \Big\{
    \frac{14}{3} \!-\! 10 r_1 \!\pm\! \t_1
    \Big\}
    ,
    S_{V_{0s}} \!=\!
    \Big\{
    \frac{65}{6} \!-\! 7 r_2 \!\pm\! \t_2
    \Big\}
    ,
    S_{V_{ss}} \!=\!
    \Big\{
    \frac{83}{6}
    \!-\! 10 r_1 \!-\! 7 r_2
    \!\pm\! (\t_1 \!+\! \t_2)
    \!+\! \frac{j}{3}
    \Big\}_{j=0}^4
    \,,
\end{align}
which precisely matches that of the monopole-like singlets appearing in the magnetic superpotential. The dual gauge ranks $\WN_1=6\cdot 10+5\cdot 7-13=82$ and $\WN_2=5\cdot 10+6\cdot 7-8=84$ also agree with the expected values.

\medskip
\noindent{\bf Generic cases}.
In general, the partition functions of the dual theories match as
\begin{equation}
    \begin{tikzpicture}[thick,scale=0.8, every node/.style={scale=0.8}]
  \tikzset{vertex/.style={circle,fill=white!25,minimum size=12pt,inner sep=2pt}}
  \tikzset{every loop/.style={}}
    \node at (-2.2,2.2) { (e)};
    \node[vertex] (eN1) at (0,0) [shape=circle,draw=black,minimum size=2em] {$N_1$};
    \node[vertex] (eN2) at (2,0) [shape=circle,draw=black,minimum size=2em] {$N_2$};
    \node[vertex] (eNf1t) at (-1.5,1.3) [shape=rectangle,draw=black,minimum height=2em, minimum width=2em] {$F_1$};
    \node[vertex] (eNf1b) at (-1.5,-1.3) [shape=rectangle,draw=black,minimum height=2em, minimum width=2em] {$F_1$};
    \node[vertex] (eNf2t) at (3.5,1.3) [shape=rectangle,draw=black,minimum height=2em, minimum width=2em] {$F_2$};
    \node[vertex] (eNf2b) at (3.5,-1.3) [shape=rectangle,draw=black,minimum height=2em, minimum width=2em] {$F_2$};

    \draw[->-=.5] ([yshift= 2pt] eN1.east) to ([yshift= 2pt] eN2.west);
    \draw[->-=.5] ([yshift= -2pt] eN2.west) to ([yshift= -2pt] eN1.east);
    \draw[->-=.5] ([yshift= 2pt] eNf1b.east) to ([xshift= 5pt,yshift= -10pt] eN1.west);
    \draw[->-=.5] ([xshift= 5pt,yshift= 10pt] eN1.west) to ([yshift= -2pt] eNf1t.east);
    \draw[->-=.5] ([xshift= -5pt,yshift= -10pt] eN2.east) to ([yshift= 2pt] eNf2b.west);
    \draw[->-=.5] ([yshift= -2pt] eNf2t.west) to ([xshift= -5pt,yshift= 10pt] eN2.east);

    \node at (1,0.5) {\scriptsize $\D$};
    \node at (-0.4,1) {\scriptsize $r_1$};
    \node at (-0.4,-1) {\scriptsize $r_1$};
    \node at (2.4,-1) {\scriptsize $r_2$};
    \node at (2.4,1) {\scriptsize $r_2$};
    \node at (0,-0.6) {\scriptsize $\t_1$};
    \node at (2,-0.6) {\scriptsize $\t_2$};
    
    \node at (5.2,0) {$=\!\!=$};

    \node at (9-2.2,2.2) { (m)};
    \node[vertex] (mN1) at (9,0) [shape=circle,draw=black,minimum size=2em] {\!$\WN_1$\!};
    \node[vertex] (mN2) at (11,0) [shape=circle,draw=black,minimum size=2em] {\!$\WN_2$\!};
    \node[vertex] (mNf1t) at (7.5,1.3) [shape=rectangle,draw=black,minimum height=2em, minimum width=2em] {$F_1$};
    \node[vertex] (mNf1b) at (7.5,-1.3) [shape=rectangle,draw=black,minimum height=2em, minimum width=2em] {$F_1$};
    \node[vertex] (mNf2t) at (12.5,1.3) [shape=rectangle,draw=black,minimum height=2em, minimum width=2em] {$F_2$};
    \node[vertex] (mNf2b) at (12.5,-1.3) [shape=rectangle,draw=black,minimum height=2em, minimum width=2em] {$F_2$};

    \draw[->-=.5] ([yshift= 2pt] mN1.east) to ([yshift= 2pt] mN2.west);
    \draw[->-=.5] ([yshift= -2pt] mN2.west) to ([yshift= -2pt] mN1.east);
    \draw[->-=.5] ([xshift= 5pt,yshift= -10pt] mN1.west) to ([yshift= 2pt] mNf1b.east);
    \draw[->-=.5] ([yshift= -2pt] mNf1t.east) to ([xshift= 5pt,yshift= 10pt] mN1.west);
    \draw[->-=.5] ([yshift= 2pt] mNf2b.west) to ([xshift= -5pt,yshift= -10pt] mN2.east);
    \draw[->-=.5] ([xshift= -5pt,yshift= 10pt] mN2.east) to ([yshift= -2pt] mNf2t.west);
    \draw[dashed,->-=.5] (mNf1b.north) to (mNf1t.south);
    \draw[dashed,->-=.5] (mNf2t.south) to (mNf2b.north);
    \draw[dashed,->-=.5] ([yshift= 4pt] mNf2t.west) to ([yshift= 4pt] mNf1t.east);
    \draw[dashed,->-=.5] ([yshift= -4pt] mNf1b.east) to ([yshift= -4pt] mNf2b.west);

    \node at (10,0.5) {\scriptsize $\D$};
    \node at (9-0.3,1) {\scriptsize $\D\texttt{-}r_1$};
    \node at (9-0.3,-1) {\scriptsize $\D\texttt{-}r_1$};
    \node at (11.3,-1) {\scriptsize $\D\texttt{-}r_2$};
    \node at (11.3,1) {\scriptsize $\D\texttt{-}r_2$};
    \node at (10,1.8) {\tiny $r_1\texttt{+}r_2\texttt{+}(2j\texttt{+}1)\D;j\texttt{=}0,\!\cdots\!,k\texttt{-}1$};
    \node at (10,-1.8) {\tiny $r_1\texttt{+}r_2\texttt{+}(2j\texttt{+}1)\D;j\texttt{=}0,\!\cdots\!,k\texttt{-}1$};
    \node at (6.8,0.2) {\tiny $2r_1\texttt{+}2\D j$};
    \node at (6.8,-0.2) {\tiny $j\texttt{=}0,\!\cdots\!,k$};
    \node at (13.2,0.2) {\tiny $2r_2\texttt{+}2\D j$};
    \node at (13.2,-0.2) {\tiny $j\texttt{=}0,\!\cdots\!,k$};
    \node at (9,-0.6) {\scriptsize $\texttt{-}\t_1$};
    \node at (11,-0.6) {\scriptsize $\texttt{-}\t_2$};

    \node at (10,2.3) {$+S_{\rm m}$};

\end{tikzpicture}
\end{equation}
with $\D = \frac{1}{k+1}$, and the contributions of the $2k+4$ monopole-like singlets
\begin{align}
    S_{\rm m}
    &=
    \Big\{
    F_1(1-r_1) + k\D N_2 - N_1 + 1 \pm \t_1
    ,
    F_2(1-r_2) + k\D N_1 - N_2 + 1 \pm \t_2
    \Big\}
    \nonumber\\
    &
    \qquad +
    \Big\{
    F_1(1-r_1) + F_2(1-r_2)
    -\D(N_1+N_2-2) +2\D j
    \pm (\t_1 + \t_2)
    \Big\}_{j=0}^{k-1}
    \,.
\end{align}
We have verified the matching of the partition functions for $k=2,\ldots,8$, using 30 randomly chosen sets of parameters $(N_1,N_2,F_1,F_2)$ for each $k$, with {\tt Quiver Lab} introduced in~\cite{Kim:2026zun}. This provides strong support for the proposed 3d $\CN=2$ two-node duality in~\eqref{eq: new dual}. We leave a proof for generic $(k,N_1,N_2,F_1,F_2)$, together with extensions to larger multi-node dualities with adjoint matters, for future work~\cite{futurework}.
 
\section{Intriligator-Leigh-Strassler duality from nodewise dualities}\label{secrelation}

\noindent
In this section, we derive the ILS duality of 4d $\CN=1$ two-node gauge theories via {\it three} successive nodewise dualities: Seiberg, Kutasov–Schwimmer, and another Seiberg duality. The intermediate dualities generate mesonic interactions that induce Higgsing and mass terms, allowing us to proceed to the next duality. By carefully tracking the Higgs mechanism, integrating out massive fields, and performing field redefinitions, we recover the expected magnetic gauge groups, matter content, and superpotential precisely.

More specifically, for $k=1$, the nodewise sequence consists of three ordinary Seiberg dualities, accompanied by integrating out massive fields. For $k\geq2$, the first Seiberg duality generates an adjoint chiral field at the adjacent node, where we then apply the KS duality. The resulting Higgsing and integrating out massive fields allow us to perform the final Seiberg duality at the beginning node. Integrating out the remaining massive fields then gives the expected magnetic theory. We assume all the intermediate gauge ranks throughout the derivation are positive.

\subsection{\texorpdfstring{$k=1$}{k=1} duality}
\noindent
\begin{figure}[H]
\centering
\begingroup
\def\KOneCol{10}
\def\KOneRow{7}
\tikzset{
  kone title/.style={font=\small,align=center,text width=7.5cm},
  kone step/.style={font=\fontsize{8.5}{9.5}\selectfont,align=center,text width=2.2cm},
  kone gauge/.style={circle,draw,minimum size=13mm,inner sep=0pt,
    font=\fontsize{10}{11}\selectfont},
  kone flavor/.style={rectangle,draw,minimum size=11mm,
    inner sep=1pt,font=\fontsize{8.5}{9.5}\selectfont},
  kone arrow/.style={-{Stealth[length=1.6mm,width=1.1mm]},line width=.45pt},
  kone label/.style={fill=white,inner sep=1.3pt,
    font=\fontsize{8.5}{9.5}\selectfont},
}
\newcommand{\KOneNodes}[2]{%
  \node[kone gauge] (L) at (0,0) {$SU(#1)$};
  \node[kone gauge] (R) at (4,0) {$SU(#2)$};
  \node[kone flavor] (A) at (0,-2.65) {$SU(F_1)$};
  \node[kone flavor] (B) at (4,-2.65) {$SU(F_2)$};
  \draw[kone arrow] (L) to[bend left=5]
    node[kone label,above] {$1/2$} (R);
  \draw[kone arrow] (R) to[bend left=5] (L);
}
\newcommand{\KOneLeftPair}[1]{%
  \draw[kone arrow] (L) to[bend left=5] (A);
  \draw[kone arrow] (A) to[bend left=5]
    node[kone label,left,pos=.5] {$#1$} (L);
}
\newcommand{\KOneRightPair}[1]{%
  \draw[kone arrow] (R) to[bend left=5] (B);
  \draw[kone arrow] (B) to[bend left=5]
    node[kone label,left,pos=.5] {$#1$} (R);
}
\newcommand{\KOneDiagonalOne}[1]{%
  \draw[kone arrow] (A) to[bend left=5] (R);
  \draw[kone arrow] (R) to[bend left=5]
    node[kone label,above,pos=.29] {$#1$} (A);
}
\newcommand{\KOneFlavorPair}{%
  \draw[kone arrow] (A) to[bend left=5] (B);
  \draw[kone arrow] (B) to[bend left=5]
    node[kone label,below,pos=.5] {$\frac12+r_1+r_2$} (A);
}
\newcommand{\KOneLoopOne}{%
  \draw[kone arrow] (A.240)
    .. controls +(-.2,-.85) and +(.2,-.85) .. (A.300);
  \node[kone label] at (0,-3.55) {$2r_1$};
}
\newcommand{\KOneExtraLoopOne}{%
  \draw[kone arrow,blue] (A.190)
    .. controls +(-1.05,-.3) and +(-1.05,.3) .. (A.170);
  \node[kone label,text=blue,anchor=east] at (-.9,-2.05) {$2r_1+1$};
}
\newcommand{\KOneLoopTwo}{%
  \draw[kone arrow,blue] (B.300)
    .. controls +(.2,-.85) and +(-.2,-.85) .. (B.240);
  \node[kone label,text=blue] at (4,-3.55) {$2r_2$};
}
\newcommand{\KOneExtraLoopTwo}{%
  \draw[kone arrow,red] (B.30)
    .. controls +(1.05,.3) and +(1.05,-.3) .. (B.330);
  \node[kone label,text=red,anchor=west] at (4.9,-3.35) {$2r_2+1$};
}
\begin{subcaptiongroup}
\phantomcaption\label{alistofduals:electric}
\phantomcaption\label{alistofduals:first}
\phantomcaption\label{alistofduals:second}
\phantomcaption\label{alistofduals:first-masses}
\phantomcaption\label{alistofduals:third}
\phantomcaption\label{alistofduals:final}
\end{subcaptiongroup}
\begin{adjustbox}{max width=\linewidth,max totalheight=.84\textheight,center}
\begin{tikzpicture}[scale=0.7, every node/.style={scale=0.7}]
\begin{scope}[shift={(0,0)}]
  \KOneNodes{N_1}{N_2}
  \KOneLeftPair{r_1}\KOneRightPair{r_2}
  \node[kone title] at (2,-4.65) {(a) Electric theory};
\end{scope}
\begin{scope}[shift={(\KOneCol,0)}]
  \KOneNodes{h}{N_2}
  \KOneLeftPair{1-r_1}\KOneRightPair{r_2}
  \KOneDiagonalOne{r_1+\frac12}\KOneLoopOne
  \node[kone title] at (2,-4.65) {(b) First Seiberg dual};
\end{scope}
\begin{scope}[shift={(0,-\KOneRow)}]
  \KOneNodes{h}{m}\KOneRightPair{1-r_2}
  \KOneDiagonalOne{\frac12-r_1}
  \draw[kone arrow,blue] (L) to[bend left=5] (B);
  \draw[kone arrow,blue] (B) to[bend left=5]
    node[kone label,above,pos=.71,text=blue] {$\frac12+r_2$} (L);
  \KOneFlavorPair\KOneLoopOne\KOneExtraLoopOne\KOneLoopTwo
  \node[kone title] at (2,-4.65) {(d) After integrating out\\massive fields};
\end{scope}
\begin{scope}[shift={(\KOneCol,-\KOneRow)}]
  \KOneNodes{h}{m}\KOneRightPair{1-r_2}
  \draw[kone arrow] (L.265) to[bend right=5] (A.95);
  \draw[kone arrow] (A.75) to[bend right=5] (L.285);
  \draw[kone arrow] (L.250) to[bend right=5]
    node[kone label,left,pos=.5] {$1-r_1$} (A.110);
  \draw[kone arrow] (A.85) to[bend right=5] (L.275);
  \KOneDiagonalOne{\frac12-r_1}
  \draw[kone arrow,blue] (L) to[bend left=5] (B);
  \draw[kone arrow,blue] (B) to[bend left=5]
    node[kone label,above,pos=.71,text=blue] {$\frac12+r_2$} (L);
  \KOneFlavorPair\KOneLoopOne\KOneExtraLoopOne\KOneLoopTwo
  \node[kone label,anchor=east] at (-.35,-.88) {$1+r_1$};
  \node[kone title] at (2,-4.65) {(c) Second Seiberg dual};
\end{scope}
\begin{scope}[shift={(0,{-2*\KOneRow})}]
  \KOneNodes{\ell}{m}
  \draw[kone arrow] (R.265) to[bend right=5] (B.95);
  \draw[kone arrow] (B.110) to[bend left=5] (R.250);
  \draw[kone arrow,red] (R.285) to[bend left=5]
    node[kone label,right,pos=.5,text=red] {$1+r_2$} (B.75);
  \draw[kone arrow,red] (B.85) to[bend right=5] (R.275);
  \KOneDiagonalOne{\frac12-r_1}
  \draw[kone arrow,blue] (L) to[bend left=5]
    node[kone label,above,pos=.29,text=blue] {$\frac12-r_2$} (B);
  \draw[kone arrow,blue] (B) to[bend left=5] (L);
  \KOneFlavorPair\KOneLoopOne\KOneExtraLoopOne
  \KOneLoopTwo\KOneExtraLoopTwo
  \node[kone label,anchor=west] at (4.35,-1.83) {$1-r_2$};
  \node[kone title] at (2,-4.65) {(e) Third Seiberg dual};
\end{scope}
\begin{scope}[shift={(\KOneCol,{-2*\KOneRow})}]
  \KOneNodes{m}{\ell}
  \draw[kone arrow] (L) to[bend left=5]
    node[kone label,right,pos=.5] {$\frac12-r_1$} (A);
  \draw[kone arrow] (A) to[bend left=5] (L);
  \draw[kone arrow] (R) to[bend left=5]
    node[kone label,right,pos=.5] {$\frac12-r_2$} (B);
  \draw[kone arrow] (B) to[bend left=5] (R);
  \KOneFlavorPair\KOneLoopOne\KOneExtraLoopOne
  \KOneLoopTwo\KOneExtraLoopTwo
  \node[kone title] at (2,-4.65) {(f) Final magnetic quiver};
\end{scope}
\draw[-{Stealth[length=2mm]},thick] (6.3,-1.5)--({\KOneCol-2.1},-1.5);
\node[kone step] at ({(\KOneCol+4.2)/2},-.65) {Seiberg$_1$};
\draw[-{Stealth[length=2mm]},thick]
  ({\KOneCol+2},-5.3)--({\KOneCol+2},{-\KOneRow+1});
\node[kone label,anchor=west] at ({\KOneCol+2.35},{-.5*\KOneRow-2.15}) {Seiberg$_2$};
\draw[-{Stealth[length=2mm]},thick]
  ({\KOneCol-2.1},{-\KOneRow-1.5})--(6.3,{-\KOneRow-1.5});
\node[kone step] at ({(\KOneCol+4.2)/2},{-\KOneRow-.55})
  {Integrating out\\massive fields};
\draw[-{Stealth[length=2mm]},thick]
  (2,{-\KOneRow-5.3})--(2,{-2*\KOneRow+1});
\node[kone label,anchor=west] at (2.35,{-1.5*\KOneRow-2.15}) {Seiberg$_1$};
\draw[-{Stealth[length=2mm]},thick]
  (6.3,{-2*\KOneRow-1.5})--({\KOneCol-2.1},{-2*\KOneRow-1.5});
\node[kone step] at ({(\KOneCol+4.2)/2},{-2*\KOneRow-.4})
  {Integrating out\\massive fields\\and reordering};
\end{tikzpicture}
\end{adjustbox}
\endgroup
\caption{Sequence of Seiberg dualities for the ILS duality at $k=1$. Here $h=F_1+N_2-N_1$, $m=2F_1+F_2-N_1$, and
$\ell=F_1+2F_2-N_2$. For simplicity, pairs of flavor nodes are merged. The labels next to arrows and loops indicate the  $R$-charges of the corresponding chiral multiplets. In (c), the fields of  $R$-charges $1-r_1$ and $1+r_1$ are integrated out to obtain (d). Similarly, the fields of  $R$-charges $1-r_2$ and $1+r_2$ in (e) are integrated out to obtain (f), where the positions of the two gauge nodes are exchanged. We use black, blue, and red lines, respectively, to denote the singlets introduced at the three dualities.}
\label{alistofduals}
\end{figure}
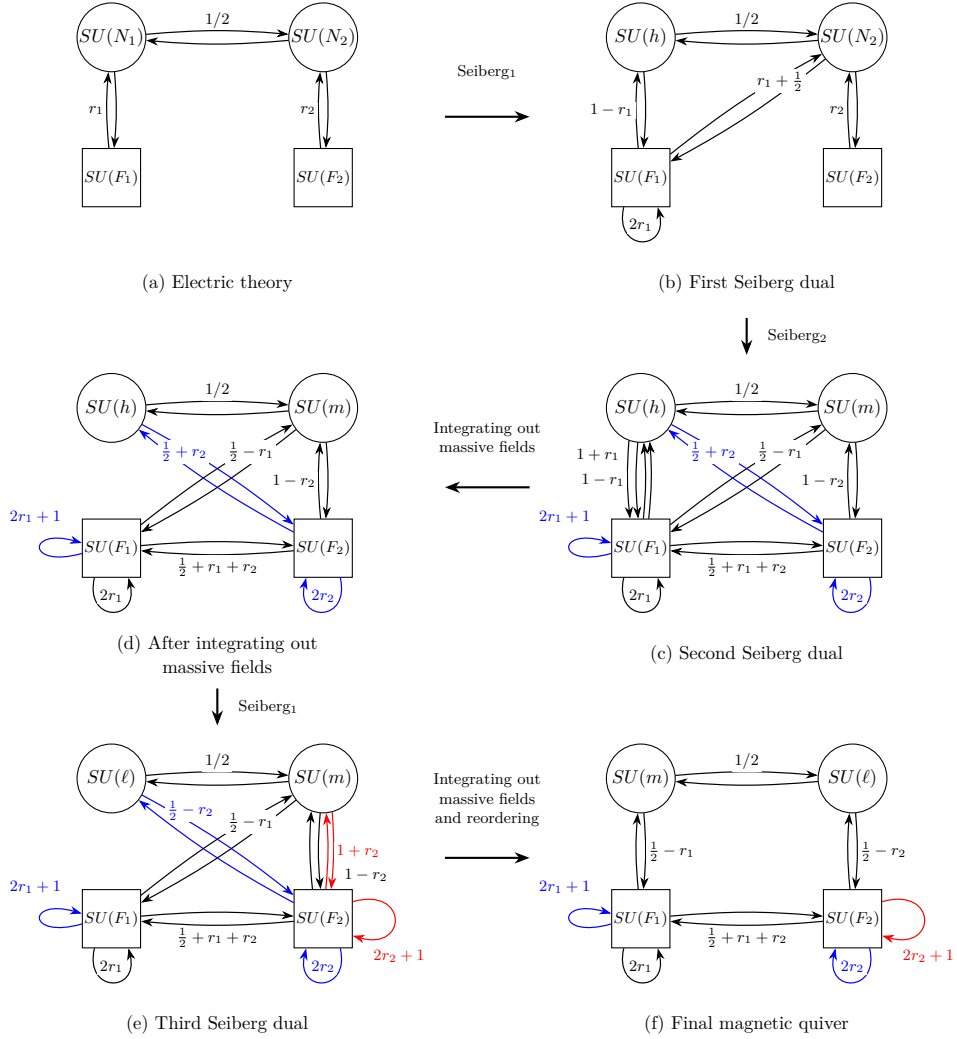 
For $k=1$, the electric superpotential is $W_E=\operatorname{Tr}(X\widetilde X)^2$. The key feature of this quartic interaction is that, upon dualizing one gauge node, it becomes a quadratic superpotential term $W=\operatorname{Tr}M^2$, where $M=\WX X$ is an adjoint meson constructed from the bi-fundamentals. This makes $M$ massive and it can be integrated out, thereby allowing us to apply subsequent Seiberg dualities. We can thus successively dualize the first, second, and first gauge nodes, with the corresponding gauge rank transformations given by
\begin{equation}
\begin{aligned}
(N_1,N_2)
&\xrightarrow{\mathrm{Seiberg}_1}
(N_2+F_1-N_1,N_2)
\\
&\xrightarrow{\mathrm{Seiberg}_2 +\text{Integrating out massive fields}}
(N_2+F_1-N_1,2F_1+F_2-N_1)
\\
&\xrightarrow{\mathrm{Seiberg}_1+\text{Integrating out massive fields}}
(F_1+2F_2-N_2,2F_1+F_2-N_1).
\end{aligned}
\end{equation}
We should exchange the positions of the two resulting gauge nodes following our convention. The dual gauge ranks are then $\WN_1=2F_1+F_2-N_1$ and $\WN_2=F_1+2F_2-N_2$, as expected.

The sequence is illustrated in Figure~\ref{alistofduals}, where pairs of flavor nodes are merged for visual simplicity. The resulting quiver (f) matches that of the magnetic theory in the ILS duality. By tracking the superpotential interactions throughout the duality sequence, we obtain the final superpotential
\begin{align}
    W_{\rm f}&=\operatorname{Tr}(x\widetilde x)^2+
M_{1,0}\widetilde q_1x\widetilde xq_1
+M_{1,1}\widetilde q_1q_1+
M_{2,0}\widetilde q_2\widetilde xxq_2
+M_{2,1}\widetilde q_2q_2
+P_{12}\widetilde q_2\widetilde xq_1
+P_{21}\widetilde q_1xq_2
\nonumber\\
&=W_M|_{k=1}
,
\end{align}
which coincides with the expected magnetic superpotential $W_M$ in \eqref{eq: ILS WM} for $k=1$. Thus, the ILS duality at $k=1$ is derived through a sequence of three Seiberg dualities. This sequence was also discussed from the brane perspective in~\cite{Brodie:1997sz}.

\subsection{Generic \texorpdfstring{$k$}{k} cases}\label{sec:main-generic-k}
\noindent
We now consider the general $k\geq2$ case. As in the $k=1$ case, the simultaneous duality can be expressed as three nodewise dualities.  We first apply the Seiberg duality to the $SU(N_1)$ node of the electric theory, generating an adjoint chiral at the neighboring $SU(N_2)$ node. We then apply the KS duality to the latter node, which generates mesonic adjoints at the original gauge node. Importantly, the resulting superpotential induces Higgsing and mass terms that eliminate these adjoints, allowing us to apply a final Seiberg duality to reach the expected magnetic theory of the ILS duality.

We define
\begin{equation}
\begin{aligned}
 h&=F_1+N_2-N_1,\,\;\;\;n=k(h+F_1+F_2)-N_2,\\
 m&=n-(k-1)h,\,\;\;\;l=m+F_2-h,
\end{aligned}
\end{equation}
and the gauge ranks evolve along the duality sequence as
\begin{equation}\label{eq:chain}
\begin{aligned}
(N_1,N_2)
&\xrightarrow{\text{Seiberg}_1}(h,N_2)\\
&\xrightarrow{\text{KS}_2}(h,n)\xrightarrow{\text{Higgs mechanism} + \text{Integrating out massive fields}}(h,m)\\
&\xrightarrow{\text{Seiberg}_1+\text{Integrating out massive fields}}(l,m).
\end{aligned}
\end{equation}
Here, $m=\WN_1$ and $l=\WN_2$ are the expected final dual gauge ranks. Below, we demonstrate each step by explicitly tracking the chiral spectrum and superpotential interactions.

\subsubsection{The first Seiberg duality}\label{sec:gkr-first}
\begin{figure}[tbp]\centering
\begin{adjustbox}{width=0.8\linewidth,center}
\begin{tikzpicture}[font=\small,>=Latex,
g/.style={circle,draw,minimum size=15mm,inner sep=2pt,align=center},
f/.style={rectangle,draw,minimum width=13mm,minimum height=13mm},
matter/.style={->,thick},
light/.style={->,draw=blue!65!black,text=blue!65!black,thick},
heavy/.style={->,draw=red!70!black,text=red!70!black,dashed,thick},
lab/.style={fill=white,inner sep=1.4pt,align=center,rotate=0}]
\begin{scope}[xshift=0cm]
 \node[font=\bfseries\fontsize{12}{14}\selectfont,align=center] at (3.7,2.35) {\(({\rm a})\) Electric theory};
 \node[g] (g1) at (0,0) {\(SU(N_1)\)};
 \node[g] (g2) at (7.4,0) {\(SU(N_2)\)};
 \node[f] (f1) at (0,-3.6) {\(F_1\)};
 \node[f] (f2) at (7.4,-3.6) {\(F_2\)};
 \draw[light] (g2) to[bend right=5]
   node[lab,above] {\(X\)} (g1);
 \draw[light] (g1) to[bend right=5]
   node[lab,below] {\(\widetilde X\)} (g2);
 \draw[light] (f1) to[bend left=5]
   node[lab,left] {\(Q_1\)} (g1);
 \draw[light] (g1) to[bend left=5]
   node[lab,right] {\(\widetilde Q_1\)} (f1);
 \draw[light] (f2) to[bend right=5]
   node[lab,right] {\(Q_2\)} (g2);
 \draw[light] (g2) to[bend right=5]
   node[lab,left] {\(\widetilde Q_2\)} (f2);
\end{scope}
\draw[transition] (9.05,-1.0) -- node[above,align=center,font=\fontsize{11}{13}\selectfont] {Seiberg} (10.65,-1.0);
\begin{scope}[xshift=12.5cm]
 \node[font=\bfseries\fontsize{12}{14}\selectfont,align=center] at (3.7,2.35) {\(({\rm b})\) First Seiberg dual};
 \node[g] (g1) at (0,0) {\(SU(h)\)};
 \node[g] (g2) at (7.4,0) {\(SU(N_2)\)};
 \node[f] (f1) at (0,-3.6) {\(F_1\)};
 \node[f] (f2) at (7.4,-3.6) {\(F_2\)};
 \draw[light] (g2) to[bend right=5]
   node[lab,above] {\(b\)} (g1);
 \draw[light] (g1) to[bend right=5]
   node[lab,below] {\(\widetilde b\)} (g2);
 \draw[light] (f1) to[bend left=5]
   node[lab,left] {\(q\)} (g1);
 \draw[light] (g1) to[bend left=5]
   node[lab,right] {\(\widetilde q\)} (f1);
 \draw[light] (f2) to[bend right=5]
   node[lab,right] {\(Q_2\)} (g2);
 \draw[light] (g2) to[bend right=5]
   node[lab,left] {\(\widetilde Q_2\)} (f2);
 \draw[light] (f1) to[bend right=10]
   node[lab,below,pos=.80] {\(P\)} (g2);
 \draw[light] (g2) to[bend right=10]
   node[lab,below,pos=.20] {\(\widetilde P\)} (f1);
 \draw[light] (g2) edge[loop above]
   node[lab,above] {\(\Phi\)} (g2);
 \draw[light] (f1) edge[loop below]
   node[lab,below] {\(M\)} (f1);
\end{scope}
\end{tikzpicture}
\end{adjustbox}
\caption{Electric theory (a) and its first Seiberg dual (b).}
\label{fig:gkr-first-comparison}\label{fig:gkr-a}\label{fig:gkr-b}
\end{figure}
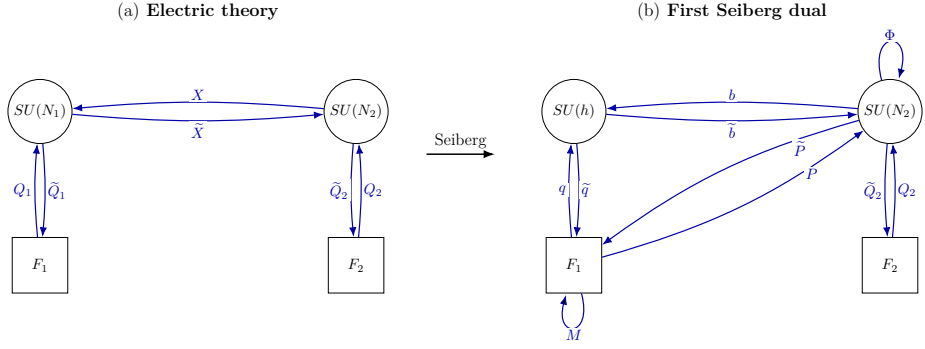

\noindent
The Seiberg duality at the $SU(N_1)$ node of the electric theory is described in Figure \ref{fig:gkr-first-comparison} where each field is labeled next to the corresponding arrows. See Table \ref{tab:gkr-first-comparison} for their charges under the symmetry group. Under the dualization, the $SU(N_1)$ node becomes $SU(h)$, while the fields are mapped as
\begin{equation}
 \WX X \to \Phi,\qquad 
 \WX Q_1\to P,\qquad
 \WQ_1 X \to \WP,\qquad 
 \WQ_1 Q_1\to M,
\end{equation}
so that the superpotential reads
\begin{equation}\label{eq:gkr-W1}
\begin{aligned}
 W_{\rm b}={}&\Tr\Phi^{k+1}+\Tr(b\Phi\widetilde b)
 +M\widetilde q q+\widetilde P\widetilde bq+\widetilde q bP.
\end{aligned}
\end{equation}

\begin{table}[tbp]
\centering
\begin{minipage}[t]{.49\linewidth}
\centering\vspace{0pt}
\parbox[t][6mm][t]{\linewidth}{\centering\small\bfseries \(({\rm a})\) Electric theory\par}\par\smallskip
\scriptsize\setlength{\tabcolsep}{1.5pt}\renewcommand{\arraystretch}{1.35}
\begin{adjustbox}{max width=\linewidth}
\begin{tabular}{@{}lcc|cccc|c@{}}\toprule
Field & $SU(N_1)$ & $SU(N_2)$ & $SU(F_1)_L$ & $SU(F_1)_R$ & $SU(F_2)_L$ & $SU(F_2)_R$ & $R$\\\midrule
$X$ & $\square$ & $\overline{\square}$ & $\mathbf1$ & $\mathbf1$ & $\mathbf1$ & $\mathbf1$ & $\frac1{k+1}$\\
$\widetilde X$ & $\overline{\square}$ & $\square$ & $\mathbf1$ & $\mathbf1$ & $\mathbf1$ & $\mathbf1$ & $\frac1{k+1}$\\
$Q_1$ & $\square$ & $\mathbf1$ & $\square$ & $\mathbf1$ & $\mathbf1$ & $\mathbf1$ & $r_1$\\
$\widetilde Q_1$ & $\overline{\square}$ & $\mathbf1$ & $\mathbf1$ & $\square$ & $\mathbf1$ & $\mathbf1$ & $r_1$\\
$Q_2$ & $\mathbf1$ & $\square$ & $\mathbf1$ & $\mathbf1$ & $\square$ & $\mathbf1$ & $r_2$\\
$\widetilde Q_2$ & $\mathbf1$ & $\overline{\square}$ & $\mathbf1$ & $\mathbf1$ & $\mathbf1$ & $\square$ & $r_2$\\
\bottomrule\end{tabular}
\end{adjustbox}
\end{minipage}
\hfill
\begin{minipage}[t]{.49\linewidth}
\centering\vspace{0pt}
\parbox[t][6mm][t]{\linewidth}{\centering\small\bfseries \(({\rm b})\) First Seiberg dual\par}\par\smallskip
\scriptsize\setlength{\tabcolsep}{1.5pt}\renewcommand{\arraystretch}{1.35}
\begin{adjustbox}{max width=\linewidth}
\begin{tabular}{@{}lcc|cccc|c@{}}\toprule
Field & $SU(h)$ & $SU(N_2)$ & $SU(F_1)_L$ & $SU(F_1)_R$ & $SU(F_2)_L$ & $SU(F_2)_R$ & $R$\\\midrule
$\Phi$ & $\mathbf1$ & $\mathrm{Adj}\oplus\mathbf1$ & $\mathbf1$ & $\mathbf1$ & $\mathbf1$ & $\mathbf1$ & $\frac2{k+1}$\\
$b$ & $\square$ & $\overline{\square}$ & $\mathbf1$ & $\mathbf1$ & $\mathbf1$ & $\mathbf1$ & $\frac{k}{k+1}$\\
$\widetilde b$ & $\overline{\square}$ & $\square$ & $\mathbf1$ & $\mathbf1$ & $\mathbf1$ & $\mathbf1$ & $\frac{k}{k+1}$\\
$q$ & $\square$ & $\mathbf1$ & $\mathbf1$ & $\overline{\square}$ & $\mathbf1$ & $\mathbf1$ & $1-r_1$\\
$\widetilde q$ & $\overline{\square}$ & $\mathbf1$ & $\overline{\square}$ & $\mathbf1$ & $\mathbf1$ & $\mathbf1$ & $1-r_1$\\
$P$ & $\mathbf1$ & $\square$ & $\square$ & $\mathbf1$ & $\mathbf1$ & $\mathbf1$ & $r_1+\frac1{k+1}$\\
$\widetilde P$ & $\mathbf1$ & $\overline{\square}$ & $\mathbf1$ & $\square$ & $\mathbf1$ & $\mathbf1$ & $r_1+\frac1{k+1}$\\
$M$ & $\mathbf1$ & $\mathbf1$ & $\square$ & $\square$ & $\mathbf1$ & $\mathbf1$ & $2r_1$\\
$Q_2$ & $\mathbf1$ & $\square$ & $\mathbf1$ & $\mathbf1$ & $\square$ & $\mathbf1$ & $r_2$\\
$\widetilde Q_2$ & $\mathbf1$ & $\overline{\square}$ & $\mathbf1$ & $\mathbf1$ & $\mathbf1$ & $\square$ & $r_2$\\
\bottomrule\end{tabular}
\end{adjustbox}
\end{minipage}
\caption{Representations and  $U(1)_R$ charges of the electric theory (a) and those after the first Seiberg duality (b).}
\label{tab:gkr-first-comparison}\label{tab:gkr-a}\label{tab:gkr-b}
\end{table}

\subsubsection{The subsequent KS duality}\label{sec:gkr-KS}
\noindent
The $SU(N_2)$ node in theory (b) contains an adjoint chiral $\Phi$ with an $A_k$-type superpotential, allowing us to apply the KS duality.\footnote{Here we employ the traceful version of KS duality~\cite{Bajeot:2023gyl}. The first Seiberg duality generates the full meson $\Phi=\widetilde X X$, transforming as $\operatorname{Adj}{SU(N_2)}\oplus\mathbf{1}$. Accordingly, we apply traceful KS duality, whose magnetic adjoint $Y$ also transforms as $\operatorname{Adj}{SU(n)}\oplus\mathbf{1}$. We retain the trace components throughout the subsequent Higgsing and superpotential reduction. Both traceful and traceless versions of Seiberg-like dualities are physically consistent.} 
\begin{figure}[H]\centering
{\small\bfseries \(({\rm c})\) KS magnetic theory\par}\smallskip
\begin{adjustbox}{max width=0.5\linewidth,max height=52mm}
\begin{tikzpicture}[font=\small,>=Latex,
g/.style={circle,draw,minimum size=15mm,inner sep=2pt,align=center},
f/.style={rectangle,draw,minimum width=13mm,minimum height=13mm},
matter/.style={->,thick},
light/.style={->,draw=blue!65!black,text=blue!65!black,thick},
heavy/.style={->,draw=red!70!black,text=red!70!black,dashed,thick},
lab/.style={fill=white,inner sep=1.4pt,align=center,rotate=0}]

 \node[g] (g1) at (0,0) {\(SU(h)\)};
 \node[g] (g2) at (7.4,0) {\(SU(n)\)};
 \node[f] (f1) at (0,-3.6) {\(F_1\)};
 \node[f] (f2) at (7.4,-3.6) {\(F_2\)};
 \draw[matter] (g2) to[bend right=5]
   node[lab,above] {\(\widetilde u\)} (g1);
 \draw[matter] (g1) to[bend right=5]
   node[lab,below] {\(u\)} (g2);
 \draw[heavy] (g1) edge[loop above]
   node[lab,above] {\(\Sigma_j^{00}\)} (g1);
 \draw[matter] (g2) edge[loop above]
   node[lab,above] {\(Y\)} (g2);
 \draw[heavy] (f1) to[bend left=10]
   node[lab,left] {\(q,\Sigma_j^{01}\)} (g1);
 \draw[heavy] (g1) to[bend left=10]
   node[lab,right] {\(\widetilde q,\Sigma_j^{10}\)} (f1);
 \draw[matter] (f2) to[bend right=10]
   node[lab,right] {\(v_2 \)} (g2);
 \draw[matter] (g2) to[bend right=10]
   node[lab,left] {\(\widetilde v_2 \)} (f2);
 \draw[matter] (f1) to[bend right=13]
   node[lab,below,pos=.90] {\(v_1 \)} (g2);
 \draw[matter] (g2) to[bend right=5]
   node[lab,below,pos=.10] {\(\widetilde v_1 \)} (f1);
 \draw[matter] (f2) to[bend left=10]
   node[lab,above,pos=.7] {\(\textcolor{blue!65!black}{\Sigma_0^{02}},\textcolor{red!70!black}{\Sigma_{j\ge1}^{02}}\)} (g1);
 \draw[matter] (g1) to[bend left=10]
   node[lab,above,pos=.4] {\(\textcolor{blue!65!black}{\Sigma_0^{20}},\textcolor{red!70!black}{\Sigma_{j\ge1}^{20}}\)} (f2);
 \draw[light] (f1) edge[loop below]
   node[lab,below] {\(M,\Sigma_j^{11}\)} (f1);
 \draw[light] (f2) edge[loop below]
   node[lab,below] {\(\Sigma_j^{22}\)} (f2);
 \draw[light] (f2) to[bend right=5]
   node[lab,above] {\(\Sigma_j^{12}\)} (f1);
 \draw[light] (f1) to[bend right=5]
   node[lab,below] {\(\Sigma_j^{21}\)} (f2);
\end{tikzpicture}\end{adjustbox}
\caption{KS magnetic theory $(\rm c)$ after KS duality. Here the index runs over $0\le j\leq k-1$.}
\label{fig:gkr-c}
\end{figure}
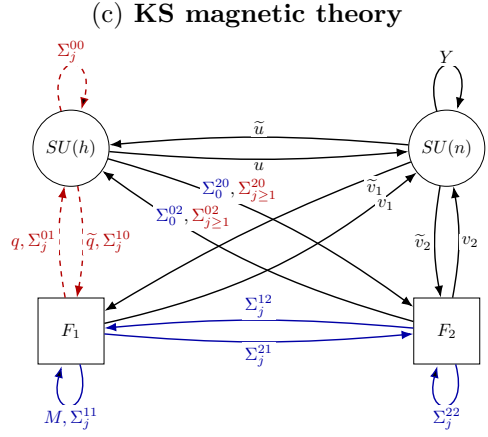
\begin{table}[H]\centering
{\small\bfseries \(({\rm c})\) KS magnetic theory\par}\smallskip
\footnotesize\setlength{\tabcolsep}{3.5pt}\renewcommand{\arraystretch}{1}
\begin{adjustbox}{max width=\linewidth}
\begin{tabular}{@{}lcc|cccc|c@{}}\toprule
Field & $SU(h)$ & $SU(n)$ & $SU(F_1)_L$ & $SU(F_1)_R$ & $SU(F_2)_L$ & $SU(F_2)_R$ & $R$\\\midrule
$Y$ & $\mathbf1$ & $\mathrm{Adj}\oplus\mathbf1$ & $\mathbf1$ & $\mathbf1$ & $\mathbf1$ & $\mathbf1$ & $\frac2{k+1}$\\
$u$ & $\overline{\square}$ & $\square$ & $\mathbf1$ & $\mathbf1$ & $\mathbf1$ & $\mathbf1$ & $\frac{2-k}{k+1}$\\
$\widetilde u$ & $\square$ & $\overline{\square}$ & $\mathbf1$ & $\mathbf1$ & $\mathbf1$ & $\mathbf1$ & $\frac{2-k}{k+1}$\\
$v_1$ & $\mathbf1$ & $\square$ & $\mathbf1$ & $\overline{\square}$ & $\mathbf1$ & $\mathbf1$ & $\frac1{k+1}-r_1$\\
$\widetilde v_1$ & $\mathbf1$ & $\overline{\square}$ & $\overline{\square}$ & $\mathbf1$ & $\mathbf1$ & $\mathbf1$ & $\frac1{k+1}-r_1$\\
$v_2$ & $\mathbf1$ & $\square$ & $\mathbf1$ & $\mathbf1$ & $\mathbf1$ & $\overline{\square}$ & $\frac2{k+1}-r_2$\\
$\widetilde v_2$ & $\mathbf1$ & $\overline{\square}$ & $\mathbf1$ & $\mathbf1$ & $\overline{\square}$ & $\mathbf1$ & $\frac2{k+1}-r_2$\\
$\Sigma_j^{00}$ & $\mathrm{Adj}\oplus\mathbf1$ & $\mathbf1$ & $\mathbf1$ & $\mathbf1$ & $\mathbf1$ & $\mathbf1$ & $2+\frac{2(j-1)}{k+1}$\\
$\Sigma_j^{01}$ & $\square$ & $\mathbf1$ & $\square$ & $\mathbf1$ & $\mathbf1$ & $\mathbf1$ & $1+r_1+\frac{2j}{k+1}$\\
$\Sigma_j^{02}$ & $\square$ & $\mathbf1$ & $\mathbf1$ & $\mathbf1$ & $\square$ & $\mathbf1$ & $r_2+\frac{k+2j}{k+1}$\\
$\Sigma_j^{10}$ & $\overline{\square}$ & $\mathbf1$ & $\mathbf1$ & $\square$ & $\mathbf1$ & $\mathbf1$ & $1+r_1+\frac{2j}{k+1}$\\
$\Sigma_j^{11}$ & $\mathbf1$ & $\mathbf1$ & $\square$ & $\square$ & $\mathbf1$ & $\mathbf1$ & $2r_1+\frac{2(j+1)}{k+1}$\\
$\Sigma_j^{12}$ & $\mathbf1$ & $\mathbf1$ & $\mathbf1$ & $\square$ & $\square$ & $\mathbf1$ & $r_1+r_2+\frac{2j+1}{k+1}$\\
$\Sigma_j^{20}$ & $\overline{\square}$ & $\mathbf1$ & $\mathbf1$ & $\mathbf1$ & $\mathbf1$ & $\square$ & $r_2+\frac{k+2j}{k+1}$\\
$\Sigma_j^{21}$ & $\mathbf1$ & $\mathbf1$ & $\square$ & $\mathbf1$ & $\mathbf1$ & $\square$ & $r_1+r_2+\frac{2j+1}{k+1}$\\
$\Sigma_j^{22}$ & $\mathbf1$ & $\mathbf1$ & $\mathbf1$ & $\mathbf1$ & $\square$ & $\square$ & $2r_2+\frac{2j}{k+1}$\\
$q$ & $\square$ & $\mathbf1$ & $\mathbf1$ & $\overline{\square}$ & $\mathbf1$ & $\mathbf1$ & $1-r_1$\\
$\widetilde q$ & $\overline{\square}$ & $\mathbf1$ & $\overline{\square}$ & $\mathbf1$ & $\mathbf1$ & $\mathbf1$ & $1-r_1$\\
$M$ & $\mathbf1$ & $\mathbf1$ & $\square$ & $\square$ & $\mathbf1$ & $\mathbf1$ & $2r_1$\\
\bottomrule\end{tabular}\end{adjustbox}
\caption{Representations and  $U(1)_R$ charges of KS magnetic theory $(\rm c)$. Every $j$ runs over $0\le j \leq k-1$. }
\label{tab:gkr-c}
\end{table}

The resulting magnetic gauge group is $SU(n)$, with quarks $u$, $v_1$, and $v_2$, and anti-quarks $\Wu$, $\Wv_1$, and $\Wv_2$, as illustrated in Figure~\ref{fig:gkr-c}, with their charges listed in Table~\ref{tab:gkr-c}. 
The duality also introduces mesons $\Sigma_j^{IJ}$ corresponding to composite operators in theory (b):
\begin{equation}\label{eq:gkr-mesonmap}
\begin{gathered}
 \begin{pmatrix}
 b\Phi^j\widetilde b&b\Phi^jP&b\Phi^jQ_2\\
 \widetilde P\Phi^j\widetilde b&\widetilde P\Phi^jP&\widetilde P\Phi^jQ_2\\
 \widetilde Q_2\Phi^j\widetilde b&\widetilde Q_2\Phi^jP&\widetilde Q_2\Phi^jQ_2
 \end{pmatrix}
 \longrightarrow
 \begin{pmatrix}
 \Sigma_j^{00}&\Sigma_j^{01}&\Sigma_j^{02}\\
 \Sigma_j^{10}&\Sigma_j^{11}&\Sigma_j^{12}\\
 \Sigma_j^{20}&\Sigma_j^{21}&\Sigma_j^{22}
 \end{pmatrix},\text{ for $j=0,\ldots,k-1$.}\\
 \Sigma_j^{IJ}:d_I\times d_J\; \text{matrix with}\;\;(d_0,d_1,d_2)=(h,F_1,F_2).
\end{gathered}
\end{equation}
 The resulting magnetic superpotential is
\begin{equation}\label{eq:gkr-W2}
\begin{aligned}
 &W_{\rm c}=\Tr Y^{k+1}+\Tr\Sigma_1^{00}
 +M\widetilde qq+\Sigma_0^{10}q+\widetilde q\Sigma_0^{01}\\
 &+\sum_{j=0}^{k-1}
 \bigg(
 \Tr(\Sigma_j^{00}\widetilde uY^{k\texttt{-}1\texttt{-}j}u)
 +\sum_{I=1}^2\bigl[
 \Sigma_j^{0I}\widetilde v_IY^{k\texttt{-}1\texttt{-}j}u
 +\Sigma_j^{I0}\widetilde uY^{k\texttt{-}1\texttt{-}j}v_I\bigr]
 +\sum_{I,J=1}^2
 \Sigma_j^{IJ}\widetilde v_JY^{k\texttt{-}1\texttt{-}j}v_I
 \bigg).
\end{aligned}
\end{equation}
Importantly, the linear term $\Tr\Sigma_1^{00}$, inherited from $W_{\rm b}$ for all $k\geq2$, induces Higgsing from its F-term equations, as we demonstrate below.

\subsubsection{Higgs mechanism}
\noindent
For $k\geq2$, the F-term equations for $\S_j^{00}$ in $W_{\rm c}$ impose
\begin{equation}\label{eq}
\widetilde uY^su=-\delta_{s,k-2}\mathbf1_h,
\qquad 0\le s\le k-1.
\end{equation}
These conditions admit a supersymmetric vacuum with non-zero vacuum expectation value (VEV) of $u,\widetilde u$ for $k\geq 2$, and also of $Y$ for $k>2$. The resulting Higgsing breaks the gauge group as follows:
\begin{equation}
 SU(h)\times SU(n)\longrightarrow SU(h)\times SU(m),
 \qquad
 m=n-(k-1)h
 \,,
\end{equation}
The explicit vacuum solution, gauge fixing,  elimination of
massive fields, and superpotentials $W_d,W_e$ are illustrated in Appendix~\ref{app:gkr-higgs}.

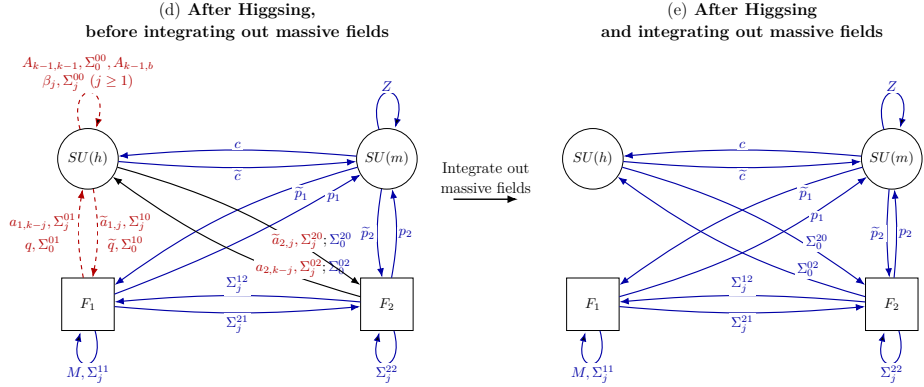
\begin{figure}[H]\centering
\begin{adjustbox}{width=0.8\linewidth,center}
\begin{tikzpicture}[font=\small,>=Latex,
g/.style={circle,draw,minimum size=15mm,inner sep=2pt,align=center},
f/.style={rectangle,draw,minimum width=13mm,minimum height=13mm},
matter/.style={->,thick},
light/.style={->,draw=blue!65!black,text=blue!65!black,thick},
heavy/.style={->,draw=red!70!black,text=red!70!black,dashed,thick},
lab/.style={fill=white,inner sep=1.4pt,align=center,rotate=0}]
\begin{scope}[xshift=0cm]
 \node[font=\bfseries\fontsize{12}{14}\selectfont,align=center] at (3.7,3.4) {\(({\rm d})\) After Higgsing,\\before integrating out massive fields};
 \node[g] (g1) at (0,0) {\(SU(h)\)};
 \node[g] (g2) at (7.4,0) {\(SU(m)\)};
 \node[f] (f1) at (0,-3.6) {\(F_1\)};
 \node[f] (f2) at (7.4,-3.6) {\(F_2\)};
 \draw[light] (g2) to[bend right=5]
   node[lab,above] {\(c \)} (g1);
 \draw[light] (g1) to[bend right=5]
   node[lab,below] {\(\widetilde c \)} (g2);
 \draw[heavy] (g1) edge[loop above]
   node[lab,above] {\(A_{k-1,k-1},\Sigma_0^{00},A_{k-1,b}\)\\\(\beta_j,\Sigma_j^{00}\ (j\ge1)\)} (g1);
 \draw[light] (g2) edge[loop above]
   node[lab,above] {\(Z\)} (g2);
 \draw[heavy] (f1) to[bend left=10]
   node[lab,left] {\(\begin{gathered}a_{1,k-j},\Sigma_j^{01}\\q,\Sigma_0^{01}\end{gathered}\)} (g1);
 \draw[heavy] (g1) to[bend left=10]
   node[lab,right] {\(\begin{gathered}\widetilde a_{1,j},\Sigma_j^{10}\\\widetilde q,\Sigma_0^{10}\end{gathered}\)} (f1);
 \draw[light] (f2) to[bend right=10]
   node[lab,right] {\(p_2 \)} (g2);
 \draw[light] (g2) to[bend right=10]
   node[lab,left] {\(\widetilde p_2 \)} (f2);
 \draw[light] (f1) to[bend right=5]
   node[lab,below,pos=.90] {\(p_1 \)} (g2);
 \draw[light] (g2) to[bend right=10]
   node[lab,below,pos=.20] {\(\widetilde p_1 \)} (f1);
 \draw[matter] (f2) to[bend left=10]
   node[lab,above,pos=.2] {\(\textcolor{red!70!black}{a_{2,k-j},\Sigma_j^{02}};\textcolor{blue!65!black}{\Sigma_0^{02}}\)} (g1);
 \draw[matter] (g1) to[bend left=10]
   node[lab,above,pos=.8] {\(\textcolor{red!70!black}{\widetilde a_{2,j},\Sigma_j^{20}};\textcolor{blue!65!black}{\Sigma_0^{20}}\)} (f2);
 \draw[light] (f1) edge[loop below]
   node[lab,below] {\(M,\Sigma_j^{11}\)} (f1);
 \draw[light] (f2) edge[loop below]
   node[lab,below] {\(\Sigma_j^{22}\)} (f2);
 \draw[light] (f2) to[bend right=5]
   node[lab,above] {\(\Sigma_j^{12}\)} (f1);
 \draw[light] (f1) to[bend right=5]
   node[lab,below] {\(\Sigma_j^{21}\)} (f2);
\end{scope}
\draw[transition] (9.05,-1.0) -- node[above,align=center,font=\fontsize{11}{13}\selectfont] {Integrate out\\massive fields} (10.65,-1.0);
\begin{scope}[xshift=12.5cm]
 \node[font=\bfseries\fontsize{12}{14}\selectfont,align=center] at (3.7,3.4) {\(({\rm e})\) After Higgsing\\and integrating out massive fields};
 \node[g] (g1) at (0,0) {\(SU(h)\)};
 \node[g] (g2) at (7.4,0) {\(SU(m)\)};
 \node[f] (f1) at (0,-3.6) {\(F_1\)};
 \node[f] (f2) at (7.4,-3.6) {\(F_2\)};
 \draw[light] (g2) to[bend right=5]
   node[lab,above] {\(c \)} (g1);
 \draw[light] (g1) to[bend right=5]
   node[lab,below] {\(\widetilde c \)} (g2);
 \draw[light] (g2) edge[loop above]
   node[lab,above] {\(Z\)} (g2);
 \draw[light] (f2) to[bend right=5]
   node[lab,right] {\(p_2 \)} (g2);
 \draw[light] (g2) to[bend right=5]
   node[lab,left] {\(\widetilde p_2 \)} (f2);
 \draw[light] (f1) to[bend right=10]
   node[lab,below,pos=.80] {\(p_1 \)} (g2);
 \draw[light] (g2) to[bend right=10]
   node[lab,below,pos=.20] {\(\widetilde p_1 \)} (f1);
 \draw[light] (f2) to[bend left=10]
   node[lab,above,pos=.2] {\(\Sigma_0^{02}\)} (g1);
 \draw[light] (g1) to[bend left=10]
   node[lab,above,pos=.8] {\(\Sigma_0^{20}\)} (f2);
 \draw[light] (f1) edge[loop below]
   node[lab,below] {\(M,\Sigma_j^{11}\)} (f1);
 \draw[light] (f2) edge[loop below]
   node[lab,below] {\(\Sigma_j^{22}\)} (f2);
 \draw[light] (f2) to[bend right=5]
   node[lab,above] {\(\Sigma_j^{12}\)} (f1);
 \draw[light] (f1) to[bend right=5]
   node[lab,below] {\(\Sigma_j^{21}\)} (f2);
\end{scope}
\end{tikzpicture}
\end{adjustbox}
\caption{After Higgsing, before \(({\rm d})\) and after \(({\rm e})\) integrating out massive fields. The gauge group is $SU(h)\times SU(m)$ in both theories. Red dashed arrows denote fields eliminated by the superpotential masses. An unrestricted meson index runs over $0\le j<k$; the upper-quark pairs have $1\le j<k$, and the blocks $A_{k-1,b}$ have $1\le b\le k-2$. The block $A_{k-1,k-1}$ is displayed separately.}
\label{fig:gkr-higgs-comparison}\label{fig:gkr-d}\label{fig:gkr-f}
\end{figure}

After Higgsing and integrating out the massive fields, we find that the adjoint components of $\Sigma_j^{00}$ are completely eliminated, as described in theory (e) of Figure~\ref{fig:gkr-d} and Table \ref{tab:gkr-d}. There is no adjoint of $SU(h)$ gauge node. Remarkably, this allows us to apply Seiberg duality once again to the $SU(h)$ gauge node.
\begin{table}[tbp]
\centering
\begin{minipage}[t]{.49\linewidth}
\centering\vspace{0pt}
\parbox[t][10mm][t]{\linewidth}{\centering\small\bfseries \(({\rm d})\) After Higgsing,\\before integrating out massive fields\par}\par\smallskip
\scriptsize\setlength{\tabcolsep}{1.5pt}\renewcommand{\arraystretch}{1.35}
\begin{adjustbox}{max width=\linewidth}
\begin{tabular}{@{}lcc|cccc|c@{}}\toprule
Field & $SU(h)$ & $SU(m)$ & $SU(F_1)_L$ & $SU(F_1)_R$ & $SU(F_2)_L$ & $SU(F_2)_R$ & $R $\\\midrule
$A_{k-1,k-1}$ & $\mathrm{Adj}\oplus\mathbf1$ & $\mathbf1$ & $\mathbf1$ & $\mathbf1$ & $\mathbf1$ & $\mathbf1$ & $\frac2{k+1}$\\
$A_{k-1,b}$ & $\mathrm{Adj}\oplus\mathbf1$ & $\mathbf1$ & $\mathbf1$ & $\mathbf1$ & $\mathbf1$ & $\mathbf1$ & $\frac{2(k-b)}{k+1}$\\
$Z$ & $\mathbf1$ & $\mathrm{Adj}\oplus\mathbf1$ & $\mathbf1$ & $\mathbf1$ & $\mathbf1$ & $\mathbf1$ & $\frac2{k+1}$\\
$c$ & $\square$ & $\overline{\square}$ & $\mathbf1$ & $\mathbf1$ & $\mathbf1$ & $\mathbf1$ & $\frac{k}{k+1}$\\
$\widetilde c$ & $\overline{\square}$ & $\square$ & $\mathbf1$ & $\mathbf1$ & $\mathbf1$ & $\mathbf1$ & $\frac{k}{k+1}$\\
$\beta_b$ & $\mathrm{Adj}\oplus\mathbf1$ & $\mathbf1$ & $\mathbf1$ & $\mathbf1$ & $\mathbf1$ & $\mathbf1$ & $\frac{2(1-b)}{k+1}$\\
$a_{1,a}$ & $\square$ & $\mathbf1$ & $\mathbf1$ & $\overline{\square}$ & $\mathbf1$ & $\mathbf1$ & $\frac1{k+1}-r_1+t_a$\\
$\widetilde a_{1,a}$ & $\overline{\square}$ & $\mathbf1$ & $\overline{\square}$ & $\mathbf1$ & $\mathbf1$ & $\mathbf1$ & $\frac1{k+1}-r_1-t_a$\\
$p_1$ & $\mathbf1$ & $\square$ & $\mathbf1$ & $\overline{\square}$ & $\mathbf1$ & $\mathbf1$ & $\frac1{k+1}-r_1$\\
$\widetilde p_1$ & $\mathbf1$ & $\overline{\square}$ & $\overline{\square}$ & $\mathbf1$ & $\mathbf1$ & $\mathbf1$ & $\frac1{k+1}-r_1$\\
$a_{2,a}$ & $\square$ & $\mathbf1$ & $\mathbf1$ & $\mathbf1$ & $\mathbf1$ & $\overline{\square}$ & $\frac2{k+1}-r_2+t_a$\\
$\widetilde a_{2,a}$ & $\overline{\square}$ & $\mathbf1$ & $\mathbf1$ & $\mathbf1$ & $\overline{\square}$ & $\mathbf1$ & $\frac2{k+1}-r_2-t_a$\\
$p_2$ & $\mathbf1$ & $\square$ & $\mathbf1$ & $\mathbf1$ & $\mathbf1$ & $\overline{\square}$ & $\frac2{k+1}-r_2$\\
$\widetilde p_2$ & $\mathbf1$ & $\overline{\square}$ & $\mathbf1$ & $\mathbf1$ & $\overline{\square}$ & $\mathbf1$ & $\frac2{k+1}-r_2$\\
$\Sigma_j^{00}$ & $\mathrm{Adj}\oplus\mathbf1$ & $\mathbf1$ & $\mathbf1$ & $\mathbf1$ & $\mathbf1$ & $\mathbf1$ & $2+\frac{2(j-1)}{k+1}$\\
$\Sigma_j^{01}$ & $\square$ & $\mathbf1$ & $\square$ & $\mathbf1$ & $\mathbf1$ & $\mathbf1$ & $1+r_1+\frac{2j}{k+1}$\\
$\Sigma_j^{02}$ & $\square$ & $\mathbf1$ & $\mathbf1$ & $\mathbf1$ & $\square$ & $\mathbf1$ & $r_2+\frac{k+2j}{k+1}$\\
$\Sigma_j^{10}$ & $\overline{\square}$ & $\mathbf1$ & $\mathbf1$ & $\square$ & $\mathbf1$ & $\mathbf1$ & $1+r_1+\frac{2j}{k+1}$\\
$\Sigma_j^{11}$ & $\mathbf1$ & $\mathbf1$ & $\square$ & $\square$ & $\mathbf1$ & $\mathbf1$ & $2r_1+\frac{2(j+1)}{k+1}$\\
$\Sigma_j^{12}$ & $\mathbf1$ & $\mathbf1$ & $\mathbf1$ & $\square$ & $\square$ & $\mathbf1$ & $r_1+r_2+\frac{2j+1}{k+1}$\\
$\Sigma_j^{20}$ & $\overline{\square}$ & $\mathbf1$ & $\mathbf1$ & $\mathbf1$ & $\mathbf1$ & $\square$ & $r_2+\frac{k+2j}{k+1}$\\
$\Sigma_j^{21}$ & $\mathbf1$ & $\mathbf1$ & $\square$ & $\mathbf1$ & $\mathbf1$ & $\square$ & $r_1+r_2+\frac{2j+1}{k+1}$\\
$\Sigma_j^{22}$ & $\mathbf1$ & $\mathbf1$ & $\mathbf1$ & $\mathbf1$ & $\square$ & $\square$ & $2r_2+\frac{2j}{k+1}$\\
$q$ & $\square$ & $\mathbf1$ & $\mathbf1$ & $\overline{\square}$ & $\mathbf1$ & $\mathbf1$ & $1-r_1$\\
$\widetilde q$ & $\overline{\square}$ & $\mathbf1$ & $\overline{\square}$ & $\mathbf1$ & $\mathbf1$ & $\mathbf1$ & $1-r_1$\\
$M$ & $\mathbf1$ & $\mathbf1$ & $\square$ & $\square$ & $\mathbf1$ & $\mathbf1$ & $2r_1$\\
\bottomrule\end{tabular}
\end{adjustbox}
\end{minipage}
\hfill
\begin{minipage}[t]{.49\linewidth}
\centering\vspace{0pt}
\parbox[t][10mm][t]{\linewidth}{\centering\small\bfseries \(({\rm e})\) After Higgsing\\and integrating out massive fields\par}\par\smallskip
\scriptsize\setlength{\tabcolsep}{1.5pt}\renewcommand{\arraystretch}{1.35}
\begin{adjustbox}{max width=\linewidth}
\begin{tabular}{@{}lcc|cccc|c@{}}\toprule
Field & $SU(h)$ & $SU(m)$ & $SU(F_1)_L$ & $SU(F_1)_R$ & $SU(F_2)_L$ & $SU(F_2)_R$ & $R $\\\midrule
$Z$ & $\mathbf1$ & $\mathrm{Adj}\oplus\mathbf1$ & $\mathbf1$ & $\mathbf1$ & $\mathbf1$ & $\mathbf1$ & $\frac2{k+1}$\\
$c$ & $\square$ & $\overline{\square}$ & $\mathbf1$ & $\mathbf1$ & $\mathbf1$ & $\mathbf1$ & $\frac{k}{k+1}$\\
$\widetilde c$ & $\overline{\square}$ & $\square$ & $\mathbf1$ & $\mathbf1$ & $\mathbf1$ & $\mathbf1$ & $\frac{k}{k+1}$\\
$p_1$ & $\mathbf1$ & $\square$ & $\mathbf1$ & $\overline{\square}$ & $\mathbf1$ & $\mathbf1$ & $\frac1{k+1}-r_1$\\
$\widetilde p_1$ & $\mathbf1$ & $\overline{\square}$ & $\overline{\square}$ & $\mathbf1$ & $\mathbf1$ & $\mathbf1$ & $\frac1{k+1}-r_1$\\
$p_2$ & $\mathbf1$ & $\square$ & $\mathbf1$ & $\mathbf1$ & $\mathbf1$ & $\overline{\square}$ & $\frac2{k+1}-r_2$\\
$\widetilde p_2$ & $\mathbf1$ & $\overline{\square}$ & $\mathbf1$ & $\mathbf1$ & $\overline{\square}$ & $\mathbf1$ & $\frac2{k+1}-r_2$\\
$\Sigma_0^{02}$ & $\square$ & $\mathbf1$ & $\mathbf1$ & $\mathbf1$ & $\square$ & $\mathbf1$ & $r_2+\frac{k}{k+1}$\\
$\Sigma_0^{20}$ & $\overline{\square}$ & $\mathbf1$ & $\mathbf1$ & $\mathbf1$ & $\mathbf1$ & $\square$ & $r_2+\frac{k}{k+1}$\\
$M$ & $\mathbf1$ & $\mathbf1$ & $\square$ & $\square$ & $\mathbf1$ & $\mathbf1$ & $2r_1$\\
$\Sigma_j^{11}$ & $\mathbf1$ & $\mathbf1$ & $\square$ & $\square$ & $\mathbf1$ & $\mathbf1$ & $2r_1+\frac{2(j+1)}{k+1}$\\
$\Sigma_j^{12}$ & $\mathbf1$ & $\mathbf1$ & $\mathbf1$ & $\square$ & $\square$ & $\mathbf1$ & $r_1+r_2+\frac{2j+1}{k+1}$\\
$\Sigma_j^{21}$ & $\mathbf1$ & $\mathbf1$ & $\square$ & $\mathbf1$ & $\mathbf1$ & $\square$ & $r_1+r_2+\frac{2j+1}{k+1}$\\
$\Sigma_j^{22}$ & $\mathbf1$ & $\mathbf1$ & $\mathbf1$ & $\mathbf1$ & $\square$ & $\square$ & $2r_2+\frac{2j}{k+1}$\\
\bottomrule\end{tabular}
\end{adjustbox}
\end{minipage}
\caption{Representations and  $U(1)_R$ charges of theories after Higgsing, before \(({\rm d})\) and after \(({\rm e})\) integrating out massive fields. The blocks $A_{k-1,b}$ have $1\le b\le k-2$, the block $A_{k-1,k-1}$ is listed separately.}
\label{tab:gkr-d}
\end{table}
\subsubsection{Final Seiberg duality}\label{sec:gkr-final}
\noindent
The final application of Seiberg duality to the $SU(h)$ node in theory (e) produces an $SU(l)$ node in theory (f), as given in Figure~\ref{fig:gkr-final-comparison} and Table \ref{tab:gkr-g}. The mesonic fields in (f) are mapped from the composite fields in (e) as follows:
\begin{equation}\label{eq:gkr-last-mesons}
 \widetilde cc\to\Psi,\qquad 
 \widetilde c\Sigma_0^{02}\to U,\qquad
 -\Sigma_0^{20}c\to \widetilde U,\qquad 
 -\Sigma_0^{20}\Sigma_0^{02}\to H.
\end{equation}
Using the reduced superpotential \eqref{eq:gkr-W4}, the superpotential of the theory (f) reads
\begin{equation}\label{eq:gkr-W5}
\begin{aligned}
 &W_{\rm f}= \Tr Z^{k+1}
 +\Tr\bigl[\Psi((k+1)Z+v\widetilde v-p_1M\widetilde p_1)\bigr]\\
 &+\bigl[\widetilde U(p_2+vq_2)
 +(\widetilde p_2+\widetilde q_2\widetilde v)U\bigr]
 +H\widetilde q_2q_2+\sum_{j=0}^{k-1}\sum_{I,J=1}^2
 \Sigma_j^{IJ}\widetilde p_JZ^{k-1-j}p_I.
\end{aligned}
\end{equation}
The F-term equations of the massive pairs $(Z,\Psi)$, $(p_2,\widetilde U)$,
and $(\widetilde p_2,U)$ are
\begin{equation}\label{eq:gkr-last-eoms}
 \frac{\partial W_{\rm f}}{\partial\Psi}
 =\frac{\partial W_{\rm f}}{\partial\widetilde U}
 =\frac{\partial W_{\rm f}}{\partial U}
 =\frac{\partial W_{\rm f}}{\partial Z}
 =\frac{\partial W_{\rm f}}{\partial p_2}
 =\frac{\partial W_{\rm f}}{\partial\widetilde p_2}=0
 \,,
\end{equation}
which give the solutions
\begin{equation}\label{eq:gkr-last-solution}
 Z=\frac{-v\widetilde v+q_1M\Wq_1}{k+1},\qquad
 p_2=-vq_2,\qquad\widetilde p_2=-\widetilde q_2\widetilde v
 \,.
\end{equation}
Plugging them into $W_{\rm f}$ and identifying $q_1=p_1$ and $\widetilde q_1=\widetilde p_1$ bring us to the superpotential of the theory (g) as follows
\begin{align}
    &W_{\rm g}=
 \Tr\!(\frac{-v\widetilde v+q_1M\widetilde q_1}{k+1})^{k+1}
 +H\widetilde q_2q_2\nonumber\\
 &+\sum_{j=0}^{k-1}
 \Big(
 \Sigma_j^{11}\widetilde q_1
 (\frac{-v\widetilde v+q_1M\widetilde q_1}{k+1})^{k-1-j}q_1-\Sigma_j^{12}\widetilde q_2\Wv
 (\frac{-v\widetilde v+q_1M\widetilde q_1}{k+1})^{k-1-j}q_1
 \nonumber\\
 &\qquad-\Sigma_j^{21}\widetilde q_1
 (\frac{-v\widetilde v+q_1M\widetilde q_1}{k+1})^{k-1-j}vq_2+\Sigma_j^{22}\widetilde q_2\Wv
 (\frac{-v\widetilde v+q_1M\widetilde q_1}{k+1})^{k-1-j}vq_2\Big)
 \,.
\end{align} 
To match the final form of the magnetic superpotential, we perform field redefinitions:
\begin{equation}\label{eq:gkr-singlet-redefinitions}
\begin{aligned}
&x=\rho v,\quad\tilde x=\rho \tilde v,\quad\rho=\frac{i}{\sqrt{k+1}},\\
& M_{1,0}=M,\qquad M_{2,k}=H,\\[5pt]
 & M_{1,j+1}
 =\Sigma_j^{11}
 +\frac{1}{k+1}\sum_{b=0}^{j-1}\Sigma_b^{11}\widetilde q_1Z^{j-1-b}q_1M+\frac{k-j}{(k+1)^2}M\widetilde q_1Z^jq_1M,\\[5pt]
 &M_{2,j}
  =-(k+1)\Sigma_j^{22},\\[5pt]
& P_{12,j}
  =(k+1)\rho\Big[\Sigma_j^{12}
 +\frac{1}{k+1}\sum_{b=0}^{j-1}M\widetilde q_1Z^{j-1-b}
 \bigl(q_1\Sigma_b^{12}-vq_2\Sigma_b^{22}\bigr)\Big],\\[5pt]
& P_{21,j}
  =(k+1)\rho\Big[\Sigma_j^{21}
 +\frac{1}{k+1}\sum_{b=0}^{j-1}\Sigma_b^{21}\widetilde q_1Z^{j-1-b}q_1M\Big].
\end{aligned}
\end{equation}
Substituting these into $W_{\rm g}$, we arrive at the magnetic superpotential
\begin{equation}\label{eq:gkr-WM}
\begin{aligned}
 &W_M= \Tr(x\widetilde x)^{k+1}
 +\sum_{j=0}^k
 \Big(
 M_{1,j}\widetilde q_1(x\widetilde x)^{k-j}q_1
 +M_{2,j}\widetilde q_2(\widetilde xx)^{k-j}q_2
 \Big)
 \\
 &\qquad\qquad\qquad
 +\sum_{j=0}^{k-1}
 \Big(
 P_{12,j}\widetilde q_2(\widetilde xx)^{k-1-j}\widetilde xq_1
 +P_{21,j}\widetilde q_1(x\widetilde x)^{k-1-j}xq_2
 \Big)
 \,.
\end{aligned}
\end{equation}
We depict the resulting theory before and after integrating out the massive fields in Figure~\ref{fig:gkr-final-comparison}. The corresponding representations and $R$-charges are collected in Table~\ref{tab:gkr-final-comparison}.
\begin{figure}[tbp]\centering
\begin{adjustbox}{width=0.8\linewidth,center}
\begin{tikzpicture}[font=\small,>=Latex,
g/.style={circle,draw,minimum size=15mm,inner sep=2pt,align=center},
f/.style={rectangle,draw,minimum width=13mm,minimum height=13mm},
matter/.style={->,thick},
light/.style={->,draw=blue!65!black,text=blue!65!black,thick},
heavy/.style={->,draw=red!70!black,text=red!70!black,dashed,thick},
lab/.style={fill=white,inner sep=1.4pt,align=center,rotate=0}]
\begin{scope}[xshift=0cm]
 \node[font=\bfseries\fontsize{12}{14}\selectfont,align=center] at (3.7,2.8) {\(({\rm f})\) Final Seiberg dual,\\before integrating out massive fields};
 \node[g] (g1) at (0,0) {\(SU(m)\)};
 \node[g] (g2) at (7.4,0) {\(SU(l)\)};
 \node[f] (f1) at (0,-3.6) {\(F_1\)};
 \node[f] (f2) at (7.4,-3.6) {\(F_2\)};
 \draw[light] (g2) to[bend right=5]
   node[lab,above] {\(v\)} (g1);
 \draw[light] (g1) to[bend right=5]
   node[lab,below] {\(\widetilde v\)} (g2);
 \draw[heavy] (g1) edge[loop above]
   node[lab,above] {\(Z,\Psi\)} (g1);
 \draw[light] (f1) to[bend left=5]
   node[lab,left] {\(p_1 \)} (g1);
 \draw[light] (g1) to[bend left=5]
   node[lab,right] {\(\widetilde p_1 \)} (f1);
 \draw[light] (f2) to[bend right=5]
   node[lab,right] {\(q_2\)} (g2);
 \draw[light] (g2) to[bend right=5]
   node[lab,left] {\(\widetilde q_2\)} (f2);
 \draw[heavy] (f2) to[bend left=10]
   node[lab,above,pos=.42] {\(p_2 ,U\)} (g1);
 \draw[heavy] (g1) to[bend left=10]
   node[lab,above,pos=.55] {\(\widetilde p_2 ,\widetilde U\)} (f2);
 \draw[light] (f1) edge[loop below]
   node[lab,below] {\(M,\Sigma_j^{11}\)} (f1);
 \draw[light] (f2) edge[loop below]
   node[lab,below] {\(H,\Sigma_j^{22}\)} (f2);
 \draw[light] (f2) to[bend right=5]
   node[lab,above] {\(\Sigma_j^{12}\)} (f1);
 \draw[light] (f1) to[bend right=5]
   node[lab,below] {\(\Sigma_j^{21}\)} (f2);
\end{scope}
\draw[transition] (9.05,-1.0) -- node[above,align=center,font=\fontsize{11}{13}\selectfont] {Integrate out\\and redefine} (10.65,-1.0);
\begin{scope}[xshift=12.5cm]
 \node[font=\bfseries\fontsize{12}{14}\selectfont,align=center] at (3.7,2.8) {\(({\rm g})\) Final magnetic theory};
 \node[g] (g1) at (0,0) {\(SU(m)\)};
 \node[g] (g2) at (7.4,0) {\(SU(l)\)};
 \node[f] (f1) at (0,-3.6) {\(F_1\)};
 \node[f] (f2) at (7.4,-3.6) {\(F_2\)};
 \draw[light] (g2) to[bend right=5]
   node[lab,above] {\(x\)} (g1);
 \draw[light] (g1) to[bend right=5]
   node[lab,below] {\(\widetilde x\)} (g2);
 \draw[light] (f1) to[bend left=5]
   node[lab,left] {\(q_1=p_1 \)} (g1);
 \draw[light] (g1) to[bend left=5]
   node[lab,right] {\(\widetilde q_1=\widetilde p_1 \)} (f1);
 \draw[light] (f2) to[bend right=5]
   node[lab,right] {\(q_2\)} (g2);
 \draw[light] (g2) to[bend right=5]
   node[lab,left] {\(\widetilde q_2\)} (f2);
 \draw[light] (f1) edge[loop below]
   node[lab,below] {\(M_{1,j},\ 0\le j\le k\)} (f1);
 \draw[light] (f2) edge[loop below]
   node[lab,below] {\(M_{2,j},\ 0\le j\le k\)} (f2);
 \draw[light] (f2) to[bend right=5]
   node[lab,above] {\(P_{12,j},\ 0\le j<k\)} (f1);
 \draw[light] (f1) to[bend right=5]
   node[lab,below] {\(P_{21,j},\ 0\le j<k\)} (f2);
\end{scope}
\end{tikzpicture}
\end{adjustbox}
\caption{Theories after the final Seiberg duality, before (f) and after (g) integrating out the massive fields.}
\label{fig:gkr-final-comparison}\label{fig:gkr-g}\label{fig:gkr-h}
\end{figure}
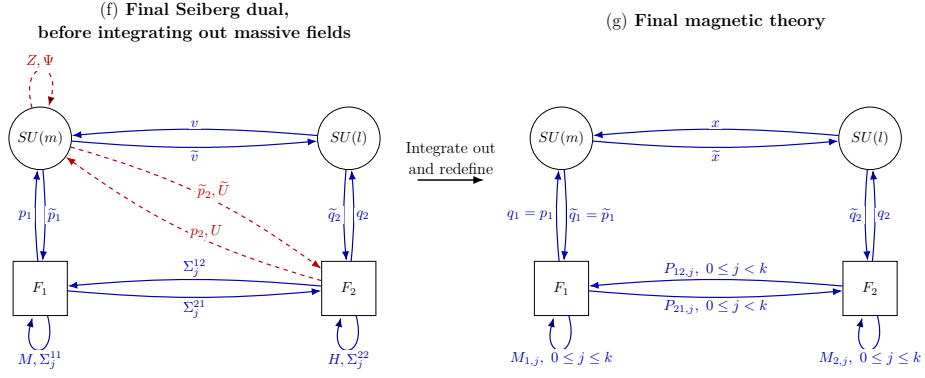
\begin{table}[H]
\centering
\begin{minipage}[t]{.49\linewidth}
\centering\vspace{0pt}
\parbox[t][10mm][t]{\linewidth}{\centering\small\bfseries \(({\rm f})\) Final Seiberg dual,\\before integrating out massive fields\par}\par\smallskip
\scriptsize\setlength{\tabcolsep}{1.5pt}\renewcommand{\arraystretch}{1.35}
\begin{adjustbox}{max width=\linewidth}
\begin{tabular}{@{}lcc|cccc|c@{}}\toprule
Field & $SU(m)$ & $SU(l)$ & $SU(F_1)_L$ & $SU(F_1)_R$ & $SU(F_2)_L$ & $SU(F_2)_R$ & $R$\\\midrule
$Z$ & $\mathrm{Adj}\oplus\mathbf1$ & $\mathbf1$ & $\mathbf1$ & $\mathbf1$ & $\mathbf1$ & $\mathbf1$ & $\frac2{k+1}$\\
$\Psi$ & $\mathrm{Adj}\oplus\mathbf1$ & $\mathbf1$ & $\mathbf1$ & $\mathbf1$ & $\mathbf1$ & $\mathbf1$ & $\frac{2k}{k+1}$\\
$v$ & $\square$ & $\overline{\square}$ & $\mathbf1$ & $\mathbf1$ & $\mathbf1$ & $\mathbf1$ & $\frac1{k+1}$\\
$\widetilde v$ & $\overline{\square}$ & $\square$ & $\mathbf1$ & $\mathbf1$ & $\mathbf1$ & $\mathbf1$ & $\frac1{k+1}$\\
$p_1$ & $\square$ & $\mathbf1$ & $\mathbf1$ & $\overline{\square}$ & $\mathbf1$ & $\mathbf1$ & $\frac1{k+1}-r_1$\\
$\widetilde p_1$ & $\overline{\square}$ & $\mathbf1$ & $\overline{\square}$ & $\mathbf1$ & $\mathbf1$ & $\mathbf1$ & $\frac1{k+1}-r_1$\\
$p_2$ & $\square$ & $\mathbf1$ & $\mathbf1$ & $\mathbf1$ & $\mathbf1$ & $\overline{\square}$ & $\frac2{k+1}-r_2$\\
$\widetilde p_2$ & $\overline{\square}$ & $\mathbf1$ & $\mathbf1$ & $\mathbf1$ & $\overline{\square}$ & $\mathbf1$ & $\frac2{k+1}-r_2$\\
$q_2$ & $\mathbf1$ & $\square$ & $\mathbf1$ & $\mathbf1$ & $\mathbf1$ & $\overline{\square}$ & $\frac1{k+1}-r_2$\\
$\widetilde q_2$ & $\mathbf1$ & $\overline{\square}$ & $\mathbf1$ & $\mathbf1$ & $\overline{\square}$ & $\mathbf1$ & $\frac1{k+1}-r_2$\\
$U$ & $\square$ & $\mathbf1$ & $\mathbf1$ & $\mathbf1$ & $\square$ & $\mathbf1$ & $r_2+\frac{2k}{k+1}$\\
$\widetilde U$ & $\overline{\square}$ & $\mathbf1$ & $\mathbf1$ & $\mathbf1$ & $\mathbf1$ & $\square$ & $r_2+\frac{2k}{k+1}$\\
$H$ & $\mathbf1$ & $\mathbf1$ & $\mathbf1$ & $\mathbf1$ & $\square$ & $\square$ & $2r_2+\frac{2k}{k+1}$\\
$M$ & $\mathbf1$ & $\mathbf1$ & $\square$ & $\square$ & $\mathbf1$ & $\mathbf1$ & $2r_1$\\
$\Sigma_j^{11}$ & $\mathbf1$ & $\mathbf1$ & $\square$ & $\square$ & $\mathbf1$ & $\mathbf1$ & $2r_1+\frac{2(j+1)}{k+1}$\\
$\Sigma_j^{12}$ & $\mathbf1$ & $\mathbf1$ & $\mathbf1$ & $\square$ & $\square$ & $\mathbf1$ & $r_1+r_2+\frac{2j+1}{k+1}$\\
$\Sigma_j^{21}$ & $\mathbf1$ & $\mathbf1$ & $\square$ & $\mathbf1$ & $\mathbf1$ & $\square$ & $r_1+r_2+\frac{2j+1}{k+1}$\\
$\Sigma_j^{22}$ & $\mathbf1$ & $\mathbf1$ & $\mathbf1$ & $\mathbf1$ & $\square$ & $\square$ & $2r_2+\frac{2j}{k+1}$\\
\bottomrule\end{tabular}
\end{adjustbox}
\end{minipage}
\hfill
\begin{minipage}[t]{.49\linewidth}
\centering\vspace{0pt}
\parbox[t][10mm][t]{\linewidth}{\centering\small\bfseries \(({\rm g})\) Final magnetic theory\par}\par\smallskip
\scriptsize\setlength{\tabcolsep}{1.5pt}\renewcommand{\arraystretch}{1.35}
\begin{adjustbox}{max width=\linewidth}
\begin{tabular}{@{}lcc|cccc|c@{}}\toprule
Field & $SU(m)$ & $SU(l)$ & $SU(F_1)_L$ & $SU(F_1)_R$ & $SU(F_2)_L$ & $SU(F_2)_R$ & $R$\\\midrule
$x$ & $\square$ & $\overline{\square}$ & $\mathbf1$ & $\mathbf1$ & $\mathbf1$ & $\mathbf1$ & $\frac1{k+1}$\\
$\widetilde x$ & $\overline{\square}$ & $\square$ & $\mathbf1$ & $\mathbf1$ & $\mathbf1$ & $\mathbf1$ & $\frac1{k+1}$\\
$q_1$ & $\square$ & $\mathbf1$ & $\mathbf1$ & $\overline{\square}$ & $\mathbf1$ & $\mathbf1$ & $\frac1{k+1}-r_1$\\
$\widetilde q_1$ & $\overline{\square}$ & $\mathbf1$ & $\overline{\square}$ & $\mathbf1$ & $\mathbf1$ & $\mathbf1$ & $\frac1{k+1}-r_1$\\
$q_2$ & $\mathbf1$ & $\square$ & $\mathbf1$ & $\mathbf1$ & $\mathbf1$ & $\overline{\square}$ & $\frac1{k+1}-r_2$\\
$\widetilde q_2$ & $\mathbf1$ & $\overline{\square}$ & $\mathbf1$ & $\mathbf1$ & $\overline{\square}$ & $\mathbf1$ & $\frac1{k+1}-r_2$\\
$M_{1,j}$ & $\mathbf1$ & $\mathbf1$ & $\square$ & $\square$ & $\mathbf1$ & $\mathbf1$ & $2r_1+\frac{2j}{k+1}$\\
$M_{2,j}$ & $\mathbf1$ & $\mathbf1$ & $\mathbf1$ & $\mathbf1$ & $\square$ & $\square$ & $2r_2+\frac{2j}{k+1}$\\
$P_{12,j}$ & $\mathbf1$ & $\mathbf1$ & $\mathbf1$ & $\square$ & $\square$ & $\mathbf1$ & $r_1+r_2+\frac{2j+1}{k+1}$\\
$P_{21,j}$ & $\mathbf1$ & $\mathbf1$ & $\square$ & $\mathbf1$ & $\mathbf1$ & $\square$ & $r_1+r_2+\frac{2j+1}{k+1}$\\
\bottomrule\end{tabular}
\end{adjustbox}
\end{minipage}
\caption{Representations and  $U(1)_R$ charges of theories (f) and (g), respectively. In (f), all $\Sigma_j^{IJ}$ have $0\leq j<k$, while in (g), the index ranges are $0\leq j\leq k$ for $M_{1,j},M_{2,j}$ and $0\leq j<k$ for $P_{12,j},P_{21,j}$.}
\label{tab:gkr-final-comparison}\label{tab:gkr-g}\label{tab:gkr-h}
\end{table}

\begin{figure}[H]\centering
\begin{adjustbox}{max width=\linewidth,max totalheight=.8\textheight,center}
\begin{tikzpicture}[
  font=\fontsize{11.5}{13.5}\selectfont,>=Latex,
  gk gauge/.style={circle,draw,minimum size=16mm,inner sep=1pt,
    font=\fontsize{11}{13}\selectfont,align=center},
  gk flavor/.style={rectangle,draw,minimum size=14mm,inner sep=1pt}
]
\begin{scope}
\node[font=\bfseries\fontsize{12}{14}\selectfont,align=center,text width=11.5cm] at (3.7,3.3) {\(({\rm a})\) Electric theory};
\node[gk gauge] (g1) at (0,0) {\(SU(N_1)\)};
\node[gk gauge] (g2) at (7.4,0) {\(SU(N_2)\)};
\node[gk flavor] (f1) at (0,-3.6) {\(F_1\)};
\node[gk flavor] (f2) at (7.4,-3.6) {\(F_2\)};
\draw[thick,draw=blue!65!black,text=blue!65!black] (g2)--node[fill=white,inner sep=1.3pt,align=center,rotate=0,above]{\(\frac1{k+1}\)} (g1);
\draw[thick,draw=blue!65!black,text=blue!65!black] (f1)--node[fill=white,inner sep=1.3pt,align=center,rotate=0,left]{\(r_1\)} (g1);
\draw[thick,draw=blue!65!black,text=blue!65!black] (f2)--node[fill=white,inner sep=1.3pt,align=center,rotate=0,right]{\(r_2\)} (g2);
 
\end{scope}

\begin{scope}[xshift=14cm]
\node[font=\bfseries\fontsize{12}{14}\selectfont,align=center,text width=11.5cm] at (3.7,3.3) {\(({\rm b})\) First Seiberg dual};
\node[gk gauge] (g1) at (0,0) {\(SU(h)\)};
\node[gk gauge] (g2) at (7.4,0) {\(SU(N_2)\)};
\node[gk flavor] (f1) at (0,-3.6) {\(F_1\)};
\node[gk flavor] (f2) at (7.4,-3.6) {\(F_2\)};
\draw[thick,draw=blue!65!black,text=blue!65!black] (g2)--node[fill=white,inner sep=1.3pt,align=center,rotate=0,above]{\(\frac{k}{k+1}\)} (g1);
\draw[thick,draw=blue!65!black,text=blue!65!black] (f1)--node[fill=white,inner sep=1.3pt,align=center,rotate=0,left]{\(1-r_1\)} (g1);
\draw[thick,draw=blue!65!black,text=blue!65!black] (f2)--node[fill=white,inner sep=1.3pt,align=center,rotate=0,right]{\(r_2\)} (g2);
\draw[thick,draw=blue!65!black,text=blue!65!black] (f1)--node[fill=white,inner sep=1.3pt,align=center,rotate=0,above,pos=.70]{\(r_1+\frac1{k+1}\)} (g2);
\draw[thick,draw=blue!65!black,text=blue!65!black,->] (g2) edge[loop above]
 node[fill=white,inner sep=1.3pt,align=center,rotate=0,above]{\(\frac2{k+1}\)} (g2);
\draw[thick,draw=blue!65!black,text=blue!65!black,->] (f1) edge[loop below]
 node[fill=white,inner sep=1.3pt,align=center,rotate=0,below]{\(2r_1\)} (f1);
 
\end{scope}

\begin{scope}[xshift=14cm,yshift=-10.4cm]
\node[font=\bfseries\fontsize{12}{14}\selectfont,align=center,text width=11.5cm] at (3.7,3.3) {\(({\rm c})\) KS magnetic theory};
\node[gk gauge] (g1) at (0,0) {\(SU(h)\)};
\node[gk gauge] (g2) at (7.4,0) {\(SU(n)\)};
\node[gk flavor] (f1) at (0,-5.0) {\(F_1\)};
\node[gk flavor] (f2) at (7.4,-5.0) {\(F_2\)};
\draw[thick] (g2)--node[fill=white,inner sep=1.3pt,align=center,rotate=0,above]{\(\frac{2-k}{k+1}\)} (g1);
\draw[thick,draw=red!70!black,text=red!70!black,dashed,->] (g1) edge[loop above]
 node[fill=white,inner sep=1.3pt,align=center,rotate=0,above]{\(2+\frac{2(j-1)}{k+1}\)} (g1);
\draw[thick,->] (g2) edge[loop above]
 node[fill=white,inner sep=1.3pt,align=center,rotate=0,above]{\(\frac2{k+1}\)} (g2);
\draw[thick,draw=red!70!black,text=red!70!black,dashed] (f1)--node[fill=white,inner sep=1.3pt,align=center,rotate=0,left]{\(\begin{gathered}1-r_1\\1+r_1+\tfrac{2j}{k+1}\end{gathered}\)} (g1);
\draw[thick] (f2)--node[fill=white,inner sep=1.3pt,align=center,rotate=0,right]{\(\frac2{k+1}-r_2\)} (g2);
\draw[thick] (f1)--node[fill=white,inner sep=1.3pt,align=center,rotate=0,below,pos=.74,yshift=-1mm]{\(\frac1{k+1}-r_1\)} (g2);
\draw[thick] (f2)--(g1);
\node[fill=white,inner sep=1.3pt,align=center] at (2.4,-1.35) {\(\begin{gathered}\textcolor{blue!65!black}{r_2+\tfrac{k}{k+1}}\\\textcolor{red!70!black}{r_2+\tfrac{k+2j}{k+1}}\end{gathered}\)};
\draw[thick,draw=blue!65!black,text=blue!65!black,->] (f1) edge[loop below]
 node[fill=white,inner sep=1.3pt,align=center,rotate=0,below]{\(\begin{gathered}2r_1\\2r_1+\tfrac{2(j+1)}{k+1}\end{gathered}\)} (f1);
\draw[thick,draw=blue!65!black,text=blue!65!black,->] (f2) edge[loop below]
 node[fill=white,inner sep=1.3pt,align=center,rotate=0,below]{\(2r_2+\frac{2j}{k+1}\)} (f2);
\draw[thick,draw=blue!65!black,text=blue!65!black] (f2)--node[fill=white,inner sep=1.3pt,align=center,rotate=0,above]{\(r_1+r_2+\frac{2j+1}{k+1}\)} (f1);
\end{scope}

\begin{scope}[yshift=-10.4cm]
\node[font=\bfseries\fontsize{12}{14}\selectfont,align=center,text width=11.5cm] at (3.7,3.8) {\(({\rm d})\) After Higgsing, before\\integrating out massive fields};
\node[gk gauge] (g1) at (0,0) {\(SU(h)\)};
\node[gk gauge] (g2) at (7.4,0) {\(SU(m)\)};
\node[gk flavor] (f1) at (0,-5.0) {\(F_1\)};
\node[gk flavor] (f2) at (7.4,-5.0) {\(F_2\)};
\draw[thick,draw=blue!65!black,text=blue!65!black] (g2)--node[fill=white,inner sep=1.3pt,align=center,rotate=0,above]{\(\frac{k}{k+1}\)} (g1);
\draw[thick,draw=red!70!black,text=red!70!black,dashed,->] (g1) edge[loop above] (g1);
\node[fill=white,inner sep=1.3pt,align=center,rotate=0,text=red!70!black] at (2.8,1.8) {\(\begin{gathered}\tfrac2{k+1},\quad \tfrac{2k}{k+1}\\\tfrac{2(1-j)}{k+1},\quad 2+\tfrac{2(j-1)}{k+1}\\\tfrac{2(k-b)}{k+1}\end{gathered}\)};
\draw[thick,draw=blue!65!black,text=blue!65!black,->] (g2) edge[loop above]
 node[fill=white,inner sep=1.3pt,align=center,rotate=0,above]{\(\frac2{k+1}\)} (g2);
\draw[thick,draw=red!70!black,text=red!70!black,dashed] (f1)--node[fill=white,inner sep=1.3pt,align=center,rotate=0,left,xshift=-4mm]{\(\begin{gathered}1-r_1\\1+r_1\\\textcolor{red!70!black}{1-r_1-\tfrac{2j}{k+1}}\\1+r_1+\tfrac{2j}{k+1}\end{gathered}\)} (g1);
\draw[thick,draw=blue!65!black,text=blue!65!black] (f2)--node[fill=white,inner sep=1.3pt,align=center,rotate=0,right]{\(\frac2{k+1}-r_2\)} (g2);
\draw[thick,draw=blue!65!black,text=blue!65!black] (f1)--node[fill=white,inner sep=1.3pt,align=center,rotate=0,below,pos=.74,yshift=-1mm]{\(\frac1{k+1}-r_1\)} (g2);
\draw[thick] (f2)--(g1);
\node[fill=white,inner sep=1pt,align=center,font=\fontsize{11.5}{13.5}\selectfont] at (2.6,-1.35) {\(\begin{gathered}\textcolor{red!70!black}{\tfrac{k+2-2j}{k+1}-r_2}\\\textcolor{red!70!black}{r_2+\tfrac{k+2j}{k+1}}\\\textcolor{blue!65!black}{r_2+\tfrac{k}{k+1}}\end{gathered}\)};
\draw[thick,draw=blue!65!black,text=blue!65!black,->] (f1) edge[loop below]
 node[fill=white,inner sep=1.3pt,align=center,rotate=0,below]{\(\begin{gathered}2r_1\\2r_1+\tfrac{2(j+1)}{k+1}\end{gathered}\)} (f1);
\draw[thick,draw=blue!65!black,text=blue!65!black,->] (f2) edge[loop below]
 node[fill=white,inner sep=1.3pt,align=center,rotate=0,below]{\(2r_2+\frac{2j}{k+1}\)} (f2);
\draw[thick,draw=blue!65!black,text=blue!65!black] (f2)--node[fill=white,inner sep=1.3pt,align=center,rotate=0,above]{\(r_1+r_2+\frac{2j+1}{k+1}\)} (f1);

\end{scope}

\begin{scope}[yshift=-22.8cm]
\node[font=\bfseries\fontsize{12}{14}\selectfont,align=center,text width=11.5cm] at (3.7,3.3) {\(({\rm e})\) After Higgsing and\\integrating out massive fields};
\node[gk gauge] (g1) at (0,0) {\(SU(h)\)};
\node[gk gauge] (g2) at (7.4,0) {\(SU(m)\)};
\node[gk flavor] (f1) at (0,-3.6) {\(F_1\)};
\node[gk flavor] (f2) at (7.4,-3.6) {\(F_2\)};
\draw[thick,draw=blue!65!black,text=blue!65!black] (g2)--node[fill=white,inner sep=1.3pt,align=center,rotate=0,above]{\(\frac{k}{k+1}\)} (g1);
\draw[thick,draw=blue!65!black,text=blue!65!black,->] (g2) edge[loop above]
 node[fill=white,inner sep=1.3pt,align=center,rotate=0,above]{\(\frac2{k+1}\)} (g2);
\draw[thick,draw=blue!65!black,text=blue!65!black] (f2)--node[fill=white,inner sep=1.3pt,align=center,rotate=0,right]{\(\frac2{k+1}-r_2\)} (g2);
\draw[thick,draw=blue!65!black,text=blue!65!black] (f1)--node[fill=white,inner sep=1.3pt,align=center,rotate=0,below,pos=.74,yshift=-1mm]{\(\frac1{k+1}-r_1\)} (g2);
\draw[thick,draw=blue!65!black,text=blue!65!black] (f2)--node[fill=white,inner sep=1.3pt,align=center,rotate=0,below,pos=.72]{\(r_2+\frac{k}{k+1}\)} (g1);
\draw[thick,draw=blue!65!black,text=blue!65!black,->] (f1) edge[loop below]
 node[fill=white,inner sep=1.3pt,align=center,rotate=0,below]{\(\begin{gathered}2r_1\\2r_1+\tfrac{2(j+1)}{k+1}\end{gathered}\)} (f1);
\draw[thick,draw=blue!65!black,text=blue!65!black,->] (f2) edge[loop below]
 node[fill=white,inner sep=1.3pt,align=center,rotate=0,below]{\(2r_2+\frac{2j}{k+1}\)} (f2);
\draw[thick,draw=blue!65!black,text=blue!65!black] (f2)--node[fill=white,inner sep=1.3pt,align=center,rotate=0,above]{\(r_1+r_2+\frac{2j+1}{k+1}\)} (f1);
\end{scope}

\begin{scope}[xshift=14cm,yshift=-22.8cm]
\node[font=\bfseries\fontsize{12}{14}\selectfont,align=center,text width=11.5cm] at (3.7,3.3) {\(({\rm f})\) Final Seiberg dual, before\\integrating out massive fields};
\node[gk gauge] (g1) at (0,0) {\(SU(m)\)};
\node[gk gauge] (g2) at (7.4,0) {\(SU(l)\)};
\node[gk flavor] (f1) at (0,-3.6) {\(F_1\)};
\node[gk flavor] (f2) at (7.4,-3.6) {\(F_2\)};
\draw[thick,draw=blue!65!black,text=blue!65!black] (g2)--node[fill=white,inner sep=1.3pt,align=center,rotate=0,above]{\(\frac1{k+1}\)} (g1);
\draw[thick,draw=red!70!black,text=red!70!black,dashed,->] (g1) edge[loop above] (g1);
\node[fill=white,inner sep=1.3pt,align=center,rotate=0,text=red!70!black] at (1.65,1.55) {\(\begin{gathered}\tfrac2{k+1}\\\tfrac{2k}{k+1}\end{gathered}\)};
\draw[thick,draw=blue!65!black,text=blue!65!black] (f1)--node[fill=white,inner sep=1.3pt,align=center,rotate=0,left]{\(\frac1{k+1}-r_1\)} (g1);
\draw[thick,draw=blue!65!black,text=blue!65!black] (f2)--node[fill=white,inner sep=1.3pt,align=center,rotate=0,right]{\(\frac1{k+1}-r_2\)} (g2);
\draw[thick,draw=red!70!black,text=red!70!black,dashed] (f2)--(g1);
\node[fill=white,inner sep=1.3pt,align=center,text=red!70!black] at (3.7,-1.1) {\(\begin{gathered}\tfrac2{k+1}-r_2\\\tfrac{2k}{k+1}+r_2\end{gathered}\)};
\draw[thick,draw=blue!65!black,text=blue!65!black,->] (f1) edge[loop below]
 node[fill=white,inner sep=1.3pt,align=center,rotate=0,below]{\(\begin{gathered}2r_1\\2r_1+\tfrac{2(j+1)}{k+1}\end{gathered}\)} (f1);
\draw[thick,draw=blue!65!black,text=blue!65!black,->] (f2) edge[loop below]
 node[fill=white,inner sep=1.3pt,align=center,rotate=0,below]{\(\begin{gathered}2r_2+\tfrac{2k}{k+1}\\2r_2+\tfrac{2j}{k+1}\end{gathered}\)} (f2);
\draw[thick,draw=blue!65!black,text=blue!65!black] (f2)--node[fill=white,inner sep=1.3pt,align=center,rotate=0,above]{\(r_1+r_2+\frac{2j+1}{k+1}\)} (f1);
\end{scope}

\begin{scope}[xshift=14cm,yshift=-34.2cm]
\node[font=\bfseries\fontsize{12}{14}\selectfont,align=center,text width=11.5cm] at (3.7,3.3) {\(({\rm g})\) Final magnetic theory};
\node[gk gauge] (g1) at (0,0) {\(SU(m)\)};
\node[gk gauge] (g2) at (7.4,0) {\(SU(l)\)};
\node[gk flavor] (f1) at (0,-3.6) {\(F_1\)};
\node[gk flavor] (f2) at (7.4,-3.6) {\(F_2\)};
\draw[thick,draw=blue!65!black,text=blue!65!black] (g2)--node[fill=white,inner sep=1.3pt,align=center,rotate=0,above]{\(\frac1{k+1}\)} (g1);
\draw[thick,draw=blue!65!black,text=blue!65!black] (f1)--node[fill=white,inner sep=1.3pt,align=center,rotate=0,left]{\(\frac1{k+1}-r_1\)} (g1);
\draw[thick,draw=blue!65!black,text=blue!65!black] (f2)--node[fill=white,inner sep=1.3pt,align=center,rotate=0,right]{\(\frac1{k+1}-r_2\)} (g2);
\draw[thick,draw=blue!65!black,text=blue!65!black,->] (f1) edge[loop below]
 node[fill=white,inner sep=1.3pt,align=center,rotate=0,below]{\(2r_1+\frac{2j}{k+1}\)} (f1);
\draw[thick,draw=blue!65!black,text=blue!65!black,->] (f2) edge[loop below]
 node[fill=white,inner sep=1.3pt,align=center,rotate=0,below]{\(2r_2+\frac{2j}{k+1}\)} (f2);
\draw[thick,draw=blue!65!black,text=blue!65!black] (f2)--node[fill=white,inner sep=1.3pt,align=center,rotate=0,above]{\(r_1+r_2+\frac{2j+1}{k+1}\)} (f1);
\end{scope}

\draw[->,line width=1pt] (9.7,-1.8)--node[above]{Seiberg} (11.7,-1.8);
\draw[->,line width=1pt] (17.7,-5.4)--(17.7,-5.9);
\node[anchor=west,align=left,font=\small,text width=5.6cm] at (18.1,-5.65)
 {KS duality\\$n=k(h+F_1+F_2)-N_2$};
\draw[->,line width=1pt] (11.7,-12.2)--node[above]{Higgsing} (9.7,-12.2);
\draw[->,line width=1pt] (3.7,-18.3)--(3.7,-18.8);
\node[anchor=west,align=left,font=\small,text width=5.6cm] at (4.1,-18.55) {Integrating out massive fields};
\draw[->,line width=1pt] (9.7,-24.6)--node[above]{Seiberg}
 node[below,align=center] {$l=m+F_2-h$\\Reorder nodes} (11.7,-24.6);
\draw[->,line width=1pt] (17.7,-29.7)--(17.7,-30.2);
\node[anchor=west,align=left,font=\small,text width=5.6cm] at (18.1,-29.95)
 {Integrating out massive fields\\and singlet redefinitions};
 
\end{tikzpicture}
\end{adjustbox}
\caption{The complete sequence of the ILS duality derivation for $k\geq 2$. For simplicity, straight lines represent pairs of bi-fundamental fields, while red dashed lines indicate massive fields. }
\label{fig:gksequenceR}
\end{figure}
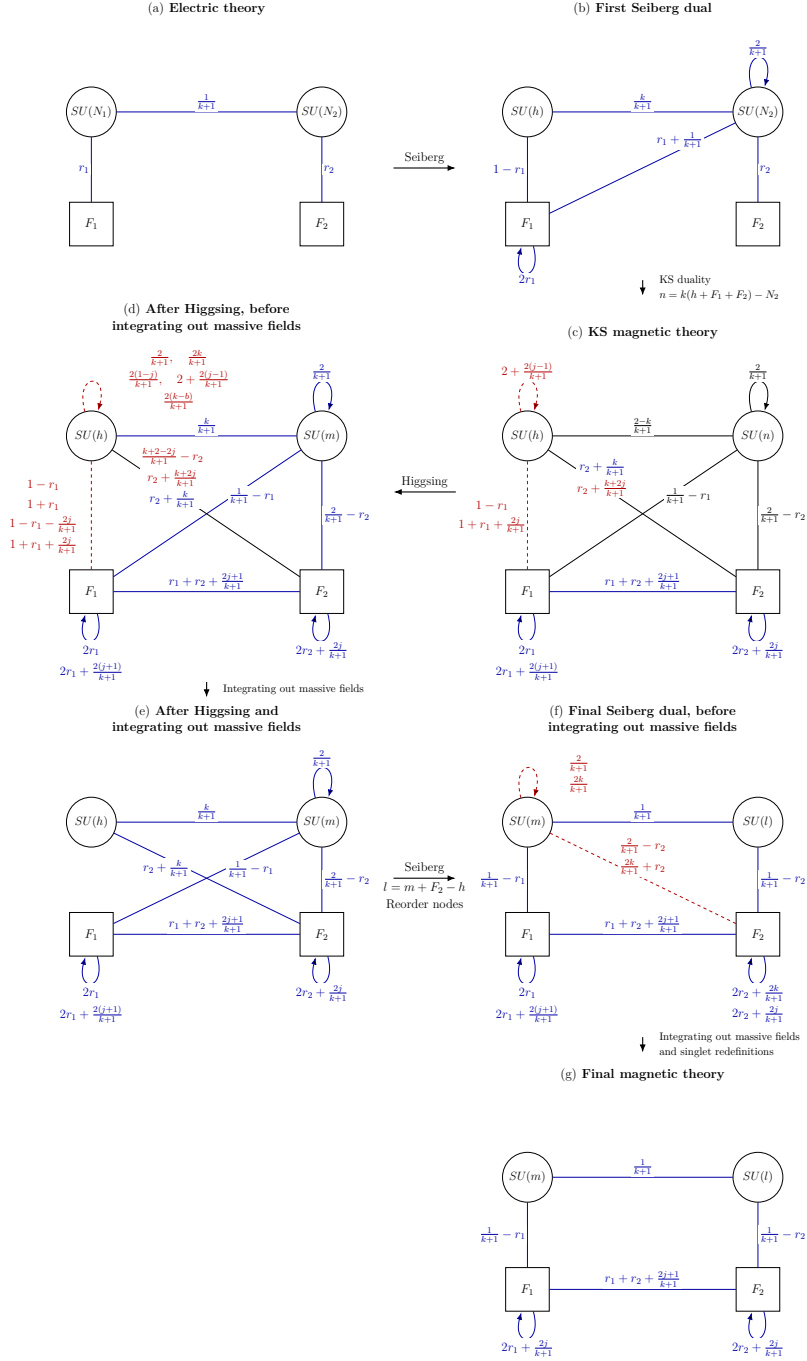

\paragraph{Summary of the derivation.} We summarize the entire procedure involving theories from (a) to (g) in Figure~\ref{fig:gksequenceR}, with  $R$-charges labeled on the quiver diagrams.

\section{Conclusion}
\noindent
In this paper, we propose a new two-node 3d $\CN=2$ duality and support it through
partition function identities from the Aharony and Benini–Benvenuti–Pasquetti dualities, while deriving the 4d ILS duality
from nodewise Seiberg and KS dualities together with Higgsing and integrating out massive fields.

There are some subsequent open questions:

\begin{itemize}
   \item {\bf Brane realization of the duality sequence}.
   Brodie and Hanany presented a perspective on the ILS duality from brane configuration for general $k$, but provided an explicit nodewise duality sequence only for $k=1$~\cite{Brodie:1997sz}. Our field theoretic derivation provides such a sequence for general $k$. Since each elementary step admits a brane realization, it would be interesting to investigate whether the entire sequence can be realized through brane dynamics and reproduce the general $k$ brane construction of Brodie and Hanany.
   
    \item {\bf Generalizations of the two-node duality}. The 3d two-node dualities studied here involve unitary gauge groups, but admit various generalizations, including $SU$ gauge groups, Chern--Simons couplings, and chiral or tensor matter. Extensions to orthogonal $SO$ and symplectic $Sp$ gauge groups are also possible. We leave a detailed investigation of these generalizations to future work~\cite{futurework}.
    
     \item {\bf Interplay between 3d and 4d duality constructions}. Our 4d derivation combines Seiberg~\cite{Seiberg:1994pq} and Kutasov--Schwimmer dualities~\cite{Kutasov:1995ss} with Higgsing. This motivates exploring an analogous derivation in 3d using Aharony~\cite{Aharony:1997gp} and Kim--Park dualities~\cite{Kim:2013cma}, while carefully tracking monopole operators and their superpotential couplings throughout the Higgsing process. Conversely, our 3d deconfinement approach suggests an alternative derivation of the 4d two-node dualities through auxiliary confining gauge sectors and successive nodewise Seiberg dualities.

     \item {\bf The missing $E_6$ and $E_8$ dualities}.
     An important open problem is to construct Seiberg-like dualities for two-adjoint SQCDs with $E_6$- and $E_8$-type superpotentials. The superpotentials of two-adjoint SQCDs are classified according to the $ADE$ series~\cite{Intriligator:2016sgx}, with corresponding dualities proposed for the $A_k$~\cite{Kutasov:1995ss,Kim:2013cma}, $D_k$~\cite{Brodie:1996vx,Hwang:2018uyj}, and $E_7$~\cite{Kutasov:2014yqa} cases. It would be particularly interesting to explore whether the adjoint deconfinement technique can solve the remaining puzzles of the $E_6$ and $E_8$ dualities in 3d, and eventually extend the results to 4d.
 
\end{itemize}
 
\section*{Acknowledgments}
\noindent
We thank Chiung Hwang and Jehyun Lee for discussions and related collaboration. We also thank Dan Xie for inspiring discussions. We acknowledge the use of AI tools for assistance with intermediate superpotential calculations. All results have been independently verified by the authors. The work of SK is supported by the National Research Foundation of Korea (NRF) Grants RS-2024-00405629 and RS-2023-NR076601. The work of YF is supported in part by the National Natural Science Foundation of China (NSFC) through the Research Fund for International Excellent Young Scientists (Grant No. W2632007).

\appendix
\section{Review of 3d \texorpdfstring{$\CN=2$}{N=2} dualities and Quivers for the ellipsoid partition function}\label{app: 3d ptf-quiver}
\noindent
In this appendix, we review two elementary 3d $\CN=2$ dualities, the s-confinement of $\CC_p[U(N)]$, and the quiver representation of ellipsoid partition functions for gauge theories with unitary groups.

\subsection{Elementary 3d \texorpdfstring{$\CN=2$}{N=2} dualities and s-confinement of \texorpdfstring{$\CC_p[U(N)]$}{Cp[U(N)]}}
\medskip\noindent{\bf Aharony duality}. The first is Aharony duality~\cite{Aharony:1997gp}, the 3d $\CN=2$ counterpart of 4d $\CN=1$ Seiberg duality~\cite{Seiberg:1994pq}, relating two unitary SQCD theories with $F$ flavors in the IR:
\begin{equation}
\begin{tikzpicture}[thick,scale=0.8, every node/.style={scale=0.8}]
  \tikzset{vertex/.style={circle,fill=white!25,minimum size=12pt,inner sep=2pt}}
  \tikzset{every loop/.style={}}
    \node[vertex] (eNc) at (-1,0) [shape=circle,draw=black,minimum size=2em] {$N$};
    \node[vertex] (eNf1) at (1,0.7) [shape=rectangle,draw=black,minimum height=2em, minimum width=2em] {$F$};
    \node[vertex] (eNf2) at (1,-0.7) [shape=rectangle,draw=black,minimum height=2em, minimum width=2em] {$F$};

    \draw[->-=.5] ([yshift= 2pt] eNc.east) to ([yshift= 0pt] eNf1.west);
    \draw[->-=.5] ([yshift= 0pt] eNf2.west) to ([yshift= -2pt] eNc.east);

    \node at (0,0.7) {\tiny$Q$};
    \node at (0,-0.7) {\tiny$\WQ$};
    \node at (0,-1.5) {$W_e=  0$};

    \node at (3.5,0.2) {$\overset{\CA}{\longleftrightarrow}$};

    \node[vertex] (mNc) at (6,0) [shape=circle,draw=black,minimum size=2em] {$N'$};
    \node[vertex] (mNf1) at (8,0.7) [shape=rectangle,draw=black,minimum height=2em, minimum width=2em] {$F$};
    \node[vertex] (mNf2) at (8,-0.7) [shape=rectangle,draw=black,minimum height=2em, minimum width=2em] {$F$};

    \draw[->-=.5] ([yshift= 0pt] mNf1.west) to ([yshift= 2pt] mNc.east);
    \draw[->-=.5] ([yshift= -2pt] mNc.east) to ([yshift= 0pt] mNf2.west);
    \draw[->-=.5] ( mNf2.north) to ( mNf1.south);

    \node at (7,0.7) {\tiny$\Wq$};
    \node at (7,-0.7) {\tiny$q$};
    \node at (8.5,0) {\tiny$M$};
    \node at (7,-1.5) {$W_m = M\, \Wq q + V^+ \hat{v}^+ + V^- \hat{v}^-$};

\end{tikzpicture}
\label{eq: Aharony quiver}
\end{equation}
with dual gauge rank $N'= F- N$. The chiral ring generators are mapped as
\begin{align}
    \WQ Q \quad &\longleftrightarrow \quad M
    \nonumber\\
    \hat V^\pm \quad &\longleftrightarrow \quad V^\pm
\end{align}
with mesonic singlets $M$, and monopole operators $\hat V^\pm$ of charges $\pm1$ on the electric side which are mapped to singlets $V^\pm$ coupled to the monopoles $\hat v^\pm$ on the magnetic side.

\medskip\noindent{\bf BBP duality}.
The second is Benini--Benvenuti--Pasquetti (BBP) duality~\cite{Benini:2017dud}, obtained by a monopole superpotential deformation of Aharony duality:
\begin{equation}
\begin{tikzpicture}[thick,scale=0.8, every node/.style={scale=0.8}]
  \tikzset{vertex/.style={circle,fill=white!25,minimum size=12pt,inner sep=2pt}}
  \tikzset{every loop/.style={}}
    \node[vertex] (eNc) at (-1,0) [shape=circle,draw=black,minimum size=2em] {$N$};
    \node[vertex] (eNf1) at (1,0.7) [shape=rectangle,draw=black,minimum height=2em, minimum width=2em] {$F$};
    \node[vertex] (eNf2) at (1,-0.7) [shape=rectangle,draw=black,minimum height=2em, minimum width=2em] {$F$};

    \draw[->-=.5] ([yshift= 2pt] eNc.east) to ([yshift= 0pt] eNf1.west);
    \draw[->-=.5] ([yshift= 0pt] eNf2.west) to ([yshift= -2pt] eNc.east);

    \node at (0,0.7) {\tiny$Q$};
    \node at (0,-0.7) {\tiny$\WQ$};
    \node at (0,-1.5) {$W_e=  \hat V^+ $};

    \node at (3.5,0.2) {$\overset{\CB}{\longleftrightarrow}$};

    \node[vertex] (mNc) at (6,0) [shape=circle,draw=black,minimum size=2em] {$N''$};
    \node[vertex] (mNf1) at (8,0.7) [shape=rectangle,draw=black,minimum height=2em, minimum width=2em] {$F$};
    \node[vertex] (mNf2) at (8,-0.7) [shape=rectangle,draw=black,minimum height=2em, minimum width=2em] {$F$};

    \draw[->-=.5] ([yshift= 0pt] mNf1.west) to ([yshift= 2pt] mNc.east);
    \draw[->-=.5] ([yshift= -2pt] mNc.east) to ([yshift= 0pt] mNf2.west);
    \draw[->-=.5] ( mNf2.north) to ( mNf1.south);

    \node at (7,0.7) {\tiny$\Wq$};
    \node at (7,-0.7) {\tiny$q$};
    \node at (8.5,0) {\tiny$M$};
    \node at (7,-1.5) {$W_m = \hat v^+ + M\, \Wq q + V^- \hat{v}^-$};

\end{tikzpicture}
\label{eq: BBP quiver}
\end{equation}
The dual gauge rank $N''=F-N-1$ is one less than in Aharony duality, with the chiral ring generators mapped as follows:
\begin{align}
    \WQ Q \quad &\longleftrightarrow \quad M
    \nonumber\\
    \hat V^+ \quad &\longleftrightarrow \quad \hat v^+
    \nonumber\\
    \hat V^- \quad &\longleftrightarrow \quad  V^-
\end{align}
where the monopole operator in $W_e$ is directly mapped to the magnetic monopole operator.

\medskip
\noindent{\bf S-confinement of $\CC_p[U(N)]$}.
For positive integers $p$ and $N$, the 3d $\CN=2$ linear quiver theory $\CC_p[U(N)]$ of~\cite{Benvenuti:2024glr} is given by:
\begin{equation}
\begin{tikzpicture}[thick,scale=0.8, every node/.style={scale=0.8}]
  \tikzset{vertex/.style={circle,fill=white!25,minimum size=12pt,inner sep=2pt}}
  \tikzset{every loop/.style={}}
    \node at (-2,0) {$\CC_p[U(N)]\quad:\quad$ };    

    \node[vertex] (A1) at (0,0) [shape=circle,draw=black,minimum size=2em] {$m_1$};
    \node[vertex] (A2) at (1.5,0) [shape=circle,draw=black,minimum size=2em] {$m_2$};
    \node[vertex] (Acd) at (3,0)  {$\cdots$};
    \node[vertex] (Ap) at (4.5,0) [shape=circle,draw=black,minimum size=2em] { \scriptsize $m_{p\texttt{-}1}\!\!$};
    \node[vertex] (AF) at (6,0) [shape=rectangle,draw=black,minimum height=2em, minimum width=2em] {$N$};

    \draw[-to, min distance=0.8cm]  (A1) edge [out=120, in=60] node {} (A1);
    \draw[-to, min distance=0.8cm]  (A2) edge [out=120, in=60] node {} (A2);
    \draw[-to, min distance=0.8cm]  (Ap) edge [out=120, in=60] node {} (Ap);
    \draw[->-=.5] ([yshift= 2pt] A1.east) to ([yshift= 2pt] A2.west);
    \draw[->-=.5] ([yshift= -2pt] A2.west) to ([yshift= -2pt] A1.east);
    \draw[->-=.5] ([yshift= 2pt] A2.east) to ([yshift= 2pt] Acd.west);
    \draw[->-=.5] ([yshift= -2pt] Acd.west) to ([yshift= -2pt] A2.east);
    \draw[->-=.5] ([yshift= 2pt] Acd.east) to ([yshift= 2pt] Ap.west);
    \draw[->-=.5] ([yshift= -2pt] Ap.west) to ([yshift= -2pt] Acd.east);
    \draw[->-=.5] ([yshift= 2pt] Ap.east) to ([yshift= 2pt] AF.west);
    \draw[->-=.5] ([yshift= -2pt] AF.west) to ([yshift= -2pt] Ap.east);
    
\end{tikzpicture}
\label{eq: Cp quiver}
\end{equation}
with the ranks
\begin{align}
    m_j = j n + \a_j
    \;\;,\quad
    j=1,\cdots,p
    \,,
\end{align}
where $n=\lfloor N/p\rfloor$ and $\vec{\a}=(\overbrace{0,\cdots,0}^{p-\a},1,2,\cdots,\a)$ with $\a:=N\!\!\mod p$, so that $m_p=pn+\a=N$. The superpotential is given by~\footnote{Unlike $\CC_p[SU(N)]$ in~\cite{Benvenuti:2024glr}, $\CC_p[U(N)]$ has no flipping field for $\hat{V}^{(1,p-1),-}$ at $\a=0$, as this operator is identified with a component of the confined $U(N)$ adjoint rather than the $SU(N)$ adjoint. The superpotential applies for $N\geq p$; for $N<p$, we use $\CC_p[U(N)]$ only as the partition function block, omitting zero-rank nodes.}
\begin{align}
    W
    =
    W_{\CN=4} 
    + 
    \sum_{i=1}^{p-1} \hat V^{(i),+}
    +
    \Tr( \phi^{\d_{ 0 \a}}
    \hat V^{(1,p-1),-})
    \,,
    \label{eq: CpN superpotential}
\end{align}
where $W_{\CN=4}$ includes the $\CN=4$-like cubic terms, $\hat V^{(i),+}$ is a bare monopole operator with magnetic charge $+\d_{ij}$ under $U(m_j)$, and $\hat V^{(1,p-1),-}$ carries magnetic charge $-1$ under each gauge group. For $\a=0$, the third term becomes a monopole operator dressed by the adjoint chiral field $\phi$.

Notably, $\CC_p[U(N)]$ is dual to an adjoint chiral multiplet $X$ of the $U(N)$ flavor symmetry:
\begin{equation}
    \begin{tikzpicture}[thick,scale=0.8, every node/.style={scale=0.8}]
      \tikzset{vertex/.style={circle,fill=white!25,minimum size=12pt,inner sep=2pt}}
      \tikzset{every loop/.style={}}
        \node[vertex] (N) at (0,0) [shape=rectangle,draw=black,minimum height=2em, minimum width=2em] {$N$};
    
        \draw[-to, min distance=1cm]  (N) edge [out=120, in=60] node {} (N);
    
        \node at (-4,0) {$\CC_p[U(N)]$};
        \node at (-2.2,0.2) {$\overset{\text{Dual}}{\longleftrightarrow}$};
        
        \node at (0,1.2) {\tiny $X$};
        \node at (1.7,0.2) {\small$\quad W=\Tr X^{p+1}$};
        \node at (3.2,-0.2) {\small$+\sum_{i=1}^{\min(\a,p\texttt{-}\a)} \eta_i \Tr X^i$};
    \end{tikzpicture}
    \label{eq: confinement}
\end{equation}
up to singlets $\eta_i$. This allows us to replace the adjoint with a $\CC_p[U(N)]$ quiver tail, a procedure known as {\it deconfinement}~\cite{Berkooz:1995km,Pouliot:1995me,Luty:1996cg}. When $U(N)$ is gauged, a suitable BF term is required for a consistent Coulomb branch spectrum, implemented by shifting the FI-parameter $\t$ by
\begin{align}
    \t_p = \t - \frac{[N-1]}{p+1}
    \,,
\end{align}
where $[l]:=l\!\!\mod p$ and $\t_p$ is the shifted FI-parameter after the deconfinement.

\subsection{Quiver expression of the ellipsoid partition function}
\noindent
The ellipsoid partition function of a 3d $\CN=2$ gauge theory with gauge group $G$ and chiral multiplets $\Phi$ of gauge weight $\r_\Phi$, flavor weight $f_\Phi$, and  $R$-charge $r_\Phi$ reads~\cite{Hama:2011ea}
\begin{align}
    \CZ_{S_b^3}(m,\xi) = \frac{1}{|W_G|} 
    \int \prod_{i=1}^{\text{rank}(G)} d \s_i
    e^{2\pi i \xi \Tr \s}
    \frac{
    \prod_{\Phi}\prod_{\r_{\Phi},f_{\Phi}} 
    s_b \big((1-r_\Phi)c_b - \rho_\Phi(\s) - f_\Phi(m) \big)
    }
    {
    \prod_{\a} s_b\big(\a(\s) + c_b\big)
    }
    \label{eq: S_b^3 ptf integral formula}
\end{align}
where $c_b:=i(b+b^{-1})/2$, $W_G$ is the Weyl group of $G$, $\a$ is a root of $G$, $\xi$ is the FI-parameter, $\s$ parametrizes the Coulomb branch, and $m$ is the real mass parameter. The double-sine function $s_b(\s)$ is defined by
\begin{align}
    s_b(\s) := \prod_{l,k\geq 0}
    \frac{l b + k b^{-1} -i c_b - i \s}{l b + k b^{-1} -i c_b + i \s} 
    \,,
\end{align}
which satisfies $s_b(\s)s_b(-\s)=1$, reflecting the cancellation of two chirals paired by a superpotential mass term.\footnote{The following partition function identities hold by meromorphic continuation in complexified masses and FI-parameters~\cite{Willett:2011gp}, with reflection pairs canceled before taking singular limits.}

\medskip\noindent{\bf Quiver expression}.
In the main text, we use a quiver representation of the ellipsoid partition function~\cite{Hwang:2024hhy,Benvenuti:2026xcv}, focusing on unitary gauge and flavor groups with bi-fundamental or adjoint chirals. The quiver components are defined as follows:
\begin{equation}
    \begin{tikzpicture}[thick,baseline={([yshift=-.5ex]current bounding box.center)}]
  \tikzset{vertex/.style={circle,fill=white!25,minimum size=12pt,inner sep=2pt}}
  \tikzset{every loop/.style={}}
    \node[vertex] (N) at (0,0) [shape=circle,draw=black,minimum size=2em] {$N$};
    \node at (0,-0.6) {\scriptsize $\t$};
    \end{tikzpicture}
    \;\;\Longleftrightarrow \;\;
    \frac{1}{N!} \int \prod_{i=1}^{N} d\s_i \; e^{2\pi i c_b \t \sum_{i=1}^{N} \s_i} \frac{1}{\prod_{i\neq j}^{N} s_b ( \s_i - \s_j +c_b)} \,,
    \label{eq: circular node}
\end{equation}
\begin{equation}
\begin{tikzpicture}[baseline={([yshift=-.5ex]current bounding box.center)}]
  \tikzset{vertex/.style={circle,fill=white!25,minimum size=12pt,inner sep=2pt}}
  \tikzset{every loop/.style={}}
    \node[vertex] (N) at (0,0) {$N$};
    \node[vertex] (M) at (1.5,0) {$M$};
    \node at (0.7,0.3) {$r$};
    \draw[->-=.55] ([yshift= 0pt] N.east) to ([yshift= 0pt] M.west);
    \end{tikzpicture}
    \;\;\Longleftrightarrow\;\;
    \prod_{i=1}^{N} \prod_{j=1}^{M} 
    s_b \Big( y_j - x_i + c_b(1-r) \Big) \,,
    \qquad\qquad\qquad\quad
    \label{eq: arrow}
\end{equation}
\begin{equation}
    +\big\{r\big\}
    \;\;\Longleftrightarrow\;\;
    s_b \Big((1-r)c_b \Big) \,.
    \qquad\qquad\qquad\qquad\qquad\qquad
    \label{eq: singlet}
\end{equation}
The circular node \eqref{eq: circular node} denotes a $U(N)$ gauge group with FI-parameter $\t$ below it. The arrow \eqref{eq: arrow} represents a bi-fundamental chiral of $U(N)$ and $U(M)$ with fugacities $x_i$ and $y_j$, while a self-loop denotes an adjoint chiral.  $R$-charges are labeled next to the arrows, and singlets are grouped in curly brackets as in \eqref{eq: singlet}. We illustrate this with a few examples:
\begin{itemize}
    \item \medskip\noindent{\bf Elementary dualities}.
The Aharony and BBP dualities in a linear quiver can be expressed as
\begin{equation}
\begin{tikzpicture}[thick,scale=0.9, every node/.style={scale=0.8}]
  \tikzset{vertex/.style={circle,fill=white!25,minimum size=12pt,inner sep=2pt}}
  \tikzset{every loop/.style={}}
    \node[vertex] (n1) at (0-1.5,0) [shape=circle,draw=black,minimum size=2em] {$M$};
    \node[vertex] (n2) at (1.5-1.5,0) [shape=circle,draw=black,minimum height=2em, minimum width=2em] {$N$};
    \node[vertex] (n3) at (3-1.5,0) [shape=circle,draw=black,minimum height=2em, minimum width=2em] {$L$};

    \draw[->-=.5] ([yshift= 2pt] n1.east) to ([yshift= 2pt] n2.west);
    \draw[->-=.5] ([yshift= -2pt] n2.west) to ([yshift= -2pt] n1.east);
    \draw[->-=.5] ([yshift= 2pt] n2.east) to ([yshift= 2pt] n3.west);
    \draw[->-=.5] ([yshift= -2pt] n3.west) to ([yshift= -2pt] n2.east);

    \node at (-0.7-1.5,0) {$\cdots$};
    \node at (3.7-1.5,0) {$\cdots$};
    \node at (0.75-1.5,0.4) {\scriptsize $r_1$};
    \node at (2.25-1.5,0.4) {\scriptsize $r_2$};
    \node at (0-1.5,-0.5) {\scriptsize $\th$};
    \node at (1.5-1.5,-0.5) {\scriptsize $\t$};
    \node at (3-1.5,-0.5) {\scriptsize $\l$};
    
    \node[vertex] (N1) at (0+5,0+1) [shape=circle,draw=black,minimum size=2em] {$M$};
    \node[vertex] (N2) at (1.5+5,0+1) [shape=circle,draw=black,minimum height=2em, minimum width=2em] {$N'$};
    \node[vertex] (N3) at (3+5,0+1) [shape=circle,draw=black,minimum height=2em, minimum width=2em] {$L$};

    \draw[-to, min distance=1cm]  (N1) edge [out=120, in=60] node {} (N1);
    \draw[-to, min distance=1cm]  (N3) edge [out=120, in=60] node {} (N3);
    \draw[->-=.5] ([yshift= 2pt] N1.east) to ([yshift= 2pt] N2.west);
    \draw[->-=.5] ([yshift= -2pt] N2.west) to ([yshift= -2pt] N1.east);
    \draw[->-=.5] ([yshift= 2pt] N2.east) to ([yshift= 2pt] N3.west);
    \draw[->-=.5] ([yshift= -2pt] N3.west) to ([yshift= -2pt] N2.east);
    \draw[->-=.5] ([xshift= 6pt,yshift= -2pt] N1.north) to [bend right=-60] ([xshift= -6pt,yshift= -2pt] N3.north);
    \draw[->-=.5] ([xshift= -9pt,yshift= -4pt] N3.north) to [bend right=60] ([xshift= 9pt,yshift= -4pt] N1.north);

    \node at (-0.7+5,0+1) {$\cdots$};
    \node at (3.7+5,0+1) {$\cdots$};
    \node at (0+5,1.2+1) {\scriptsize $2r_1$};
    \node at (3+5,1.2+1) {\scriptsize $2r_2$};
    \node at (0.75+5,0.25+1) {\scriptsize $1\texttt{-}r_1$};
    \node at (2.25+5,0.25+1) {\scriptsize $1\texttt{-}r_2$};
    \node at (1.5+5,1.2+1) {\scriptsize $r_1 \texttt{+} r_2$};
    \node at (0+5,-0.7+1.2) {\scriptsize $\th+\t$};
    \node at (1.5+5,-0.7+1.2) {\scriptsize $-\t$};
    \node at (3+5,-0.7+1.2) {\scriptsize $\l+\t$};

    \node at (6.5,0.7-0.5) {\scriptsize $+\{M(1\!-\!r_1) \!+\! L (1\!-\!r_2) \!-\!N \!+\! 1 \pm \t\}$};

    \node[vertex] (b1) at (5,0-3+1) [shape=circle,draw=black,minimum size=2em] {$M$};
    \node[vertex] (b2) at (6.5,0-3+1) [shape=circle,draw=black,minimum height=2em, minimum width=2em] {$N''$};
    \node[vertex] (b3) at (8,0-3+1) [shape=circle,draw=black,minimum height=2em, minimum width=2em] {$L$};

    \draw[-to, min distance=1cm]  (b1) edge [out=120, in=60] node {} (b1);
    \draw[-to, min distance=1cm]  (b3) edge [out=120, in=60] node {} (b3);
    \draw[->-=.5] ([yshift= 2pt] b1.east) to ([yshift= 2pt] b2.west);
    \draw[->-=.5] ([yshift= -2pt] b2.west) to ([yshift= -2pt] b1.east);
    \draw[->-=.5] ([yshift= 2pt] b2.east) to ([yshift= 2pt] b3.west);
    \draw[->-=.5] ([yshift= -2pt] b3.west) to ([yshift= -2pt] b2.east);
    \draw[->-=.5] ([xshift= 6pt,yshift= -2pt] b1.north) to [bend right=-60] ([xshift= -6pt,yshift= -2pt] b3.north);
    \draw[->-=.5] ([xshift= -9pt,yshift= -4pt] b3.north) to [bend right=60] ([xshift= 9pt,yshift= -4pt] b1.north);

    \node at (4.3,0-3+1) {$\cdots$};
    \node at (8.7,0-3+1) {$\cdots$};
    \node at (5,1.2-3+1) {\scriptsize $2r_1$};
    \node at (8,1.2-3+1) {\scriptsize $2r_2$};
    \node at (5.75,0.25-3+1) {\scriptsize $1\texttt{-}r_1$};
    \node at (7.25,0.25-3+1) {\scriptsize $1\texttt{-}r_2$};
    \node at (1.5+5,1.2-3+1) {\scriptsize $r_1 \texttt{+} r_2$};
    \node at (5,-0.7-3+1.2) {\scriptsize $\th\!+\!\t \!-\! r_1$};
    \node at (6.5,-0.7-3+1.2) {\scriptsize $\!-\!\t \!+\! 1$};
    \node at (8,-0.7-3+1.2) {\scriptsize $\l\!+\!\t \!-\! r_2$};

    \node at (1.4+5,1.7-3+0.8) {\scriptsize $+\{2- 2\t\}$};
    
    
    \node at (3,1-0.4) 
    {\begin{tikzpicture}
      \node [rotate=45] {\textcolor{black}{$=\!\!=$}};    
    \end{tikzpicture}
    };
    \node at (2.9,1.3-0.4) {\scriptsize Aharony};
    
    \node at (3,-1.5+0.4) 
    {\begin{tikzpicture}
      \node [rotate=135] {\textcolor{black}{$=\!\!=$}};    
    \end{tikzpicture}
    };
    \node at (3,-1.2+0.4) {\scriptsize BBP};
    \node at (2.9,-2+0.4) {\scriptsize if \eqref{eq: B cond}};

\end{tikzpicture}
\label{eq: basic duality in quiver}
\end{equation}
with $N' = M+L-N$ and $N''=M+L-N-1$. Importantly, BBP is applicable only when the monopole superpotential term has  $R$-charge $2$.
\begin{align}
    M(1-r_1) + L(1-r_2) - N + 1 + \t = 2
    \,.
    \label{eq: B cond}
\end{align}

\item \medskip\noindent{\bf The confinement}. 
The s-confinement of $\CC_p[U(N)]$ can also be expressed as
\begin{equation}
\begin{tikzpicture}[thick,scale=1.1, every node/.style={scale=0.8}]
\tikzset{vertex/.style={circle,fill=white!25,minimum size=12pt,inner sep=2pt}}
  \tikzset{every loop/.style={}}
    \node[vertex] (A1) at (0,0) [shape=circle,draw=black,minimum size=2em] {$m_1$};
    \node[vertex] (A2) at (1.2,0) [shape=circle,draw=black,minimum size=2em] {$m_2$};
    \node[vertex] (Acd) at (2.4,0)  {$\cdots$};
    \node[vertex] (Ap) at (3.6,0) [shape=circle,draw=black,minimum size=2em] { \scriptsize $m_{p\texttt{-}1}\!\!$};
    \node[vertex] (AF) at (4.8,0) [shape=rectangle,draw=black,minimum height=2em, minimum width=2em] {$N$};

    \draw[-to, min distance=0.8cm]  (A1) edge [out=120, in=60] node {} (A1);
    \draw[-to, min distance=0.8cm]  (A2) edge [out=120, in=60] node {} (A2);
    \draw[-to, min distance=0.8cm]  (Ap) edge [out=120, in=60] node {} (Ap);
    \draw[->-=.5] ([yshift= 2pt] A1.east) to ([yshift= 2pt] A2.west);
    \draw[->-=.5] ([yshift= -2pt] A2.west) to ([yshift= -2pt] A1.east);
    \draw[->-=.5] ([yshift= 2pt] A2.east) to ([yshift= 2pt] Acd.west);
    \draw[->-=.5] ([yshift= -2pt] Acd.west) to ([yshift= -2pt] A2.east);
    \draw[->-=.5] ([yshift= 2pt] Acd.east) to ([yshift= 2pt] Ap.west);
    \draw[->-=.5] ([yshift= -2pt] Ap.west) to ([yshift= -2pt] Acd.east);
    \draw[->-=.5] ([yshift= 2pt] Ap.east) to ([yshift= 2pt] AF.west);
    \draw[->-=.5] ([yshift= -2pt] AF.west) to ([yshift= -2pt] Ap.east);

    \node at (0.6,0.4) {\scriptsize $\D$};
    \node at (1.8,0.4) {\scriptsize $\D$};
    \node at (3,0.4) {\scriptsize $\D$};
    \node at (4.2,0.4) {\scriptsize $\D$};
    \node at (0,1.1) {\scriptsize $2\texttt{-}2\D$};
    \node at (1.2,1.1) {\scriptsize $2\texttt{-}2\D$};
    \node at (3.6,1.1) {\scriptsize $2\texttt{-}2\D$};
    \node at (5.2,1) {$+\big\{2\D j\big\}_{j=1}^{\min(\a,p-\a)}$};

    \node at (0,-0.5) {\scriptsize $\t_1$};
    \node at (1.2,-0.5) {\scriptsize $\t_2$};
    \node at (3.6,-0.5) {\scriptsize $\t_{p-1}$};

    \node[vertex] (FF) at (7.5,0) [shape=rectangle,draw=black,minimum height=2em, minimum width=2em] {$N$};
    \draw[-to, min distance=0.8cm]  (FF) edge [out=120, in=60] node {} (FF);

    \node at (7.5,1.1) {\scriptsize $2\D$};


    \node at (6.2,0) 
    {\begin{tikzpicture}
      \node [rotate=0] {\Large\textcolor{black}{$=\!\!=$}};    
    \end{tikzpicture}
    };
    
\end{tikzpicture}
\end{equation}
with $\D\equiv\frac{1}{p+1}$. The FI-parameters are fixed as $\t_j = (2-\d_{j,p-\a}\,p)\D$.

\end{itemize}
\section{Details of the Higgsing procedure}\label{app:gkr-higgs}
\noindent
In this appendix, we detail the Higgsing and integrating out massive fields that connect the theory with superpotential $W_{\rm c}$ to that with $W_{\rm e}$: 
\[
W_{\rm c}\xrightarrow{\text{Higgsing}} W_{\rm d}\xrightarrow{\text{Integrating out massive fields}} W_{\rm e}.
\]
We first construct a supersymmetric vacuum and identify the unbroken gauge group. We then fix the broken gauge transformations to isolate the uneaten chiral fluctuations and derive their superpotential $W_{\rm d}$. Next, we determine the $U(1)_R$ symmetry preserved by the vacuum and verify its compatibility with $W_{\rm d}$. Finally, we integrate out the massive chiral fields to obtain $W_{\rm e}$. The subsequent Seiberg duality is discussed in Section~\ref{sec:gkr-final}.

\paragraph{Vacuum and unbroken gauge group.}
To determine the pattern of gauge symmetry breaking, we begin by examining the F-term equations for the $h\times h$ matrices $\Sigma_j^{00}$ in $W_c$:
\begin{equation}\label{eq:gkr-moments}
 \widetilde uY^su=-\delta_{s,k-2}\mathbf{1}_h,\qquad 0\le s\le k-1.
\end{equation}
These conditions require non-zero VEVs of $u$ and $\widetilde u$ for $k\geq 2$,
and also of $Y$ for $k>2$. Let $J_{k-1}$ denote the $(k-1)\times(k-1)$ nilpotent Jordan matrix with ones on the superdiagonal, and let $e_a$ denote the standard basis column vectors. We choose the following vacuum configuration:
\begin{equation}\label{eq:gkr-vacuum}
\begin{gathered}
 Y_0=\begin{pmatrix}J_{k-1}\otimes\mathbf{1}_h&0\\0&0_m\end{pmatrix},\;\;
 u_0=\begin{pmatrix}e_{k-1}\otimes\mathbf{1}_h\\0_{m\times h}\end{pmatrix},\;\;
 \widetilde u_0=-\begin{pmatrix}e_1^T\otimes\mathbf{1}_h&0_{h\times m}\end{pmatrix},
\end{gathered}
\end{equation}
where $m=n-h(k-1)$ and all other fields vanish. The identities $e_1^TJ_{k-1}^se_{k-1}=\delta_{s,k-2}$ and $Y_0^k=0$ ensure that the F-term equations \eqref{eq} are satisfied. This vacuum also satisfies the D-term equations:
\begin{equation}
 [Y_0,Y_0^\dagger]+u_0u_0^\dagger-
 \widetilde u_0^\dagger\widetilde u_0=0,\qquad
 u_0^\dagger u_0-\widetilde u_0\widetilde u_0^\dagger=0.
\end{equation}
Thus, we identify the unbroken gauge group by requiring that the vacuum configuration $(Y_0,u,\tilde u)$ remains invariant under gauge transformations:
\begin{equation}
 g_2=\operatorname{diag}(\underbrace{g_1,\ldots,g_1}_{k-1},g),
 \qquad g_1\in SU(h),\quad g\in SU(m).
\end{equation}
Namely, the unbroken gauge groups are given by
\begin{equation}
 SU(h)\times SU(n)\longrightarrow SU(h)\times SU(m).
\end{equation}
The $SU(h)$ factor acts identically on each of the $k-1$ occupied blocks, while the remaining factor is $SU(m)$.

\paragraph{General fluctuations and gauge fixing.}
Having identified the unbroken gauge group, we now determine the chiral fields that survive the Higgs mechanism.

We begin by parameterizing the general fluctuations around the chosen vacuum as
\begin{equation}
    \begin{aligned}
 &      Y=Y_0+\left(
 \begin{array}{ccc|c}
  A_{11}&\cdots&A_{1,k-1}&C_{1}\\
  \vdots&\ddots&\vdots&\vdots\\
  A_{k-1,1}&\cdots&A_{k-1,k-1}&C_{k-1}\\ \hline
  D_{1}&\cdots&D_{k-1}&Z
 \end{array}\right)   , u=u_0+\begin{pmatrix}
            \alpha_1\\
            \vdots\\
            \alpha_{k-1}\\
           \alpha_R
        \end{pmatrix},\tilde u=\tilde u_0+\begin{pmatrix}
          \beta_1&\cdots&\beta_{k-1}&\beta_R
        \end{pmatrix}.
    \end{aligned}
\end{equation}
Here, $A_{ab}$, $\alpha_a$, and $\beta_a$ are $h\times h$ matrices; $C_a$ and $\beta_R$ are $h\times m$ matrices; $D_b$ and $\alpha_R$ are $m\times h$ matrices; and $Z$ is an $m\times m$ matrix. To identify the fluctuations that can be removed by gauge fixing, we consider complexified gauge transformations with $g_1\in SL(h,\mathbb C)$ and $g_2\in SL(n,\mathbb C)$.\footnote{Since we focus on the holomorphic superpotential and F-flatness, we use complexified gauge transformations to impose holomorphic gauge-fixing conditions for simplicity~\cite{LutyTaylor:1996}.}
\[
 Y\longmapsto g_2Yg_2^{-1},\qquad
 u\longmapsto g_2ug_1^{-1},\qquad
 \widetilde u\longmapsto g_1\widetilde ug_2^{-1}.
\]
Expanding $g_1=\id_h+\eta+\cdots$ and $g_2=\id_n+\epsilon+\cdots$, with $\Tr\eta=\Tr\epsilon=0$, the infinitesimal gauge transformations of the fluctuations around the vacuum are given by ($1\leq a,b\leq k-1$):
\begin{equation}\label{eq:gkr-gauge-shifts}
\begin{aligned}
 \delta_{\rm g}\alpha_a&=\epsilon_{a,k-1}-\delta_{a,k-1}\eta,&
 \delta_{\rm g}\beta_b&=\epsilon_{1b}-\delta_{b1}\eta,\\
 \delta_{\rm g}\alpha_R&=\epsilon_{R,k-1},&
 \delta_{\rm g}\beta_R&=\epsilon_{1R},\\
 \delta_{\rm g}A_{ab}&=\epsilon_{a,b-1}-\epsilon_{a+1,b},&
 \delta_{\rm g}Z&=0,\\
 \delta_{\rm g}C_a&=-\epsilon_{a+1,R}\quad(a<k-1),&
 \delta_{\rm g}C_{k-1}&=0,\\
 \delta_{\rm g}D_b&=\epsilon_{R,b-1}\quad(b>1),&
 \delta_{\rm g}D_1&=0.
\end{aligned}
\end{equation}
A convenient gauge choice consistent with these transformations is\footnote{We briefly justify this gauge choice. For instance, under the transformation
\[
\alpha_R \longmapsto
\alpha_R+\epsilon_{R,k-1},
\]
we can set $\alpha_R=0$ by choosing $\epsilon_{R,k-1}=-\alpha_R$. In contrast, $\delta_g Z=0$ implies that $Z$ is invariant under these gauge transformations and thus cannot be removed by gauge fixing. We can therefore impose the gauge conditions described above, retaining $Z$, $C_{k-1}$, and $D_1$, together with the $\beta_b$ fluctuations discussed below.

Importantly, a non-zero shift of an individual component does not guarantee that it can be removed independently, as the same gauge parameter may act on multiple components. In particular, although the $\beta_b$ have non-zero shifts in~\eqref{eq:gkr-gauge-shifts}, the corresponding gauge parameters are already used to impose the gauge conditions on $\alpha_a$ and $A_{ab}$. Hence, the $\beta_b$ must also be retained at this stage.}
\begin{equation}\label{eq:gkr-gauge-conditions}
\begin{gathered}
 \underbrace{\alpha_a=0}_{(k-1)h^2},\qquad
 \underbrace{\alpha_R=0}_{hm},\qquad
 \underbrace{\beta_R=0}_{hm},\\[8pt]
 \underbrace{A_{ab}=0\quad(a<k-1)}_{(k-2)(k-1)h^2},\qquad
 \underbrace{C_a=0\quad(a<k-1)}_{(k-2)hm},\qquad
 \underbrace{D_b=0\quad(b>1)}_{(k-2)hm}
 \,,
\end{gathered}
\end{equation}
where the underbraces indicate the numbers of complex chiral components removed by gauge fixing. To verify that these account for all broken gauge directions, we sum their contributions:
\begin{equation}\label{eq:gkr-eaten-count}
\begin{aligned}
 N_{\rm eaten}
 &=(k-1)h^2+2hm+(k-2)(k-1)h^2+2(k-2)hm\\
 &=n^2-m^2
 \,,
\end{aligned}
\end{equation}
which equals $\dim[SU(h)\times SU(n)]-\dim[SU(h)\times SU(m)]$, precisely the number of broken gauge generators. Thus, the gauge-fixing conditions remove exactly the chiral components absorbed by the massive vector multiplets. The remaining fields can be organized as follows:
\begin{equation}\label{eq:gkr-unconstrained-slice}
\begin{gathered}
 Y=\left(
 \begin{array}{ccccc|c}
  0&\id_h&0&\cdots&0&0\\
  0&0&\id_h&\ddots&\vdots&\vdots\\
  \vdots&\vdots&\ddots&\ddots&0&\vdots\\
  0&0&\cdots&0&\id_h&0\\
  A_{k-1,1}&A_{k-1,2}&\cdots&A_{k-1,k-2}&A_{k-1,k-1}&c\\ \hline
  \widetilde c&0&\cdots&0&0&Z
 \end{array}\right),\\[6pt]
 u=u_0=\begin{pmatrix}e_{k-1}\otimes\id_h\\0_{m\times h}\end{pmatrix},
 \qquad
 \widetilde u=\left(
 \begin{array}{cccc|c}
  -\id_h+\beta_1&\beta_2&\cdots&\beta_{k-1}&0_{h\times m}
 \end{array}\right).
\end{gathered}
\end{equation}

\paragraph{Superpotential and uneaten fields.} We now express the superpotential $W_{\rm d}$ in terms of the remaining uneaten chiral fields. The ordinary quarks can be decomposed into the same occupied blocks and the remaining $m$-dimensional block.
\begin{equation}
 v_I=(a_{I,1},\ldots,a_{I,k-1},p_I)^T,\qquad
 \widetilde v_I=(\widetilde a_{I,1},\ldots,\widetilde a_{I,k-1},\widetilde p_I),
 \quad I=1,2
 \,,
\end{equation}
where the matrices $a_{I,a}$ and $p_I$ have dimensions $h\times F_I$ and $m\times F_I$, respectively, while their conjugates $\tilde a_{I,a}$ and $\tilde p_I$ have dimensions $F_I\times h$ and $F_I\times m$. Since the VEVs of $v_I$ and $\tilde v_I$ vanish, these fields are not eaten by the Higgs mechanism. Substituting the block decompositions of the fields into $W_{\rm c}$ and imposing the gauge-fixing conditions, we obtain the superpotential

\enlargethispage{2\baselineskip}

{\allowdisplaybreaks[4]
\begin{align}
W_{\rm d}={}&\Tr_m Z^{k+1}
 +(k+1)\Tr_h\!\left[(A_{k-1,k-1}c+cZ)\widetilde c\right]\notag\\
&+\sum_{\ell=1}^{k+1}\frac{k+1}{\ell}
 \sum_{\substack{1\le b_1,\ldots,b_\ell\le k-1\\
                 \sum_{\nu=1}^\ell(k-b_\nu)=k+1}}
\Tr_h\!\left(A_{k-1,b_1}\cdots A_{k-1,b_\ell}\right)
 -\Tr_h\!\left(\Sigma_0^{00}A_{k-1,k-1}\right)\notag\\
&+M\widetilde qq+\Sigma_0^{10}q   +\widetilde q\Sigma_0^{01} 
+\sum_{I=1}^2\Tr_h\!\left(\Sigma_0^{0I}\widetilde p_I\widetilde c\right)\notag\\
\displaybreak[2]
&+\sum_{j=0}^{k-1}\sum_{a=1}^{k-1}\sum_{\ell=0}^{k-1}
 \sum_{\substack{1\le b_1,\ldots,b_\ell\le k-1\\
                 \sum_{\nu=1}^\ell(k-b_\nu)=a-j}}
\Tr_h\!\left[
 \left(\Sigma_j^{00}\beta_a
       +\sum_{I=1}^2\Sigma_j^{0I}\widetilde a_{I,a}\right)
 A_{k-1,b_1}\cdots A_{k-1,b_\ell}\right]\notag\\
&+\sum_{j=1}^{k-1}\sum_{I=1}^2\sum_{a=1}^j
 \left(\Sigma_j^{I0}(-\delta_{a1}\id_h+\beta_a)
       +\sum_{J=1}^2\Sigma_j^{IJ}\widetilde a_{J,a}\right)
 a_{I,a+k-1-j}\notag\\
\displaybreak[1]
&+\sum_{j=0}^{k-2}\sum_{I=1}^2
 \sum_{\substack{j+1\le a\le k-1\\0\le t\le a-j-1}}
 \sum_{\ell=0}^{a-j-1-t}
 \sum_{\substack{1\le b_1,\ldots,b_\ell\le k-1\\
                 \sum_{\nu=1}^\ell(k-b_\nu)=a-j-1-t}}
\left(\Sigma_j^{I0}(-\delta_{a1}\id_h+\beta_a)
       +\sum_{J=1}^2\Sigma_j^{IJ}\widetilde a_{J,a}\right)\notag\\
&\qquad\times A_{k-1,b_1}\cdots A_{k-1,b_\ell}\left[
       \sum_{b=t+1}^{k-1}A_{k-1,b-t}a_{I,b}
       +cZ^tp_I
       +\sum_{b=1}^t cZ^{t-b}\widetilde c\,a_{I,b}
                 \right]\notag\\
&+\sum_{j=0}^{k-1}\sum_{I,J=1}^2
 \Sigma_j^{IJ}\widetilde p_J \left(Z^{k-1-j}p_I
       +\sum_{b=1}^{k-1-j}Z^{k-1-j-b}\widetilde c\,a_{I,b}\right).
\label{eq: Wd}
\end{align}
}
This contains additional massive fields to be eliminated. Before doing so, we determine the  $R$-charges after Higgsing.

\paragraph{The unbroken $U(1)_R$ symmetry after Higgsing.}

To identify the $U(1)_R$ symmetry preserved by the chosen vacuum, we combine the original $U(1)_R$ transformation with a compensating gauge transformation~\cite{FarrarWeinberg:1983R,CordovaDumitrescu:2024Higgs}, requiring that the combined transformation leaves the VEVs $Y_0$, $u_0$, and $\widetilde u_0$ invariant. Fields with vanishing VEVs may carry non-zero  $R$-charges, while gauge singlets retain their original  $R$-charges.
 
Let $\theta$ denote the constant transformation parameter of the original $U(1)_R$ symmetry, acting as
\begin{align}
    Y\longmapsto e^{\frac{2i\theta}{(k+1)}}Y,\qquad
 u\longmapsto e^{\frac{(2-k)i\theta}{(k+1)}}u,\qquad
 \widetilde u\longmapsto e^{\frac{(2-k)i\theta}{(k+1)}}\widetilde u.
\end{align}
To compensate for these phases, we take $g_1=\id_h\in SU(h)$ and $g_2=\gamma(\theta)\in SU(n)$ in the form
\begin{equation}\label{eq:gamma}
  \gamma(\theta)=
\begin{pmatrix}
 e^{i\theta t_1}\mathbf1_h & 0 & \cdots & 0 \\
 0 & \ddots & & \vdots \\
 \vdots & & e^{i\theta t_{k-1}}\mathbf1_h & 0 \\
 0 & \cdots & 0 & \mathbf1_m
\end{pmatrix},\qquad t_a\in\mathbb R.
\end{equation}
The phases $t_a$ are constrained only by the non-zero vacuum entries, which must remain invariant for arbitrary $\theta$. We therefore determine $t_a$ by imposing the invariance of $Y_0$, $u_0$, and $\tilde u_0$ as follows.

We first use the constraint of $Y_0$, which transforms as
\begin{align}
    Y_0\longmapsto e^{\frac{2i\theta}{(k+1)}}\gamma Y_0\gamma^{-1}
 \,.
\end{align}
Since $(Y_0)_{a,a+1}=\id_h$, its $(a,a+1)$ block acquires the phase
\begin{align}
    e^{i\theta[\,\frac{2}{(k+1)}+t_a-t_{a+1}\,]}
 \,,
\end{align}
and preserving this block for arbitrary $\theta$ requires
\begin{equation}\label{eq:ta}
 t_{a+1}-t_a=\frac2{k+1},\qquad 1\le a\le k-2.
\end{equation}
This gives a recursion relation between adjacent phases. The endpoint values $t_1$ and $t_{k-1}$ are then fixed by requiring the invariance of the quark VEVs $u_0$ and $\tilde u_0$. Since $R(u)=R(\widetilde u)=\frac{(2-k)}{(k+1)}$, the combined transformations take the form
\begin{align}
    u_0\longmapsto e^{\frac{i\theta(2-k)}{(k+1)}}\gamma u_0,\qquad
 \widetilde u_0\longmapsto
 e^{\frac{i\theta(2-k)}{(k+1)}}\widetilde u_0\gamma^{-1}
 \,,
\end{align}
and for their non-zero entries $(u_0)_{k-1}=\id_h$ and
$(\widetilde u_0)_1=-\id_h$,
\begin{align}
    &(u_0)_{k-1}\longmapsto e^{i\theta[\,\frac{(2-k)}{(k+1)}+t_{k-1}\,]}\id_h,
    \nonumber
    \\
 &(\widetilde u_0)_1\longmapsto -e^{i\theta[\,\frac{(2-k)}{(k+1)}-t_1\,]}\id_h.
\end{align}
Requiring these entries to remain invariant for arbitrary $\theta$, we obtain
\begin{equation}\label{eq:tk+1}
 \frac{2-k}{k+1}+t_{k-1}=0,\qquad
 \frac{2-k}{k+1}-t_1=0.
\end{equation}
Consequently, we determine
\[
 t_a=\frac{2a-k}{k+1},
 \qquad 1\le a\le k-1.
\]
Substituting these into \eqref{eq:gamma} gives the compensating transformation
\begin{equation}
  \gamma(\theta)=
\begin{pmatrix}
 e^{i\theta \frac{2-k}{k+1}}\mathbf1_h & 0 & \cdots & 0 \\
 0 & \ddots & & \vdots \\
 \vdots & & e^{i\theta \frac{k-2}{k+1}}\mathbf1_h & 0 \\
 0 & \cdots & 0 & \mathbf1_m
\end{pmatrix}.
\end{equation}

Now, let us read the  $R$-charges of the fields after the Higgsing. For $1\le b\le k-1$, the fluctuations transform as
\begin{align}
    Z&\longmapsto e^{\frac{2i\theta}{k+1}}Z,\nonumber\\
 c&\longmapsto e^{i\theta[\,\frac{2}{k+1}+t_{k-1}\,]}c,\nonumber\\
 \widetilde c&\longmapsto
 e^{i\theta[\,\frac{2}{k+1}-t_1\,]}\widetilde c,\nonumber\\
 A_{k-1,b}&\longmapsto
 e^{i\theta[\,\frac{2}{k+1}+t_{k-1}-t_b\,]}A_{k-1,b},\nonumber\\
 \beta_b&\longmapsto
 e^{i\theta[\,\frac{2-k}{k+1}-t_b\,]}\beta_b.
\end{align}
Substituting the values of $t_a$ therefore gives:
\begin{align}
    &R(Z)=\frac2{k+1},\quad
 R(c)=\frac2{k+1}+\frac{k-2}{k+1}=\frac{k}{k+1},\quad
 R(\widetilde c)=\frac2{k+1}-\frac{2-k}{k+1}=\frac{k}{k+1},
 \nonumber\\
 &R(A_{k-1,b})=\frac{2(k-b)}{k+1},\quad
 R(\beta_b)=\frac{2(1-b)}{k+1}.
\end{align}
In Figure~\ref{fig:gkr-d} and Table~\ref{tab:gkr-d}, the $k-2$ blocks $A_{k-1,b}$ with $1\le b\le k-2$ are grouped together, while $A_{k-1,k-1}$ is listed separately. Each block transforms as $\mathrm{Adj}_{SU(h)}\oplus\mathbf1$ and as a singlet under the other gauge and flavor factors. The first set is empty at $k=2$.

The same compensating transformation also acts on the quark decomposition as
\begin{align}
    a_{I,a}&\longmapsto e^{i\theta \frac{2a-k}{k+1}}a_{I,a},&
 \widetilde a_{I,a}&\longmapsto e^{-i\theta \frac{2a-k}{k+1}}\widetilde a_{I,a},
 &&1\le a\le k-1,\nonumber\\
 p_I&\longmapsto p_I,&
 \widetilde p_I&\longmapsto\widetilde p_I.
\end{align}
Hence, the  $R$-charges of $a_{I,a}$ and $\widetilde a_{I,a}$ receive
gauge contributions $+\frac{2a-k}{k+1}$ and $-\frac{2a-k}{k+1}$, respectively. In contrast, the blocks $p_I$ and $\widetilde p_I$ receive no gauge contributions. Adding these contributions to the original charges
$R(v_I)=R(\widetilde v_I)=\frac{I}{(k+1)}-r_I$, for $I=1,2$, gives the resulting  $R$-charges
\begin{align}
    R(a_{1,a})&=\frac{2a-k+1}{k+1}-r_1,&
 R(\widetilde a_{1,a})&=\frac{k-2a+1}{k+1}-r_1,&
 R(p_1)=R(\widetilde p_1)&=\frac1{k+1}-r_1,\nonumber\\
 R(a_{2,a})&=\frac{2a-k+2}{k+1}-r_2,&
 R(\widetilde a_{2,a})&=\frac{k-2a+2}{k+1}-r_2,&
 R(p_2)=R(\widetilde p_2)&=\frac2{k+1}-r_2.
\end{align}
The resulting  $R$-charges are summarized in Table~\ref{tab:gkr-d}. Note that every term in $W_{\rm d}$ has  $R$-charge $2$ under the combined $U(1)_R$ transformation.
 
\paragraph{Integrating out massive fields.}
Now, let us return to \eqref{eq: Wd} and solve the F-term equations
\begin{equation}\label{eq:gkr-heavy-eoms}
\begin{gathered}
 \frac{\partial W_{\rm d}}{\partial\Sigma_j^{00}}=0\quad(0\le j<k),\qquad
 \frac{\partial W_{\rm d}}{\partial A_{k-1,b}}
 =\frac{\partial W_{\rm d}}{\partial\beta_b}=0\quad(1\le b<k),\\
 \frac{\partial W_{\rm d}}{\partial\Sigma_j^{0I}}
 =\frac{\partial W_{\rm d}}{\partial\Sigma_j^{I0}}=0
 \quad(1\le j<k,\ I=1,2),\\
 \frac{\partial W_{\rm d}}{\partial a_{I,b}}
 =\frac{\partial W_{\rm d}}{\partial\widetilde a_{I,b}}=0
 \quad(1\le b<k,\ I=1,2),\\
 \frac{\partial W_{\rm d}}{\partial q}
 =\frac{\partial W_{\rm d}}{\partial\widetilde q}
 =\frac{\partial W_{\rm d}}{\partial\Sigma_0^{01}}
 =\frac{\partial W_{\rm d}}{\partial\Sigma_0^{10}}=0,
\end{gathered}
\end{equation}
which admit the following solution:
\begin{equation}\label{eq:gkr-heavy-solution}
\begin{gathered}
 A_{k-1,b}=\beta_b=a_{I,b}=\widetilde a_{I,b}=0,\\
 q=cp_1,\qquad\widetilde q=-\widetilde p_1\widetilde c,\qquad
 \Sigma_0^{10}=M\widetilde p_1\widetilde c,\qquad
 \Sigma_0^{01}=-cp_1M,\\
 \Sigma_0^{00}=(k+1)c\widetilde c,\\
 \Sigma_b^{0J}=-\sum_{I=1}^2\sum_{j=0}^{b-1}
 cZ^{b-1-j}p_I\Sigma_j^{IJ},\qquad
 \Sigma_b^{I0}=\sum_{J=1}^2\sum_{j=0}^{b-1}
 \Sigma_j^{IJ}\widetilde p_JZ^{b-1-j}\widetilde c,\\
 \Sigma_b^{00}=-\sum_{I=1}^2\sum_{j=0}^{b-1}
 cZ^{b-1-j}p_I\Sigma_j^{I0},\qquad 1\le b<k.
\end{gathered}
\end{equation}
Substituting this into $W_{\rm d}$ considerably simplifies the superpotential
\begin{equation}\label{eq:gkr-W4}
\begin{aligned}
 &W_{\rm e}= \Tr_m Z^{k+1}\\
 &+(k+1)\Tr_m(Z\widetilde cc)-\Sigma_0^{20}cp_2+\widetilde p_2\widetilde c\Sigma_0^{02}
 -M\widetilde p_1\widetilde ccp_1 +\sum_{j=0}^{k-1}\sum_{I,J=1}^2\Sigma_j^{IJ}\widetilde p_JZ^{k-1-j}p_I 
 \,,
\end{aligned}
\end{equation}
which no longer contains adjoint chiral fields under the $SU(h)$ gauge group, allowing us to apply Seiberg duality in the next step, as shown in section \ref{sec:gkr-final}.

\bibliographystyle{JHEP}
\bibliography{ref}

\end{document}